\documentclass[%
reprint,
twocolumn,
 amsmath,amssymb,
 aip,
 jcp,
floatfix,
]{revtex4-2}

\usepackage[T1]{fontenc}
\usepackage{float}
\usepackage{braket}
\usepackage{graphicx}
\usepackage{dcolumn}
\usepackage{bm}
\usepackage{inputenc}
\usepackage{blindtext}
\usepackage{color}
\usepackage{booktabs,siunitx}
\usepackage{todonotes}
\usepackage{comment}
\usepackage{makecell}

\usepackage{hyperref}
\hypersetup{colorlinks,citecolor=blue}

\newcolumntype{~}{>{\global\let\currentrowstyle\relax}}
\newcolumntype{^}{>{\currentrowstyle}}

\DeclareMathAlphabet\mathbfcal{OMS}{cmsy}{b}{n}

\begin{document}

\newcommand{\eigvec}{\ensuremath{\pmb{\epsilon}}}
\newcommand{\Fnoise}{\ensuremath{\delta\mathbf{F}}}
\newcommand{\Fe}{\ensuremath{\mathcal{F}}}
\newcommand{\Ra}{\ensuremath{\mathcal{R}}}
\newcommand{\Feff}{\ensuremath{\widetilde{\mathcal{F}}}}
\newcommand{\Veff}{\ensuremath{\widetilde{V}}}
\newcommand{\Zpart}{\ensuremath{\mathcal{Z}}}
\newcommand{\kBT}{\ensuremath{\mathrm{k}_\mathrm{B}\mathrm{T}}}
\newcommand{\proj}{\ensuremath{\mathcal{P}}}
\newcommand{\Proj}{\ensuremath{\mathbfcal{P}}}
\newcommand{\qroj}{\ensuremath{\mathcal{Q}}}
\newcommand{\Qroj}{\ensuremath{\mathbfcal{Q}}}
\newcommand{\bbl}{\ensuremath{\big(}}
\newcommand{\bbr}{\ensuremath{\big)}}
\newcommand{\liou}{\ensuremath{\mathcal{L}}}
\renewcommand\vec{\mathbf}
\newcommand{\qvec}{\vec{q}}
\newcommand{\umass}{\tilde{u}}
\newcommand{\pmass}{\tilde{p}}
\newcommand{\fmass}{\tilde{f}}
\newcommand*{\bigchi}{\raisebox{2pt}{\mbox{\large$\chi$}}}

\preprint{AIP/123-QED}

\title{Mode-coupling theory of lattice dynamics for classical and quantum crystals}


\author{Alo\"is Castellano}
\author{J. P. Alvarinhas Batista}
\author{Matthieu J. Verstraete}

\affiliation{Nanomat group, QMAT center, CESAM research unit and European Theoretical Spectroscopy Facility, Université de Liège, allée du 6 août, 19, B-4000 Liège, Belgium}



\begin{abstract}
The dynamical properties of nuclei, carried by the concept of phonon quasiparticles (QP), are central to the field of condensed matter.
While the harmonic approximation can reproduce a number of properties observed in real crystals, the inclusion of anharmonicity in lattice dynamics is essential to accurately predict properties such as heat transport or thermal expansion.
For highly anharmonic systems, non perturbative approaches are needed, which result in renormalized theories of lattice dynamics.
In this article, we apply the Mori-Zwanzig projector formalism to derive an exact generalized Langevin equation describing the quantum dynamics of nuclei in a crystal.
By projecting this equation on quasiparticles in reciprocal space, and with results from linear response theory, we obtain a formulation of vibrational spectra that fully accounts for the anharmonicity.
Using a mode-coupling approach, we construct a systematic perturbative expansion in which each new order is built to minimize the following ones.
With a truncation to the lowest order, we show how to obtain a set of self-consistent equations that can describe the lineshapes of quasiparticles.
The only inputs needed for the resulting set of equations are the static Kubo correlation functions, which can be computed using (fully quantum) path-integral molecular dynamics or approximated with (classical or ab initio) molecular dynamics.
We illustrate the theory with an application on fcc \textsuperscript{4}He, an archetypal quantum crystal with very strong anharmonicity.
\end{abstract}

\maketitle

\section{Introduction}
\label{sec:Introduction}

In the condensed phases of matter, many physical properties are profoundly related to the vibrations of nuclei.
The understanding of thermodynamical, dynamical or transport phenomena in crystals requires the inclusion of the correlated motion of atoms around their equilibrium positions.
To describe such dynamics, the cornerstone of modern theories is the harmonic approximation~\cite{Born1954}.
Using a second-order Taylor expansion of the Born-Oppenheimer surface (BOS), this model describes the lattice vibrations as a non-interacting quantum gas of quasiparticles called phonons.
Despite its simplicity, the harmonic approximation has been tremendously successful in explaining a large number of phenomena observed in the solid state, including excitation spectra, the phase stability of various materials, elastic properties or even zero point motion~\cite{Dove1993,Fultz2010}.
Consequently, it naturally became one of the most important tools in the field of condensed matter.
Nevertheless, the early truncation of the BOS Taylor expansion can be a severe limitation to the accuracy of the harmonic approximation~\cite{Klein1972,Fultz2010,Knoop2020}.
In some cases these inaccuracies are only quantitative, but many phenomena cannot be described even qualitatively with a purely harmonic theory of lattice dynamics.
Indeed, the higher order terms of the Taylor expansion, which are denoted as anharmonic, are responsible for a number of properties that cannot be predicted at all with the non-interacting phonon picture.
Two of the most striking examples of these limitations are probably the dynamical stabilisation of some structures~\cite{Grimvall2012}, and  the computation of the thermal conductivity, which diverges in the harmonic approximation~\cite{Dove1993}.
These limitations are a good illustration of the intrinsic problem with the non-interacting phonon picture.
Indeed, within the harmonic approach, phonons are quasiparticles with temperature-independent energies and infinite lifetimes.
Such features are incompatible with the frequency changes associated with a dynamical stabilisation, or the finite lifetimes associated with a finite thermal conductivity.
In order to reconcile the temperature-dependent frequencies with a harmonic approximation, the quasi-harmonic approximation is often used~\cite{Fultz2010,Allen2019}, in which the temperature evolution is included indirectly through a volume dependence.
Unfortunately, this approach is biased in the type of anharmonicity it includes\cite{Glensk2015}, is often insufficient to accurately describe the frequency evolution, and is still unable, by construction, to incorporate a linewidth~\cite{Kim2020}.

Nevertheless, a quasiparticle picture still seems to be an adequate representation of the finite temperature vibrational excitations for most solids.
As a matter of fact, in most vibrational spectroscopy measurements on pure crystals one can observe reasonably sharp Lorentzian-like resonances~\cite{Allen2015}.
These resonances can be associated with vibrational quasiparticles, and it is these excitations that are approximated with harmonic phonons.
However, in contrast with the harmonic picture, these quasiparticles have temperature-dependent frequencies $\Omega_{\nu}(T)$, as well as a linewidth $\Gamma_{\nu}(T)$, which can be interpreted as a manifestation of their finite lifetimes.

In recent years, a number of frameworks have been developed to account for the anharmonicity of materials.
For example, it is nowadays common to use perturbation theory~\cite{Maradudin1962,Pathak1965,Cowley1968} to describe the effects of finite temperature on phonons within an \textit{ab initio} setting~\cite{Broido2005,Togo2015,Li2014,Ravichandran2020}.
In this approach, the anharmonic contributions to the BOS are responsible for phonon-phonon interactions, which leads to temperature-dependent frequencies as well as the finite lifetimes missing in the harmonic approximation.
However, if perturbation theory can improve the agreement between harmonic phonons and more realistic vibrational QP, it is based on the assumption that the higher order anharmonic contributions will be of progressively lower importance, the (T = 0K) harmonic contribution being dominant.
This hypothesis is often too drastic, and perturbative approaches are consequently known to fail in some cases~\cite{Sun2010a,Sun2010b,Xia2020,Yang2022b}.
Moreover, a significant drawback of such approaches is that they can not be applied at all in strongly anharmonic systems with unstable harmonic-level phonons.
These failures suggest that approaches beyond standard perturbation theory are necessary to obtain a satisfactory description of vibrational properties of materials.

Starting from this observation, a significant number of works have focused on developing non-perturbative methods to treat anharmonicity without the drawbacks of the perturbative expansion.
Most of these approaches are constructed on the idea of renormalized phonons and phonon-phonon interactions.
The foremost non-perturbative theory of anharmonicity is the Self-Consistent Harmonic Approximation (SCHA)~\cite{Koehler1966,Werthamer1970,Ravichandran2018,Esfarjani2020,Souvatzis2009,Tadano2015,Tadano2018,Bianco2017,Monacelli2021,van_Roekeghem2021}, also known as self consistent phonons, first imagined by Born and since then rederived several times~\cite{Klein1972}.
The idea behind the SCHA is that interactions between harmonic phonons renormalize their frequencies.
Using the Gibbs-Bogoliubov free energy inequality~\cite{Choquard1967,Werthamer1970,Monacelli2021}, a Dyson equation for the Green function~\cite{Tadano2015} or even a path-integral formalism~\cite{Samathiyakanit1973}, the SCHA can be derived as a set of self-consistent equations which uses the density matrix of an effective harmonic Hamiltonian.
In essence, the renormalized phonons are obtained from an average computed on the distribution associated with the converged Hamiltonian.
However, the quadratic form of the effective Hamiltonian means that the average is always computed on a multivariate Gaussian distribution, which can differ substantially from the true distribution.
For instance, the renormalized phonons obtained as a result of the self-consistent equations still need a correction to approximate the QP frequencies~\cite{Choquard1967,Bianco2017}.

Fortunately, there exists a clear way to lift the multivariate Gaussian distribution approximation.
Indeed, given a potential $V(\vec{R})$, which can be computed using \textit{ab initio} methods such as DFT or a classical model, it is possible to sample the exact canonical distribution in the classical limit for a given temperature $T$ with methods such as molecular dynamics (MD) or Markov-chain Monte-Carlo.
Consequently, it seems natural to use these methods to obtain a non-perturbative description of lattice dynamics to go beyond the effective harmonic distribution of the SCHA.
For instance, it has been proposed to use the displacements covariance tensor computed from MD to map it on an effective harmonic model through effective interatomic force constants (IFC)~\cite{Levy1984,Kong2009,Kong2011}.
However, the prevailing MD-based lattice dynamics method is probably the temperature dependent effective potential (TDEP)~\cite{Hellman2011,Hellman2013a,Hellman2013b}.
This method uses configurations extracted from MD to fit an effective harmonic potential as well as third and sometimes fourth order anharmonic terms~\cite{Hellman2013b}.
An important feature of this potential is that each new order is fitted on the residual forces from the previous order~\cite{Klarbring2020}.
The goal of this procedure is to minimize the importance of high order contributions compared to the effective harmonic one.
By using perturbation theory on top of this effective anharmonic Hamiltonian, TDEP has proven its ability to describe anharmonic properties with a great number of applications on the free energy and transport of materials~\cite{Hellman2011,Hellman2013a,Hellman2013b,Mei2015,Romero2015,Chaney2021}.
Unfortunately, the renormalized anharmonic Hamiltonian at the heart of TDEP lacks a theoretical justification.
This feature hinders a clear vision of its limits and raises questions on the justification for the use of perturbation theory with the effective Hamiltonian.

For instance, an important question raised by this approach is the easily overlooked problem of going from the classical description of MD to the quantum description of perturbation theory.
Indeed, the QP description of a many-body system is intrinsically constructed on correlation functions.
However, while the definition of classical correlation functions is unambiguous, this is not the case for a quantum system, where there exist an infinite number of ways to define them~\cite{Hele2017}.
Consequently, the classical to quantum transformation of correlation functions is a non-trivial task that needs strong justifications~\cite{Schofield1960,Egorov1999,Ramirez2004}.
It would seem that to dispose of this problem, a solution could be to use the true quantum distribution of displacements using for example path-integral molecular dynamics.
Unfortunately, the freedom of choice in how to define quantum correlation functions is still an obstacle to a clear vision of how to apply TDEP.
Indeed, from all the different types of quantum correlation functions, which one should be used as input for the least-squares problem ?
In a recent publication, it has been proposed to use the standard correlation functions~\cite{Geng2022}.
However, this choice was justified with the goal of having a good description of the BOS, which unfortunately doesn't answer the remaining questions on how to approximate the QP with the effective Hamiltonian.

A more satisfying answer comes from the recent method introduced by Morresi \textit{et al}~\cite{Morresi2021,Morresi2022} which uses the static Kubo correlation function (KCF) as input to obtain the QP.
With simple manipulations, the authors were able to show that QPs computed this way match the exact zero frequency of the Matsubara QP Green's function.
While their QPs feature an infinite lifetime, this result is a strong suggestion that the KCF might be an answer to the questions raised previously.
From this type of correlation functions, the missing step is how to obtain the dynamical QP properties, but the use of the KCF points to the field of linear response theory, in which established theoretical tools exist to solve such problems.

In this work, we build a foundation for an exact non-perturbative theory of lattice dynamics that can describe strongly anharmonic quantum systems.
With this objective, we derive a dynamical theory for solids based on the Mori-Zwanzig projector formalism of linear response theory~\cite{Mori1965,Zwanzig1961,Kubo1966,Hijn2010,Carof2014,Li2015,Fiorentino2022}.
This approach projects the dynamics on the slow variables of the system, resulting in a generalized Langevin equation (GLE) and thus splitting the equation of motion for the KCF into a conservative and in a memory-dependent dissipative part.
After defining a projection on quasiparticle operators in reciprocal space, we use linear response theory to obtain an exact formulation of the spectral function.
By using a mode-coupling theory~\cite{Bosse1978,Reichman2005,Markland2012,Janssen2018} to approximate the memory kernel inherent to the Mori-Zwanzig formalism, we derive a theory in which the renormalized interatomic force constants of all orders appear naturally.
This approximation allows us to obtain a set of self-consistent equations to compute vibrational spectra, for which we also propose a simplification based on a scattering approximation.
Moreover, we show how our findings can justify a theory of temperature dependent phonons, thus giving an explanation of the success of renormalized phonon methods.
We also discuss the implications of the method, showing for instance how it can be used to approximate nuclear quantum effects on classical simulations, while alleviating ambiguities related to the multiple definitions of quantum correlation functions.
We show that well established methods such as perturbation theory or the SCHA can be recovered as approximations of our mode-coupling theory, and that TDEP, used with (PI)MD, is a direct application of our approach. 
Another interesting result of the present theory is that it shows that the often-used Lorentzian approximation of spectral lineshapes can never be an exact representation.

The paper is organized as follows.
In section \ref{sec:Derivation}, we introduce the notation used throughout the paper as well as the Kubo correlation functions (KCF), before deriving the generalized Langevin equation for the displacement KCF, using the Mori-Zwanzig projector formalism.
We then construct the projection in reciprocal space in section \ref{sec:Quasiparticle space}, before using the GLE to derive exact equations describing the spectral properties in section \ref{sec:Vibrational spectral}.
Using a mode-coupling approach, we then derive in section \ref{sec:Mode-coupling} a set of approximations to the memory kernel appearing in the GLE.
Next, we discuss how to compute from \textit{ab initio} calculations the quantities needed as input for the method in section \ref{sec:Simulations}.
Section \ref{sec:Discussion} is devoted to the discussion of the classical limit as well as the comparison to other theories of anharmonicity.
We illustrate the theory with an application on \textsuperscript{4}He in section \ref{sec:Applications} before concluding in section \ref{sec:Conclusion}.

\section{Derivation of the Generalized Langevin Equation}
\label{sec:Derivation}

\subsection{Notations}
In this work, we consider a three-dimensional crystal in the Born-Oppenheimer approximation, so that the Hamiltonian governing the dynamics of the system is taken with the general form
\begin{equation}
\label{eq:Hamiltonian}
    H = \sum_{i} \frac{p_{i}^2}{2M_i} + V(\vec{R})
\end{equation}
where $\vec{R}$ and $\vec{p}$ are respectively the position and momentum operators, and $V(\vec{R})$ is the potential.
In this work, we will consider that there is no atomic diffusion in the system, so that we can use the displacements $\vec{u}_i = \vec{R}_i - \braket{\vec{R}_i}$ as dynamical variables.
In this definition, $\braket{O(t)}$ is the thermal average of the operator $O$, meaning that $\braket{\vec{R}}$ are the equilibrium positions.
It should be noted that with this definition, the equilibrium positions, and consequently the definitions of displacements, are temperature dependent.
While it is common to expand the potential in eq.(\ref{eq:Hamiltonian}) in a Taylor expansion, here we make no assumptions on its form, apart from its time independence.

To describe the time evolution of the system, we use the Heisenberg representation, where the time dependence and derivatives of operators are obtained with
\begin{equation}
\begin{split}
    A(t) =& e^{\frac{i}{\hbar}Ht} A e^{-\frac{i}{\hbar}Ht} = e^{i\liou t} A \\
    \dot{A}(t) =& \frac{i}{\hbar} \big[ H, A(t) \big] = i\liou A(t)
\end{split}
\end{equation}
where $A = A(0)$ and $\liou = \frac{1}{\hbar}\big[H,\cdot]$ is the quantum Liouville operator.
We will denote $p_i(t) = M_i \dot{u}_i(t)$ the time derivative of the displacement of atom $i$ with mass $M_i$.
Furthermore, we will define mass scaled coordinates $\umass_i = \sqrt{M_i} u_i$ and $\pmass_i = \frac{1}{\sqrt{M_i}} p_i$.

\subsection{Kubo correlation function}
Given the many-body nature of atomic vibrations in a crystal, a natural tool to describe the dynamics are correlation functions.
In this work, we will be using the Kubo correlation function (KCF), which, for two operators $A$ and $B$, is defined as
\begin{equation}
\label{eq:KT corr}
\begin{split}
    G_{AB}(t) &= \kBT\int_0^{\beta} d\lambda \braket{A^{\dagger}(i\lambda\hbar) B(t)} \\
    &= \frac{\kBT}{\Zpart} \int_0^{\beta} d\lambda \mathrm{Tr}\bigg[ e^{-(\beta-\lambda)H} A^{\dagger} e^{-\lambda H} B(t) \bigg] \\
    &= \bbl A, B(t)\bbr
\end{split}
\end{equation}
where $\beta=\frac{1}{\kBT}$ is the inverse temperature and $\Zpart$ is the partition function.
This correlation function, which can be derived from linear response theory~\cite{Kubo1966,Kubo1991}, has the advantage of being easily approximated in the path-integral molecular dynamics formalism~\cite{Craig2004,Morresi2021}.
Moreover, the KCF has a number of properties which simplify its manipulation.
First, similarly to classical correlation functions, it is a real and even function of time and it obeys the relations $G_{AB}(t) = G_{AB}(-t)$ as well as $G_{AB}(t) = G_{BA}(t)$.
Using the Lehmann representation, it can also be shown that derivatives of the static Kubo correlation function are related~\cite{Brown1994}
\begin{align}
\label{eq:prop2}
    \frac{d^k}{dt^k} \bbl A, B(t) \bbr \vert_{t=0} =& (-1)^k \frac{d^k}{dt^k} \bbl B, A(t) \bbr\vert_{t=0} \\
\label{eq:prop3}
    \frac{d^k}{dt^k} \bbl A, A(t) \bbr \vert_{t=0} =& 0 ~~ \forall k=1,3,5,\hdots
\end{align}
Naturally, all the different quantum correlation functions are related by simple relations~\cite{Kubo1966,Ramirez2004}, so that knowledge of any type of correlation function allows to obtain any other one.
Finally, we will note that the KCF, as the standard or any other quantum correlation function, reduces to the classical correlation function in the limit $\hbar\rightarrow 0$.

In this work, the KCF of interest is the mass-weighted displacement KCF, which is a $3 N_{\mathrm{at}} \times 3 N_{\mathrm{at}}$ symmetric matrix defined with
\begin{equation}
\begin{split}
    G_{ij}(t) =& \sqrt{M_i M_j} \frac{\kBT}{\Zpart} \int_0^\beta d\lambda \mathrm{Tr} \bigg[ e^{-(\beta-\lambda)H} u_i^{\dagger}e^{-\lambda H} u_j(t)\bigg] \\
    =& \sqrt{M_i M_j} \bbl u_i, u_j(t) \bbr \\
    =& \bbl \umass_i, \umass_j(t) \bbr
\end{split}
\end{equation}
with an equation of motion given by
\begin{equation}
\label{eq:eom correlation}
\begin{split}
    \ddot{G}_{ij}(t) =& \sqrt{M_i M_j}\bbl u_i, \ddot{u}_j(t) \bbr \\
    =& \bbl \umass_i, \dot{\pmass}_j(t) \bbr
\end{split}
\end{equation}
Our goal in this work is to derive a solution for this equation of motion.

\subsection{The Generalized Langevin Equation}
\label{sec:Mori-Zwanzig}
Presented in the form of eq.(\ref{eq:eom correlation}), solving for the dynamics of the many-body system is an intractable problem.
Fortunately, the projection operator formalism derived by Mori~\cite{Mori1965} and Zwanzig~\cite{Zwanzig1961} allows to express eq.(\ref{eq:eom correlation}) in a form amenable to approximations.
This formalism is based on operators $\proj$ and $\qroj$ which project an observable $O(t)$ onto the subspace of ``slow" dynamical variables of interest.
In this work, we will use the mass scaled coordinates as ``slow" variables, so that the projectors are defined as
\begin{align}
\label{eq:proj}
    \proj O(t) =& \sum_{ij} \frac{\bbl \umass_j, O(t) \bbr}{\bbl \umass_i, \umass_j \bbr} \umass_i +  \sum_{ij} \frac{\bbl \pmass_j, O(t) \bbr}{\bbl \pmass_i, \pmass_j \bbr} \pmass_i\\
    \qroj O(t) =& (1 - \proj) O(t)
\end{align}
By construction, it should be noted that the projectors $\proj$ and $\qroj$ are orthogonal, in the sense that they follow $\proj + \qroj = 1$ and $\proj \qroj = \qroj \proj = 0$.
Being projection operators, it is natural that $\proj$ and $\qroj$ are idempotent, so that $\proj\proj = \proj$ and $\qroj \qroj = \qroj$.
Moreover, it can be shown that these projectors, as the Liouville operators, are hermitian with respect to the static Kubo correlation function~\cite{Carof2014}, which results in the properties
\begin{align}
    \bbl X, \proj Y \bbr =& \bbl \proj X, Y \bbr \\ 
    \bbl X, \qroj Y \bbr =& \bbl \qroj X, Y \bbr
\end{align}
\begin{figure}
    \centering
    \includegraphics[width=\columnwidth]{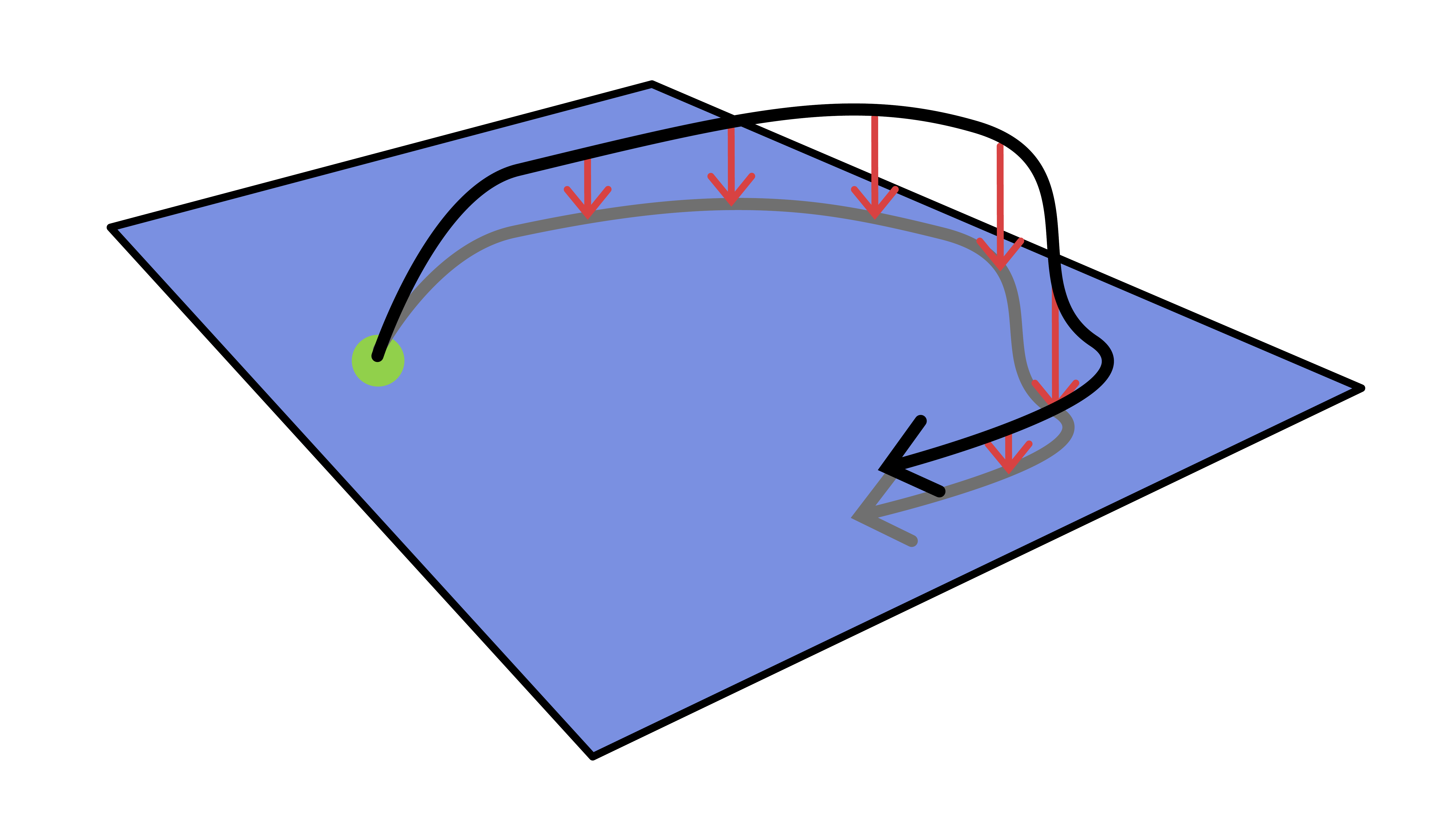}
    \caption{Schematic illustration of the Mori-Zwanzig projection operator formalism. Within this formalism, the dynamics of a dynamical variable, represented here with a black arrow, happen in the full Hilbert space. 
    The operators $\proj$, represented as the red arrows, projects this dynamic onto a ``slow" subspace depicted here as a the blue rectangle. By construction, the starting point of the dynamics, visualized with the green dot, is located on the projected subspace.}
    \label{fig:projection}
\end{figure}

Equipped with these operators, we can now express the equation of motion of the mass scaled momentum of atom $i$
\begin{equation}
\label{eq:eom momentum}
\begin{split}
    \dot{\pmass}_i(t) =& e^{i\liou t} i \liou \pmass_i \\
    =& e^{i\liou t} (\proj + \qroj) i \liou \pmass_i \\
    =& e^{i\liou t} \proj i \liou \pmass_i + e^{i\qroj\liou t} \qroj i \liou \pmass_i \\
    & + \int_0^t ds e^{i\liou (t-s)}\proj i \liou e^{i\qroj \liou s}\qroj i \liou \pmass_i
\end{split}
\end{equation}
where we used the Dyson identity~\cite{Reichman2005,Carof2014} to expand the $\qroj$ projection.

To simplify the notation of eq.(\ref{eq:eom momentum}), we introduce the frequency matrix $\vec{\Theta}$ (note that it is not in mass scaled coordinates)
\begin{equation}
\label{eq:theta}
    \Theta_{ij} = -\sum_{k} \frac{\bbl u_k, f_i \bbr}{\bbl u_j, u_k \bbr}
\end{equation}
where $f_i =  M_i \liou^2 u_i $ is the force acting on atom $i$, as well as $\delta \vec{f}(t)$, which is often called the random force,
\begin{equation}
\label{eq:random forces}
    \delta f_i(t) = e^{i\qroj \liou t} \big(f_i + \sum_{j} \Theta_{ij}u_j \big)
\end{equation}
and the memory kernel $\vec{K}(t)$
\begin{equation}
\label{eq:memory kernel}
    K_{ij}(t) = \frac{\bbl \delta f_i, \delta f_j(t) \bbr}{\sqrt{M_i M_j}}
\end{equation}
so that the equations of motion for the momentum can be written as~\cite{Bosse1978,Reichman2005,Markland2012,Janssen2018}
\begin{equation}
\label{eq:eom momentum mori-zwanzig}
\begin{split}
    \dot{\pmass}_i(t) =& -\sum_{j} \frac{\Theta_{ij}}{\sqrt{M_i M_j}} \umass_j(t) - \frac{\delta f_i(t)}{\sqrt{M_i}}  \\
    &- \beta \sum_j \int_0^t ds K_{ij}(s) \pmass_j(t-s)
\end{split}
\end{equation}
Formally, the Mori-Zwanzig projection splits the evolution of the dynamical variables into a ``slow" contribution, driven by the frequency matrix and a ``fast" one, given by the random force and the memory kernel.
As depicted in Fig.\ref{fig:projection}, the ``slow" part of the dynamics represents a subspace of the full Hilbert space of the dynamical variables.
Thus, the goal of the operator $\proj$ is to project any dynamical variable into this subspace.
The random forces, which account for the effects of other degrees of freedom, evolve in the ``fast" orthogonal subspace, as indicated by the evolution operator $e^{i\qroj \liou t}$, and its orthogonal nature is recovered thanks to the property $\bbl \tilde{u}_i, \delta f_i(t) \bbr = 0$.
It should be noted that the randomness of this term is only apparent.
By contracting information through the projector $\qroj$, the time evolution of this term is more difficult to apprehend and its dynamics appears ``random"~\cite{Carof2014}, though this is not the case.

We can now inject $\dot{\pmass}_i(t)$ into the equations of motion of eq.(\ref{eq:eom correlation}), to obtain the generalized Langevin equation of motion of the mass weighted displacement KCF
\begin{equation}
\label{eq:GLE real space}
    \ddot{\vec{G}}(t) = -\frac{\vec{\Theta}}{\sqrt{\vec{M}^T\vec{M}}} \vec{G}(t) - \beta \int_0^t ds \vec{K}(s) \dot{\vec{G}}(t - s)
\end{equation}
An interesting result of this formulation is the separation of the dynamics into two parts.
The first one, driven by the static frequency matrix is conservative, while the second part, driven by the memory kernel is dissipative~\cite{Janssen2018}.
The formulation of the memory kernel as the autocorrelation of the random forces is a result of the equilibrium state of the system and is known as the second fluctuation-dissipation theorem~\cite{Kubo1966}.
It is important to note that there is no approximation leading to eq.(\ref{eq:GLE real space}), so that it is an exact description of the dynamics of the system.
However, the main difficulty in solving this equation comes from the memory kernel of eq.(\ref{eq:memory kernel}), for which the dynamics is not generated by the Liouville operator, but by its orthogonal projection.

\section{The GLE in QP space}
\label{sec:Quasiparticle space}

Up to this point, we have been working in real space.
To formulate the problem in reciprocal space, there is a freedom of choice for the basis set into which we will project the dynamical variables.
A natural projection is one where the conservative part of the GLE is block diagonal in momentum and completely diagonal in vibrational modes, all while taking advantage of the periodicity of the crystal.
Such a transformation can be achieved using concepts from the toolkit of harmonic phonon theory.

With this goal, we define a generalized dynamical matrix $\mathbf{D}(\vec{q})$ as the Fourier transform of the mass weighted $\vec{\Theta}$
\begin{equation}
\label{eq:dynamical matrix}
    \mathbf{D}_{ij}(\vec{q}) = \sum_{\tau} \frac{\Theta_{0, i}^{\tau, j}}{\sqrt{M_i M_j}} e^{i\vec{q} (\braket{\vec{R}_{\tau,j}} - \braket{\vec{R}_{0,i}})}
\end{equation}
where we introduced a unit cell based notation $i\rightarrow (\tau,i)$, so that $\vec{\Theta}_{\delta,i}^{\tau,j}$ is the frequency matrix relating atom $i$ in the unit cell $\delta$ to the atom $j$ in unit cell $\tau$.
Similarly to the harmonic phonon case, we can now obtain QP polarization vectors $\eigvec^\nu(\vec{q})$ and frequencies $\Omega_{\vec{q},\nu}$ by diagonalizing $\vec{D}(\vec{q})$
\begin{equation}
\label{eq:dyn mat}
    \vec{D}(\vec{q}) \eigvec^{\nu}(\qvec) = \Omega_{\nu}^2(\qvec) \eigvec^{\nu}(\qvec)
\end{equation}
where $\nu$ is a QP mode.
Without loss of generality, we can now express the mass-weighted atomic displacements as
\begin{equation}
\label{eq:temp displ proj}
    \umass_i(t) = \sum_{\qvec,\lambda}\rho_{\qvec,\lambda}\epsilon_i^\lambda(\qvec)A_\lambda(\qvec, t)
\end{equation}
where $A_\lambda(\qvec,t)$ are unitless quasiparticle operators with the property $\bbl A_\lambda(\qvec), A_{\lambda'}(\qvec') \bbr = \frac{2\kBT}{\hbar \Omega_\lambda(\vec{q})} \delta_{\lambda\lambda'}\delta_{\qvec,\qvec'}$\footnote{While the choice of prefactor here  can seem arbitrary, we use it to simplify comparison with other methods of anharmonic lattice dynamics} and $\rho_{\qvec,\lambda}$ is a prefactor determined below.
Using the static solution of eq.(\ref{eq:GLE real space}) as well as the property $\ddot{G}_{ij}(0) = -\kBT/M_i \delta_{ij}$~\cite{Braams2006,Rossi2014}, we can express the static mass-weighted displacement covariance tensor as
\begin{equation}
    \bbl \umass_i, \umass_j \bbr = \kBT \sqrt{M_i M_j} \Theta^{-1}_{ij}
\end{equation}
Injecting eq.(\ref{eq:temp displ proj}) into this equation and with some reorganisation, it can be shown that $\rho_{\qvec,\lambda}=\sqrt{\hbar / \Omega_\lambda(\qvec)}$, so that the displacements can be expressed as
\begin{equation}
\label{eq:displ qp operator}
    u_i(t) = \sqrt{\frac{\hbar}{2 M_i}} \sum_{\qvec,\lambda}\frac{\epsilon_i^\lambda(\qvec)}{\sqrt{\Omega_\lambda(\qvec)}} A_\lambda(\qvec, t)
\end{equation}

We can now define a Kubo transformed QP Green's function
\begin{equation}
\label{eq:G(0) QP}
    \vec{G}_{\lambda,\mu}(\qvec,t) = \bbl A_\lambda(\qvec), A_\mu(\qvec,t) \bbr
\end{equation}
To obtain the expression of the memory kernel projected on the QP, we note that from eq.(\ref{eq:displ qp operator}), the force acting on nucleus $i$ can be expressed as
\begin{equation}
\label{eq:forces qp operator}
    f_i = \sqrt{\frac{\hbar M_i}{2}} \sum_{\vec{q}\lambda} \frac{\epsilon_i^\lambda(\qvec)}{\sqrt{\Omega_{\lambda}(\qvec)}} \ddot{A}_{\lambda}(\qvec)
\end{equation}

The part of the random forces involving the frequency matrix has a structure resembling mass-scaled harmonic forces, which can be written in terms of quasiparticle operators as
\begin{equation}
    \sum_{j \beta} \frac{\Theta_{ij}^{\alpha\beta}}{\sqrt{M_i M_j}} u_j^\beta = \sqrt{\frac{\hbar M_i}{2}} \sum_{\vec{q}\lambda}  \frac{\epsilon_i^{\lambda}(\qvec)}{\sqrt{\Omega_{\lambda}(\qvec)}} \Omega_{\lambda}^2(\qvec) A_{\lambda}(\qvec)
\end{equation}
Consequently, we can define the random forces in terms of QP operators as
\begin{equation}
\begin{split}
    \delta f_i =& \sqrt{\frac{\hbar M_i}{2}} \sum_{\qvec,\lambda} \frac{\epsilon_i^\lambda(\qvec)}{\sqrt{\Omega_\lambda(\qvec)}}\bigg(\ddot{A}_\lambda(\qvec) - \Omega_\lambda^2(\qvec) A_\lambda(\qvec) \bigg)\\
    =& \sqrt{\frac{\hbar M_i}{2}} \sum_{\qvec,\lambda} \frac{\epsilon_i^\lambda(\qvec)}{\sqrt{\Omega_\lambda(\qvec)}} \delta A_\lambda(\qvec)
\end{split}
\end{equation}
where we defined the QP random forces operator $\delta A_\lambda(\qvec) = \ddot{A}_\lambda(\qvec) - \Omega_{\lambda}^2(\qvec) A_\lambda(\qvec)$, which follows orthogonal time dynamics
\begin{equation}
    \delta A_\lambda(\qvec, t) = e^{i\qroj \liou t} \delta A_\lambda(\qvec)
\end{equation}
We are now able to define the memory kernel in reciprocal space
\begin{equation}
    \vec{K}(\vec{q}, t) = \frac{\hbar}{\kBT} \vec{\Omega}(\qvec)^{-1} \bbl \delta \vec{A}(\qvec), \delta \vec{A}(\qvec,t) \bbr
\end{equation}
which allows us to write the GLE for the Kubo transformed Green's function as
\begin{equation}
\label{eq:GLE phonon space}
    \ddot{\vec{G}}(\vec{q},t) = -\vec{\Omega}^2(\qvec) \vec{G}(\vec{q},t) - \int_0^t ds \vec{K}(\vec{q},s) \dot{\vec{G}}(\vec{q},t - s)
\end{equation}
The dynamical properties of the many-body system can now be obtained by solving eq.(\ref{eq:GLE phonon space}) with the initial conditions
\begin{align}
\label{eq:init cond 1}
    G_{\lambda\mu}(\qvec, 0) =& \frac{2\kBT}{\hbar\Omega_\lambda(\vec{q})}\delta_{\lambda,\mu} \\
\label{eq:init cond 2}
    \dot{G}_{\lambda\mu}(\qvec, 0) =& 0
\end{align}

\section{Vibrational spectra from the Kubo correlation function}
\label{sec:Vibrational spectral}

From linear response theory, it is known that the correlations of an unperturbed system are related to its response to a perturbation~\cite{Kubo1966,Kubo1991,Zwanzig2001}.
For a perturbation applied to an observable $A$, the response of the system through the observable $B$ can be computed using the response function $\Phi_{AB}$, that we define as
\begin{equation}
\label{eq:response function}
    \Phi_{AB}(t) = -\frac{i}{\hbar} \braket{[A, B(t)]}
\end{equation}
where the average of the commutator of $A$ and $B(t)$ is called the symmetrized correlation function.
The dynamical susceptibility of the system is then simply given by the Laplace transform of the response function
\begin{equation}
    \bigchi_{AB}(\omega) = \int_0^\infty dt \Phi_{AB}(t) e^{-i\omega t}
\end{equation}
It should be noted that the use of a Laplace transform with the definition eq.(\ref{eq:response function}) allows to include causality in the response of the system.
This formulation is equivalent to more common definitions of the susceptibility where causality is introduced with the use of the Heaviside function.
To connect the preceding derivation to the response function, we need to express $\Phi_{AB}(t)$ as a function of the KCF.
This can be achieved using the cyclic property of the trace, which gives a classical-like formulation of the response function
\begin{equation}
\begin{split}
    \Phi_{AB}(t) =& -\frac{i}{\hbar} \frac{1}{\Zpart} \bigg(\mathrm{Tr}\big[e^{-\beta H}A B(t)]\big] - \mathrm{Tr}\big[e^{-\beta H}B(t) A]\big] \bigg)\\
    =& \frac{i}{\hbar} \frac{1}{\Zpart} \bigg(\mathrm{Tr}\big[e^{-\beta H} e^{\beta H} Ae^{-\beta H}B(t) ]\big]\\
    &- \mathrm{Tr}\big[e^{-\beta H}A B(t)]\big]  \bigg)\\
    =& \int_0^\beta d\lambda \frac{1}{\Zpart} \mathrm{Tr}\big[e^{-\beta H} \dot{A}(i\hbar\lambda)B(t)] \\
    =& \beta \dot{G}_{AB}(t)
\end{split}
\end{equation}
where $\dot{A} = i[H, A]/\hbar$ is the time derivative of $A$ at $t=0$.

From this response function, the dynamical susceptibility of the system is simply obtained by applying a Laplace transform
\begin{equation}
\begin{split}
\label{eq:dyn susc Kubo}
    \bigchi_{AB}(\omega) =& \beta \int_0^\infty dt \bbl \dot{A}, B(t) \bbr e^{-i\omega t} \\
    =& \beta \widetilde{\dot{G}}_{AB}(i\omega) \\
    =& i \beta \omega \widetilde{G}_{AB}(i\omega) - \beta G_{AB}(0)
\end{split}
\end{equation}
with $\widetilde{G}_{AB}(i\omega)$ the Laplace transform of the KCF of $A$ and $B$, and $\widetilde{\dot{G}}_{AB}(i\omega)$ is the Laplace transform of its time derivative.
The dynamical susceptibility can be decomposed into a real and imaginary part $\bigchi_{AB}(\omega) = \bigchi'_{AB}(\omega) + i \bigchi''_{AB}(\omega)$, where $\bigchi_{AB}'(\omega)$ and $\bigchi_{AB}''(\omega)$ are related by a Kramers-Kronig transform, meaning that the knowledge of one is sufficient to recover the dynamical response of the system.
From eq.~(\ref{eq:dyn susc Kubo}), we see that the imaginary part of the dynamical susceptibility, which corresponds to the dissipative part of the response, is simply proportional to the real part of $\widetilde{G}_{AB}(\omega)$.
Since the KCF is an even and real function of time, this real part is proportional to the power spectrum $G_{AB}(\omega)$, which corresponds to the Fourier transform of $G_{AB}(t)$, so that the dissipative part of the response, also called the spectral function, can be written as
\begin{equation}
\label{eq:fluctuation-dissipation theorem}
    \bigchi_{AB}''(\omega) = \frac{\omega}{2\kBT} G_{AB}(\omega)
\end{equation}

This equation is simply the fluctuation-dissipation theorem, written using the Kubo transformed correlation function.
The trace of the dissipation $\mathbf{\bigchi}''(\qvec,\omega)$ is particularly important since this quantity can be linked to the cross section $d^2\sigma/d\Omega dE$ from spectroscopy measurements, allowing thus a direct comparison between theory and experiment.
For the QP KCF, using the initial conditions eq.~(\ref{eq:init cond 1}) and (\ref{eq:init cond 2}), the Laplace transform is expressed as
\begin{equation}
\label{eq:KCF Laplace}
    \widetilde{\vec{G}}(\vec{q}, i\omega) =  \frac{2 \kBT}{\hbar\vec{\Omega}(\vec{q})}\frac{-i\omega -  \tilde{\vec{K}}(\vec{q}, i\omega)}{\omega^2 - \vec{\Omega}^2(\vec{q}) - i \omega \tilde{\vec{K}}(\vec{q},i\omega)}
\end{equation}
To obtain a formula for the Fourier transform of $\vec{G}(\qvec, \omega)$, we start by separating the memory kernel into a real and an imaginary part as $\tilde{\vec{K}}(\qvec, i\omega) = \tilde{\vec{\Gamma}}(\qvec, \omega) - i\tilde{\vec{\Delta}}(\qvec, \omega)$.
Since $\vec{K}(\vec{q},t)$ is a real and even function of time, the real part $\tilde{\vec{\Gamma}}(\vec{q}, \omega)$ of its Laplace transform is proportional to its Fourier transform $\vec{\Gamma}(\vec{q}, \omega)$, with $\tilde{\vec{\Gamma}}(\vec{q},\omega) = 2 \vec{\Gamma}(\vec{q},\omega)$.
Moreover, $\tilde{\vec{\Gamma}}(\vec{q},\omega)$ and $\tilde{\vec{\Delta}}(\vec{q},\omega)$ are not independent, and one can recover the other using Kramers-Kronig relations~\cite{Berne1970}.
For instance, $\tilde{\vec{\Delta}}(\qvec, \omega)$ can be computed with
\begin{equation}
\label{eq:Kramers}
    \tilde{\vec{\Delta}}(\vec{q}, \omega) = \frac{1}{\pi} \int d\omega' \frac{\tilde{\vec{\Gamma}}(\vec{q}, \omega')}{\omega' - \omega}
\end{equation}
where the absence of the usual minus sign comes from the previous definition of the memory kernel.
This allows us to take the real part of eq.(\ref{eq:KCF Laplace}) and obtain the power spectrum as
\begin{equation}
\label{eq:power spectrum}
    \vec{G}(\vec{q},\omega) = \frac{\kBT}{\pi\hbar}\frac{8\vec{\Omega}(\qvec) \vec{\Gamma}(\vec{q},\omega)}{(\omega^2 - \vec{\Omega}^2(\qvec) - 2\omega \vec{\Delta}(\vec{q},\omega))^2 +  4\omega^2\vec{\Gamma}^2(\vec{q},\omega)}
\end{equation}
Finally, the dissipative part of the response function can be computed using the fluctuation-dissipation theorem of eq.(\ref{eq:fluctuation-dissipation theorem}), giving
\begin{equation}
    \vec{\bigchi}''(\vec{q},\omega) =\frac{1}{\pi\hbar} \frac{4 \omega \vec{\Omega}(\qvec) \vec{\Gamma}(\vec{q},\omega)}{(\omega^2 - \vec{\Omega}^2(\qvec) - 2\omega \vec{\Delta}(\vec{q},\omega))^2 + 4\omega^2\vec{\Gamma}^2(\vec{q},\omega)}
\end{equation}

In the end, the dynamical response of the system can be entirely characterised using $\vec{\Omega}(\qvec)$, $\vec{\Gamma}(\qvec, \omega)$ and $\vec{\Delta}(\qvec, \omega)$.
Consequently, at this point the only missing ingredient needed to compute the dynamical properties of the system is the real part of the memory kernel.

\section{Mode-coupling approximation for the memory kernel}
\label{sec:Mode-coupling}

For the moment, the formulation is exact and consists only in a rewriting of the dynamical problem, by framing the unsolvable part into a memory kernel.
The exact computation of $\vec{K}(t)$ is a formidable task and we have to resort to approximations to be able to apply the theory to real simulations.
To obtain a tractable formulation of the memory kernel, we use a cluster expansion of the random forces.
With this goal, we define the $n$-th order projection operators $\proj_n$~\cite{Szamel2007}, which project an operator $O(t)$ on the subspace of $n$ QP operators, as well as the corresponding orthogonal projection $\qroj_n$
\begin{align}
\label{eq:proj n}
    \proj_n O(t) =& \frac{1}{n!} \sum_{\substack{\lambda_1\hdots\lambda_n \\ \lambda'_1 \hdots \lambda'_n}} \frac{\bbl A_{\lambda'_1} \hdots A_{\lambda'_n}, O(t) \bbr}{\bbl A_{\lambda_1}\hdots A_{\lambda_n}, A_{\lambda'_1}\hdots A_{\lambda'_n} \bbr} A_{\lambda_1}\hdots A_{\lambda_n} \\
    \qroj_n =& 1 - \proj_n
\end{align}
where the prefactor $1 / n!$ is here to account for all the permutations of $A_{\lambda}$ operators, so that $\proj_n\proj_n = \proj_n$.
Going up to order $N$ and inserting several times the identity $\proj_n + \qroj_n = 1$, we can rewrite the random forces
\begin{align}
    \begin{split}
    \delta A_{\lambda} =& (\proj_2 + \qroj_2) \delta A_{\lambda} \\
    =& \proj_2 \delta A_{\lambda} + (\proj_3 + \qroj_3)\qroj_2 \delta A_{\lambda} \\
    =& \sum_n^N \proj_{n} \delta_{n-1} A_{\lambda} + \delta_{N} A_{\lambda}
    \end{split} \\
    \delta_{N} A_{\lambda} =& \qroj_N \qroj_{N-1} \hdots \qroj_2 \delta A_{\lambda}
\end{align}
where each successive application of $\qroj_n$ will produce an ever smaller  $\delta_nA_\lambda$, and $\delta_NA_\lambda$ is the final residual random force we will neglect.
We can now inject this expression into the memory kernel to obtain
\begin{equation}
\label{eq:memory kernel2}
\begin{split}
    K^{\lambda,\lambda'}(t) =& \bbl \delta A_{\lambda}, \delta A_{\lambda'}(t) \bbr \\
    =& \sum_{n=2}^{N}\sum_{n'=2}^{N} \bbl \proj_n \delta_{n-1} A_{\lambda}, \proj_{n'}\delta_{n'-1} A_{\lambda'}(t) \bbr \\
    &+ \bbl \delta_N A_{\lambda'}, \delta_N A_{\lambda'}(t) \bbr \\
    =& \sum_{n=2}^N \sum_{n'=2}^N K_{nn'}^{\lambda,\lambda'}(t) + \bbl \delta_N A_{\lambda}, \delta_N A_{\lambda'}(t) \bbr 
\end{split}
\end{equation}
where we define the $(n,n')$ order memory kernel
\begin{equation}
\label{eq:int memory kernel}
\begin{split}
    K_{n, n'}^{\lambda,\lambda'}(t) =& \bbl \proj_n \delta_{n-1} A_{\lambda}, \proj_{n'} \delta_{n'-1} A_{\lambda'}(t) \bbr \\
    =& \sum_{\substack{\lambda_1,\hdots,\lambda_n \\ \lambda'_1,\hdots,\lambda'_{n'} }} \Psi_n(\lambda, \lambda_1,\hdots, \lambda_n) \Psi_{n'}(\lambda', \lambda'_1,\hdots, \lambda'_{n'}) \\
    & \times \bbl A_{\lambda_1} \hdots A_{\lambda_n}, e^{i\qroj\liou t} A_{\lambda'_1} \hdots A_{\lambda'_{n'}} \bbr
\end{split} 
\end{equation}
as well as the vertex
\begin{equation}
\label{eq:renormalized int}
    \Psi_{n}(\lambda_1, \hdots, \lambda_n) = \frac{1}{n!} \sum_{\mu_1,\hdots,\mu_n} \frac{\bbl A_{\mu_1} \hdots A_{\mu_n}, \delta_{n-1} A_{\lambda} \bbr} {\bbl A_{\lambda_1} \hdots A_{\lambda_n}, A_{\mu_1} \hdots A_{\mu_n} \bbr }
\end{equation}
The preceding manipulations serve to shift the orthogonal space time dependence from the random forces into multiple displacement Kubo correlation functions.
By subtracting the contributions up to order $n$, the $\qroj_n$ projectors minimize the corresponding contribution of the random forces $\delta_{n} A_\lambda$.
Consequently, this ``clustering" procedure ensures that the contribution of each new order in the memory kernel is less important than the previous one.
While this is still impossible to solve in practice, this rewriting opens a way to approximate $\vec{K}(\vec{q},t)$.

As a first approximation, we truncate the expansion of the memory kernel at some order $N$ and we neglect the $N$-th order correlation of the random forces.
The second approximation consists in decoupling the $n$ point correlations in orthogonal time into $n/2$ products of $2$ point correlations in real time~\cite{Bosse1978,Reichman2005,Janssen2018}.
For instance, a four point correlation function is given by
\begin{equation}
\begin{split}
    \bbl A_{\lambda_1} A_{\lambda_2}, e^{i\qroj\liou t} A_{\lambda_3} A_{\lambda_4} \bbr \approx& \bbl A_{\lambda_1}, A_{\lambda_3}(t) \bbr \bbl A_{\lambda_2}, A_{\lambda_4}(t) \bbr \\
    &+ \bbl A_{\lambda_1}, A_{\lambda_4}(t) \bbr \bbl A_{\lambda_2}, A_{\lambda_3}(t) \bbr
\end{split}
\end{equation}
These two approximations are at the foundation of the mode-coupling theory of the liquid-glass transition~\cite{Bosse1978,Reichman2005}.

In this work, we will limit ourselves to the second order in the memory kernel expansion.
To simplify further the expression of the memory kernel, we will neglect its off-diagonal terms ($n \neq n'$, $\lambda \neq \lambda'$). We note that this approximation does not remove mode coupling (the sum in eq.(\ref{eq:int memory kernel}) runs over all the other modes) and would not prevent the inclusion of off-diagonal velocity terms in the thermal conductivity as in Ref.~\cite{Simoncelli2019}.
Finally, taking into account the conservation of crystal momentum (with the $-\lambda$), our approximation for the memory kernel is given by
\begin{equation}
\label{eq:approx memory kernel}
    K_{\lambda}(t) \approx \frac{\hbar}{2\kBT \Omega_\lambda} \sum_{\mu, \nu} \vert \Psi(-\lambda,\mu,\nu)\vert^2 G_{\mu}(t)G_{\nu}(t)
\end{equation}
The only inputs are contained in the vertex $\Psi$.
We define the third-order tensor $\Theta$ as
\begin{equation}
    \Theta_{ijk} = -\frac{1}{2} \frac{\sum_{l,m} \bbl u_l u_m, \delta f_i \bbr}{\sum_{l,m} \bbl u_j u_k, u_l u_m \bbr}
\end{equation}
Using the projection of displacements and random forces on QP, the vertex can then be expressed as
\begin{equation}
\begin{split}
    \Psi(\lambda, \mu, \nu) =& \sqrt{\frac{\hbar}{2} \frac{\Omega_\lambda}{\Omega_\mu \Omega_\nu}} \sum_{i,j,k} \frac{\eigvec_i^\lambda \eigvec_j^\mu \eigvec_k^\nu}{\sqrt{M_i M_j M_k}} \Theta_{ijk} \\
    &\times e^{i(\vec{R}_i\vec{q}_\lambda + \vec{R}_j\vec{q}_\mu + \vec{R}_k\vec{q}_\nu)} \delta_{\vec{G}}(\vec{q}_\lambda + \vec{q}_\mu + \vec{q}_\nu) \\
\end{split}
\end{equation}
where $\delta_\vec{G}(\qvec_\lambda + \qvec_\mu + \qvec_\nu)=1$ if $\qvec_\lambda + \qvec_\mu + \qvec_\nu$ is $0$ modulo a reciprocal lattice vector $\vec{G}$.

With the preceding approximation of $\vec{K}$, we now have all the ingredients to compute the power spectrum.
Fourier transforming eq.(\ref{eq:approx memory kernel}) in frequency domain, and using eq.(\ref{eq:Kramers}) and eq.(\ref{eq:power spectrum}), we obtain the following set of self-consistent equations
\begin{align}
\label{eq:scf gamma}
    \Gamma_\lambda(\omega) =& \frac{\hbar}{2\kBT\Omega_\lambda} \sum_{\mu,\nu}\vert \Psi(-\lambda,\mu,\nu)\vert^2 (G_\mu \ast G_\nu)(\omega) \\
\label{eq:scf delta}
    \Delta_\lambda(\omega) =& \frac{1}{\pi} \int d\omega' \frac{\Gamma_\lambda(\omega')}{\omega' - \omega} \\
\label{eq:scf g}
    G_\lambda(\omega) =& \frac{\kBT}{\pi\hbar}\frac{ 8\Omega_{\lambda} \Gamma_\lambda(\omega)}{(\omega^2 - \Omega_{\lambda}^2 - 2\omega \Delta_\lambda(\omega))^2 + 4\omega^2 \Gamma_\lambda^2(\omega)}
\end{align}
In order to obtain a starting point to the self-consistent cycle, we introduce an approximation inspired by scattering theory.
We begin by defining a bare Green's function as the solution of a memory-less GLE
\begin{equation}
    \ddot{G}_{\lambda,0}(t) = -\Omega^2_\lambda G_{\lambda,0}(t)
\end{equation}
which is simply given by $G_{\lambda,0}(t) = \frac{\kBT}{\hbar\Omega_\lambda} \cos(\Omega_\lambda t)$.
The scattering theory approximation is obtained by replacing the Green's function in eq.(\ref{eq:scf gamma}) by $G_{\lambda,0}(\omega)$, which allows to obtain the linewidth
\begin{equation}
\label{eq:scatt gamma}
    \Gamma_\lambda^{s}(\omega) = \frac{\pi\kBT}{\hbar \Omega_\lambda \Omega_\mu \Omega_\nu}\sum_{\mu,\nu,s,s'} \vert \Psi(-\lambda,\mu,\nu)\vert^2 \delta(\omega + s\Omega_\mu + s'\Omega_\nu)
\end{equation}
with $s,s'=\pm 1$.
From this result, a first approximation to the power spectrum can be obtained by using eq.(\ref{eq:scf delta}) and (\ref{eq:scf g}) with $\Gamma_\lambda^s(\omega)$ as an input.

\section{Computing the frequencies and vertex from (PI)MD simulations}
\label{sec:Simulations}

One advantage of the formalism derived previously is that only static averages are needed as input.
This is particularly interesting for applications given that the static KCF of two variables $A$ and $B$ can be computed exactly in the path-integral molecular dynamics framework as~\cite{Craig2004,Morresi2021}:
\begin{equation}
\label{eq:kubo pimd}
\begin{split}
    G_{AB}(0) =& \lim_{P\rightarrow\infty} \lim_{t\rightarrow\infty}\int_0^t \frac{e^{-\beta H(t)}}{\Zpart} \bbl \sum_{p}^P  A_p(t) \bbr \bbl \sum_p^P B_p(t)\bbr dt\\
    \approx& \frac{1}{N} \sum_n^N \bar{A}(t_n) \bar{B}(t_n)
\end{split}
\end{equation}
where $P$ is the number of beads in the classical polymer, $\bar{A}(t) = \sum_p A_p(t)$ is the mean of the property $A$ over the classical polymer. The second line of eq.(\ref{eq:kubo pimd}) corresponds to the approximation with a finite number of configurations and beads from path integral molecular dynamics.

The static correlation function of  classical MD simulations corresponds to the limit $P=1$ and can be computed as
\begin{equation}
\begin{split}
    G_{AB}(0) \approx& \frac{1}{N} \sum_n^N A(t_n) B(t_n)
\end{split}
\end{equation}
We note that this can be used as a way to approximate the static KCF, as will be discussed on the next section.

In the limit of a finite number $N$ of configurations, the frequency matrix can be computed with
\begin{equation}
\label{eq:least squares freq}
    \Theta_{ij} = -\frac{\sum_n^N \sum_k \bar{u}_k(t_n) \bar{f}_i(t_n)}{\sum_n^N \sum_k \bar{u}_k(t_n) \bar{u}_j(t_n)}
\end{equation}
where we used the property $\bbl u_i, f_j \bbr = -\kBT \delta_{ij}$ to decouple the sums in the numerator and denominator of eq.(\ref{eq:theta}).
In this equation, one can recognize the least squares solution $\vec{\Theta}$ of the linear problem $\vec{\Theta} \vec{u} = -\vec{f}$.
This implies that the computation of the frequency matrix can be mapped to the problem of fitting an effective harmonic Hamiltonian using the distribution of forces and displacements at the studied temperature.
This corresponds to the TDEP method at second order~\cite{Hellman2011}, for which our framework offers an exact formal justification and a path for generalization.
In practice, the average needed in the computation of $\vec{\Theta}$ might necessitate long simulations which could become prohibitive.
Fortunately, $\vec{\Theta}$ is a sparse matrix, with many coefficients related by symmetry, as it must be invariant under the symmetry group of the crystal.
Moreover, since the displacements and forces are invariant with respect to global translation and rotation of the system, their static KCF should also respect these invariances.
This means that, by definition, $\vec{\Theta}$ follows the usual symmetries found in the IFC tensor of the harmonic approximation for phonons.
Consequently, imposing them on $\vec{\Theta}$ before applying eq.(\ref{eq:least squares freq}) will considerably decrease the length of the simulations needed, similarly to what is done in lattice dynamic fitting methods~\cite{Hellman2013a,Eriksson2019,Zhou2019,Bottin2020}.
A similar analysis can be done for the second order vertex in real space, which can be computed as
\begin{equation}
\label{eq:least squares vertex}
    \Theta_{ijk} = -\frac{1}{2} \sum_n^N \frac{\sum_{l,m} \bar{u}_l(t_n) \bar{u}_m(t_n) \delta\bar{f}_i(t_n)}{\sum_{l,m} \bar{u}_l(t_n) \bar{u}_m(t_n) \bar{u}_j(t_n) \bar{u}_k(t_n)}
\end{equation}
with $\delta f_i = f_i + \sum_j \Theta_{ij} u_j$.
A simple variable transformation allows to rewrite this equation as a least-squares solution of the equation $\vec{\Theta}^T \vec{U} = \delta\vec{f}$, where $\vec{U} = \vec{u}\vec{u}^T$~\cite{Hellman2013b,Bottin2020}.
As in the frequency matrix case, this means that this interaction can be constructed exactly as the third-order IFC in the TDEP method.
It can be shown that this TDEP-like construction arises naturally for every order, by definition of the $n$-th order projectors $\proj_n$.

\section{Discussion}
\label{sec:Discussion}

\subsection{Classical-like correlation functions and classical limit}
\label{sec:quantum Green's function}

Looking at the main results in the preceding derivation, it would be easy to miss the quantum character of the dynamics described, since most of the equations seem to lack the usual quantum-related prefactors.
The quantum behavior of the system is hidden in the Kubo correlation function, as can be inferred from its Lehmann representation given by
\begin{equation}
\label{eq:Lehmann Kubo}
    G_{\lambda}(\omega) = \kBT \sum_{jk} p_j \frac{\vert \bra{j}A_\lambda \ket{k}\vert^2}{\hbar \omega (n(\omega) + 1)} \delta\bigg(\omega - \frac{E_j - E_k}{\hbar}\bigg) 
\end{equation}
where $p_j=e^{-\beta E_j}/\Zpart$ is the Boltzmann weight associated with state $j$ of energy $E_j$ and $n(\omega)=1/(e^{\beta \hbar\omega} - 1)$ is the Bose-Einstein distribution.
Consequently, our approach allows to alleviate any ambiguity related to the type of quantum correlation functions one should use in PIMD simulations, all while giving a consistent formulation of the dynamics of the system.

Interestingly, the classical limit of this mode-coupling approach is completely analogous to its quantum counterpart~\cite{Markland2012}.
Indeed, the classical version of the Mori-Zwanzig formalism is obtained by simply replacing the KCF with classical correlation functions.
Using the similarity between the properties of Kubo and classical correlation functions, the same derivation can be used to obtain the classical version of the set of self-consistent equations (\ref{eq:scf gamma}), (\ref{eq:scf delta}) and (\ref{eq:scf g}).
Therefore, the only difference between the quantum and classical description comes from the type of correlation functions used to compute the frequency matrix $\vec{\Theta}$ and the vertex.
This classical-like behavior is particularly visible in the KCF formulation of the quantum fluctuation-dissipation theorem given by eq.(\ref{eq:fluctuation-dissipation theorem}), which is exactly the same as its classical limit~\cite{Kubo1991}.
This observation brings a new justification to the interpretation of classical correlation functions as approximated KCF~\cite{Ramirez2004}.
From a practical point of view, and as discussed in the previous section, this means that the same set of equations can be used in the two limits, the only difference stemming from the type of simulation used to compute the correlation functions.

Further, our approach allows for a simple inclusion of quantum behavior when doing classical simulations.
For this, we introduce the lesser Green's function $G^{<}_\lambda(t) = \braket{A^\dagger_\lambda(t) A_\lambda}$, given in the Lehmann representation by
\begin{equation}
    G^{<}_\lambda(\omega) = \sum_{jk} p_j e^{-\beta\hbar\omega}\vert \bra{j}A_\lambda \ket{k} \vert^2 \delta\bigg(\omega - \frac{E_j - E_k}{\hbar}\bigg)
\end{equation}
From this representation, it is easy to see that the Kubo and lesser Green's functions are related by $G_\lambda^{<}(\omega) = \frac{\hbar \omega}{\kBT} n(\omega) G_\lambda(\omega)$.
Replacing the Kubo Green's functions in eq.(\ref{eq:scf gamma}) with the lesser Green's functions makes the Bose-Einstein distributions visible in the equation
\begin{equation}
\begin{split}
    \Gamma_\lambda(\omega) =& \frac{\hbar}{2\kBT\Omega_\lambda} \sum_{\mu,\nu} \vert \Psi(-\lambda, \mu, \nu) \vert^2 \\
    &\times \int_{-\infty}^\infty d\omega' \frac{\kBT}{\hbar(\omega-\omega') n(\omega-\omega')} G_\lambda^{<}(\omega-\omega') \\
    &\times \frac{\kBT}{\hbar\omega' n(\omega')}  G_\lambda^{<}(\omega')
\end{split}
\end{equation}
In the scattering theory approximation, $G_\lambda^<(t) = n(\Omega_\lambda) \cos(\Omega_\lambda t)$, so that $\Gamma_\lambda^s(\omega)$ reduces to
\begin{align}
\label{eq:gamma lesser}
    \Gamma_\lambda^s(\omega) =& \frac{\hbar\pi}{4\kBT\Omega_\lambda} \sum_{\mu,\nu} \vert \Psi(-\lambda,\mu,\nu) \vert^2 F(\omega, \mu, \nu) \\
    \label{eq:F function}
    \begin{split}
    F(\omega, \mu, \nu) =& \sum_{s=\pm 1} \bigg[\big( n_\mu + n_\nu + 1\big) \delta(\omega + s\Omega_\mu + s\Omega_\nu) \\
    &+ \big( n_\mu - n_\nu\big) \delta(\omega + s\Omega_\mu - s\Omega_\nu) \bigg]
    \end{split}
\end{align}
with $n_\mu = n(\Omega_\mu)$, and where we used $\lim_{\omega\rightarrow0} 1 / [\hbar\omega n(\omega)] = \beta$.
Once $\Psi$ has been calculated, using eq.(\ref{eq:gamma lesser}) to process inputs from MD simulations is a simple zero-cost and legitimate way to approximate quantum effects without needing to do expensive PIMD simulations.
Such an approximation should be valid for systems with low anharmonicity or at high temperature.

\subsection{Interpretation of the static quasiparticles}

With the previous discussions, we focused on dynamical properties.
It is interesting to also consider static properties associated with the formalism, 
in particular the quasiparticle frequencies.
Interestingly, these temperature-dependent frequencies have often been compared with great success to experiments~\cite{Kong2009,Hellman2011,Chaney2021}, but these comparisons lacked a rigorous theoretical justification.
It has been proposed to interpret these frequencies as an approximation of the first excitation of the system~\cite{Ramrez1999,Morresi2021}.
While such an approximation would be accurate at low temperature, it doesn't hold for high temperature and completely misses the exact character associated with the quantity~\cite{Ramrez1999}.

Taking the $\omega\rightarrow0$ limit of eq.(\ref{eq:dyn susc Kubo}), the static susceptibility can be written as $\bigchi = -\beta G(0)$.
From eq.(\ref{eq:theta}) and using the property $\bbl u_i, f_j \bbr = -\kBT \delta_{ij}$, we obtain a proportionality relation between the frequency matrix and the inverse static susceptibility in real space
\begin{equation}
    \frac{\vec{\Theta}}{\sqrt{\vec{M}^T \vec{M}}} = \vec{\bigchi}^{-1}
\end{equation}
meaning that the frequency matrix is directly related to the instantaneous response of the system to an external perturbation.
By doing the transformation in section \ref{sec:Quasiparticle space}, and due to the Hermitian character of the matrices involved, we can conclude that the quasiparticle polarizations $\eigvec_\lambda(\qvec)$ are also eigenvectors of the Fourier transform of the susceptibility matrix, with eigenvalues $\hbar \Omega_\lambda^{-2}(\qvec)$.
From this observation, we can interpret the eigenvectors $\eigvec^\lambda(\qvec)$ of the quasiparticles as displacement patterns associated to a perturbation with a momentum $\qvec$.
The $A_\lambda(\qvec)$ are thus the amplitude of these displacement patterns, which can be seen as a temperature-dependent generalization of the normal modes of the harmonic theory.

To have a better understanding of the meaning of these frequencies, it is useful to focus on the second order time derivative of the KCF.
From the relation $\ddot{G}_{ij} = \kBT$, and using eqs. (\ref{eq:displ qp operator}) and (\ref{eq:forces qp operator}), we have $\ddot{G}_\lambda(\qvec) / G_{\lambda}(\qvec)  = \Omega_\lambda^2(\qvec)$.
Using then the Lehmann representation of the KCF $G_{\lambda}(\qvec)$ and $\ddot{G}_\lambda(\qvec)$
\begin{align}
    G_\lambda(\qvec) &= \kBT \sum_{jk} \frac{\vert \bra{j} A_\lambda(\qvec) \ket{k}\vert^2 (p_k - p_j)}{\omega_{jk}} \\
\begin{split}
    \ddot{G}_\lambda(\qvec) &= \kBT \sum_{jk} \frac{\bra{j} A_\lambda(\qvec) \ket{k}\bra{k} \ddot{A}_\lambda(\qvec) \ket{j}  (p_k - p_j)}{\omega_{jk}} \\
     &= \kBT \sum_{jk} \frac{\omega_{jk}^2 \vert\bra{j} A_\lambda(\qvec) \ket{k}\vert^2  (p_k - p_j)}{\omega_{jk}} \\
\end{split}
\end{align}
with $\omega_{jk} = (E_j - E_k) / \hbar$, and where we used $\bra{j}\ddot{A}_\lambda(\qvec)\ket{k} = -\omega_{jk}^2 \bra{j}A_\lambda(\qvec)\ket{k}$,
we obtain a formulation of the frequencies in terms of a complete set of eigenstates of the Hamiltonian
\begin{align}
    c_{jk}(\lambda,\qvec) &= \frac{\vert\bra{j} A_\lambda(\qvec) \ket{k}\vert^2 \big(p_k - p_j)}{\omega_{jk}} \\
    \Omega_{\lambda}(\qvec) &= \sqrt{\frac{\sum_{jk} \omega_{jk}^2 c_{jk}(\lambda,\qvec)}{\sum_{jk} c_{jk}(\lambda,\qvec)}}
\end{align}
This result shows that the quasiparticle frequencies correspond to a thermodynamic average of the transitions between states of the system projected on the displacement patterns.


It should be noted that these frequencies do not necessarily align with peaks in the spectral functions.
Indeed, the relation between the static frequencies and the spectral function is given by the sum rule
\begin{equation}
    \int_0^{\infty} d\omega \bigchi''_\lambda(\qvec, \omega) \omega  = \Omega_\lambda(\qvec)
\end{equation}
Consequently, the peaks and the static frequencies should be aligned only in the case of a symmetric lineshape, or at least in cases where the spectral weight is evenly distributed around $\Omega_\lambda$.

\subsection{Comparison with other methods}

To help understanding the implication of the mode-coupling theory of lattice dynamics, it is informative to compare it to well established methods.
In this section, we show that the perturbation theory, the SCHA and the TDEP method can be understood as different approximations of the mode-coupling approach, that we summarize in table \ref{tab:comparison}.

\begin{table*}
\caption{Summary of different approach of anharmonic lattice dynamics. \label{tab:comparison}}
\begin{ruledtabular}
\begin{tabular}{c|c c c}
 & Perturbation theory & SCHA & Mori-Zwanzig \\ \hline
 Sampling of displacements & 
 \makecell{Perturbative through \\ finite difference or DFPT} & 
 \makecell{Effective harmonic \\(multivariate Gaussian)} &
 \makecell{Exact through \\ MD (classical case) \\ or PIMD (quantum case)} \\ \hline
Static quasiparticles &    
\makecell{Harmonic phonons} &
\makecell{Effective harmonic phonons or \\Eigenstates of \\ the inverse susceptibility} &
\makecell{Eigenstates of \\ the inverse susceptibility}
\\ \hline
Vertex in the bubble contribution & perturbative & Gaussian approximation & exact 
\\ \hline
Implementations &
\makecell{Phono3py~\cite{Togo2015} \\ ShengBTE~\cite{Li2014}} &
\makecell{SSCHA~\cite{Monacelli2021} \\ Alamode~\cite{Tadano2015,Tadano2018} \\ sTDEP~\cite{Shulumba2017} \\ qSCAILD~\cite{van_Roekeghem2021}} &
\makecell{MD-TDEP~\cite{Hellman2011,Hellman2013a}} 
\end{tabular}
\end{ruledtabular}
\end{table*}

\subsubsection{The harmonic limit}

A good theory of lattice dynamics should reduce to the exactly known harmonic limit.
This can be trivially shown to be true for the mode-coupling theory.
Indeed, for a harmonic system, the frequency matrix presented in eq.(\ref{eq:theta}) is equal to the harmonic force constants, so that the random forces cancel out.
Consequently, the GLE reduces to the expected memory-less harmonic equations of motion.

\subsubsection{Perturbation theory}
More interestingly, it can be shown that the perturbative limit of the mode-coupling approach allows us to retrieve the usual linewidth formulation obtained from a diagrammatic expansion~\cite{Maradudin1962,Cowley1968}.
To show this, we suppose that the potential energy of the system can be approximated by an anharmonic potential truncated to third order
\begin{align}
    V(\vec{R}) \approx& \frac{1}{2}\sum_{ij} \Theta_{ij}^{\mathrm{pt}} u_i u_j+ \frac{1}{3!} \sum_{ijk} \Psi_{ijk}^{\mathrm{pt}} u_i u_j u_k\\
    \Theta_{ij}^{\mathrm{pt}} =& \frac{\partial^2 V(\vec{R})}{\partial R_i \partial R_j} \\
    \Psi_{ijk}^{\mathrm{pt}} =& \frac{\partial^3 V(\vec{R})}{\partial R_i \partial R_j \partial R_k}
\end{align}
where we assume that contributions from $\Psi_{ijk}^{\mathrm{pt}}$ are perturbations with respect to $\Theta_{ij}^{\mathrm{pt}}$, so that averages can be taken with respect to the harmonic contribution.
In this approximation, the frequency matrix is equal to the second order force constants $\vec{\Theta}^{\mathrm{pt}}$ and the memory kernel can be written
\begin{equation}
    K_{ij}^{\mathrm{pt}}(t) = \sum_{klmn} \Psi_{ikl}^{\mathrm{pt}} \Psi_{jmn}^{\mathrm{pt}} \bbl u_k u_l, e^{-i\qroj \liou t} u_m u_n \bbr
\end{equation}
which corresponds to the $(2, 2)$-order memory kernel defined in real space, but with third-order force constants instead of the full vertex.
From this observation, the derivation of the phonon quasiparticles follows the same route as the more general Mori-Zwanzig theory.
In the end, the main difference between the perturbation and mode-coupling theories lies in the definition of the quasiparticles and in the strength of the vertex $\Psi(\lambda, \mu, \nu)$.
In order to quantitatively reconcile experimental observations and theoretical predictions for properties such as thermal conductivity, these renormalizations are important.
Indeed, while the temperature dependence of QP frequencies is known to have a large impact on this kind of quantity~\cite{Xia2020b}, the deviations brought by scattering renormalization can be of similar magnitude and have the potential to remove the expected high temperature trend $\kappa \propto T^{-\alpha}$, as shown in Refs.~\cite{Zhu2020,Zeng2021,Yang2022}.

\begin{figure}
    \centering
    \includegraphics[width=\columnwidth]{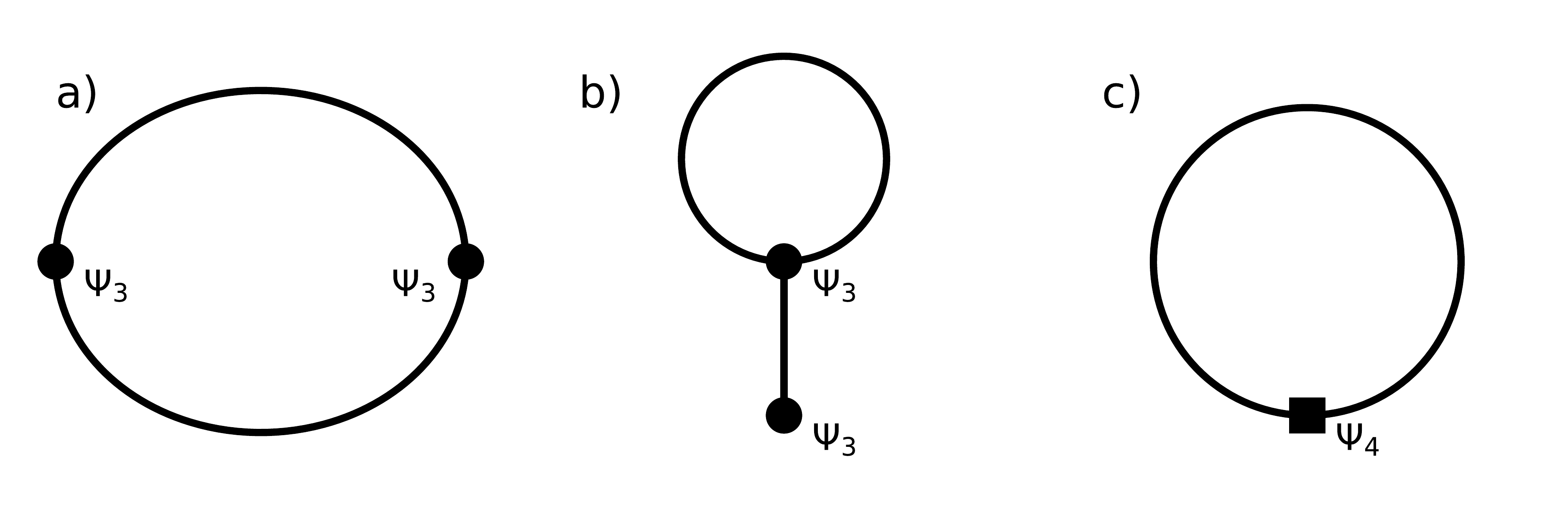}
    \caption{Feynman diagrams appearing in the self-energy in perturbation theory to lowest order.
    Each line represent a propagator while dots and squares are respectively third and fourth order vertices.
    a) The bubble b) the tadpole and c) the loop.
    In the mode-coupling approach presented here, we observe only bubble-like contributions in the memory kernel.}
    \label{fig:Diagrams}
\end{figure}

From a diagrammatic derivation to lowest-order, this dynamical contribution is given by the bubble diagram represented in Fig.\ref{fig:Diagrams} a).
Doing a naive comparison, it would seem that the other lowest-order diagrams, known as the tadpole (Fig.\ref{fig:Diagrams} b)) and the loop (Fig.\ref{fig:Diagrams} c)), are missing from the mode-coupling theory.
However, the absence of these diagrams is correct, and a consequence of some exact properties of this approach.
By construction, the Mori-Zwanzig projection formalism used in section \ref{sec:Mori-Zwanzig} results in a conservative contribution that gives the \emph{exact} static part of the Kubo Green's function.
The tadpole and loop diagrams are purely real, so that their contributions to the Green's function only result in a shift of the static terms~\cite{Maradudin1962,Tadano2018}.
From this observation, one can infer that the frequencies $\Omega_\lambda(\qvec)$ already include all static diagrams to all orders.
Adding any static diagram, such as the loop or the tadpole, would result in an over-correction, thus explaining their absence from the lowest-order mode-coupling approximation.

Even if the approaches are quite similar, we should expect the mode-coupling theory to have a wider range of validity than bare perturbation theory.
Indeed, being constructed on a truncated expansion of the Hamiltonian with respect to displacements, perturbation theory is known to fail for systems with strong anharmonicity.
In these systems, the high-order contributions to the Hamiltonian can be of similar - or even higher - magnitude than the harmonic contribution.
This is because the perturbative nature of the high order contributions is only a supposition, and there is no guarantee that this holds in general~\cite{Feng2017,Ravichandran2020,Yang2022b}.
On the contrary, the projection operators of eq.(\ref{eq:proj}) and (\ref{eq:proj n}) are constructed so that effects of higher order correlations are minimized.
Such a construction allows the mode-coupling theory to be valid even in strongly anharmonic systems.
One should note that the use of linear response theory for this kind of system is not a problem for the Mori-Zwanzig approach, as the linearity is assumed in the response of the system to a weak probe, not in the behavior of the system itself.

\subsubsection{Self-consistent harmonic approximation}
The self consistent harmonic approximation is a workhorse for the study of anharmonic crystals.
The method is based on the minimization of the Gibbs-Bogoliubov free energy $\Feff$ defined as
\begin{equation}
    \Feff = \Feff_0 + \braket{V(\vec{R}) - \Veff(\vec{R})}_0
\end{equation}
where $\Veff(\vec{R})$ is an effective potential, $\Feff_0$ is its associated free energy and $\braket{O}_0$ is an average computed within the distribution associated with $\Veff(\vec{R})$.
As the name of the method implies, the effective potential has a harmonic form
\begin{equation}
\label{eq:gb free energy}
    \Veff(\vec{R}) = (\vec{R} - \pmb{\Ra})^{\mathrm{T}} \vec{\Theta}^{\mathrm{SCHA}} (\vec{R} - \pmb{\Ra})
\end{equation}
where the effective force constants $\vec{\Theta}^{\mathrm{SCHA}}$ and equilibrium positions $\pmb{\Ra}$ are parameters to optimize.
The advantage of such a potential is that the distribution of displacements around $\vec{\Ra}$ is a multivariate Gaussian.
Consequently, the sampling of atomic positions to compute averages is greatly simplified compared to the true potential.
Using Jensen's inequality, it can be shown that the Gibbs-Bogoliubov is an upper bound to the real free energy $\Fe < \Feff$~\cite{Decoster2004}, so that optimizing the parameters means minimizing $\Feff$.
For the effective force constants, the minimum is obtained when
\begin{equation}
\label{eq:ifc scha}
    \Theta_{ij}^{\mathrm{SCHA}} = \Braket{\frac{\partial^2 V(\vec{R})}{\partial u_i \partial u_j}}_0
\end{equation}
while the effective equilibrium positions are optimized when the average difference of true and harmonic forces is zero.
In practice, the minimization is done self-consistently, and several approaches and implementations can be found in the literature~\cite{Souvatzis2009,Bianco2017,Monacelli2021,Tadano2015,Tadano2018,van_Roekeghem2021,Shulumba2017}.

After the minimization, while the effective force constants can be used to approximate quasiparticles, a more consistent result is obtained by using the Hessian $\vec{\Xi}$ of the free energy~\cite{Bianco2017,Monacelli2021}
\begin{equation}
    \Xi_{ij} = \frac{\partial^2 \Feff}{\partial \Ra_i \partial \Ra_j}
\end{equation}
By seeing the Gibbs-Bogoliubov free energy as a Landau expansion with an order parameter $\Ra$, the free energy Hessian $\vec{\Xi}$ can be understood as the inverse static susceptibility.
However, compared to the mode-coupling approach, where this quantity is exact, $\vec{\bigchi}^{-1}$ is approximated in the SCHA, as only the contribution from the first-order free energy cumulant enters in $\Feff$.

On top of the quasiparticles computed from the free energy Hessian, the dynamical susceptibility can be obtained with a bubble-like contribution~\cite{Bianco2017,Monacelli2021} in which the vertices are computed using the average of the third-order derivative of the potential energy
\begin{equation}
\label{eq:vertex SCHA}
\begin{split}
    \Psi_{ijk}^{\mathrm{SCHA}} =& \frac{1}{\sqrt{M_i M_j M_k}} \Braket{\frac{\partial^3 V(\vec{R})}{\partial R_i \partial R_j \partial R_k}}_0 \\
    =& -\frac{1}{\sqrt{M_i M_j M_k}} \sum_{lm} \frac{\braket{u_l u_m f_k}_0}{\braket{u_i u_l}_0 \braket{u_j u_m}_0}
\end{split}
\end{equation}
where the second line can be derived with an integration by parts.
This procedure can also be compared with our mode-coupling theory.
To show this, we start by assuming that we can approximate the real distribution with the Gaussian associated with $\vec{\Theta}^{\mathrm{SCHA}}$.
With this assumption, the frequency matrix is actually given by the SCHA force constants $\vec{\Theta}^{\mathrm{SCHA}}$, and the random forces are given with by $\delta f_i = f_i + \sum_j \Theta_{ij}^{\mathrm{SCHA}} u_j$.
Applying the mode-coupling decomposition of section \ref{sec:Mode-coupling} to approximate $\vec{K}^{\mathrm{SCHA}}(t)$, it is then simple to show that the vertex appearing in the $(2,2)$-order memory kernel in real space is given by eq.(\ref{eq:vertex SCHA}), as the average of an odd number of displacements cancels out with the Gaussian average.
From this observation, we can conclude that the bubble-like dynamical contribution in the SCHA is an approximation of the mode-coupling theory.

To summarize, two approximations enter in the SCHA compared to the mode-coupling theory.
The first one concerns the static frequencies, which are approximated by the effective harmonic frequencies corrected by a first-order cumulant of the free energy.
The second one is in the vertices in the bubble contribution, which are computed on the basis of a Gaussian approximation.
From a computational point-of-view, the SCHA bears the important advantage of a lower cost due to its simplified sampling compared to (PI)MD.
This is particularly true for the inclusion of nuclear quantum effects which can be added in a simple way.
However higher-order corrections and applications to strongly anharmonic (and hence non-Gaussian) systems are more delicate to justify.

\subsubsection{Average of the Hessian}

In the literature, it has been proposed to define a generalization of the harmonic force constants through an average of the potential energy Hessian~\cite{Morresi2021}
\begin{equation}
\begin{split}
    \Pi_{ij} =& \bigg(\frac{\partial^2 V(\vec{R})}{\partial R_i \partial R_j}\bigg)
\end{split}
\end{equation}
where the only difference with the SCHA force constants of eq.(\ref{eq:ifc scha}) lies in the Hamiltonian used to compute the average.
From the harmonic limit, where $\Pi_{ij}$ reduces to the harmonic force constants, and the similarity with eq.(\ref{eq:ifc scha}), this generalization seems like a sensible choice.
However, it is constructed on the equality $\Pi_{ij} = \Theta_{ij}$ that only holds in a harmonic system.

To highlight the differences between this formulation and the mode-coupling theory, the potential energy Hessian average can be expanded in term of force-force KCF~\cite{Morresi2021}
\begin{equation}
    \Pi_{ij} = \frac{\bbl f_i, f_j \bbr}{\kBT}
\end{equation}
Using this equation, the static part of memory kernel can be formulated as
\begin{equation}
    K_{ij}(0) = \Pi_{ij} - \Theta_{ij}
\end{equation}
Since $\vec{K}$ is a positive definite matrix~\cite{Berne1970}, this equation tells us that $\vec{\Pi}$ is greater that $\vec{\Theta}$, meaning that frequencies computed from the diagonalization of the average Hessian will overestimate the $\Omega_\lambda$.
This result shows that while formulations based on the SCHA and on the true distribution by means of (PI)MD are similar, they are not interchangeable.

\subsubsection{Temperature dependent effective potential}
\label{sec:Discussion TDEP}

TDEP is a method based on the construction of an effective anharmonic potential of the form (here presented up to third order)
\begin{equation}
    \Veff(\vec{R}) = \frac{1}{2} \sum_{ij} \Theta_{ij}^{\mathrm{TDEP}} u_i u_j + \frac{1}{3!} \sum_{ijk} \Psi_{ijk}^{\mathrm{TDEP}} u_i u_j u_k
\end{equation}
The temperature dependence in this potential is introduced through a successive fitting of each order.
First, using a set of positions and forces distributed according to a given temperature, the second order interatomic force constants $\Theta_{ij}$ are fit using ordinary least-squares.
Then the third order force constants $\Psi_{ijk}$ are fit on the residual of the forces.
In principle, this successive fitting can be continued to any order, though usual applications of the method truncate the potential at the third or fourth order.

In applications of TDEP, two approaches have been used to generate the set of positions and forces.
In the original works~\cite{Hellman2011,Hellman2013a}, that we will call MD-TDEP, MD simulations are used.
In the second one, sometimes called sTDEP (for stochastic TDEP), the displacements are computed from an effective harmonic model.
In the latter case, the set of forces and displacements have to be constructed self-consistently, by refitting the second order potential from the set of the previous iteration.

It is important to differentiate the approaches.
On one hand, sTDEP is simply an implementation of the SCHA, since the self-consistent least-squares fit allows to minimize the Gibbs-Bogoliubov free energy in eq.(\ref{eq:gb free energy}).
On the other hand, by using the real distribution, MD-TDEP is an implementation of the mode-coupling theory.
Indeed, as has already been discussed in section \ref{sec:Simulations}, the force constants $\vec{\Theta}^{\mathrm{TDEP}}$ and $\vec{\Psi}^{\mathrm{TDEP}}$ computed from the TDEP methods are equal to the frequency matrix $\vec{\Theta}$ and the real space vertex $\vec{\Psi}$.
Consequently, the perturbation theory usually applied to compute the spectral function is equivalent to the scattering approximation of mode-coupling theory eq.(\ref{eq:gamma lesser}) if only the bubble term is used.
As this is the case for most applications of TDEP in the literature, we have a large number of examples showing the accuracy of the mode-coupling approach.
For instance, the method has been used successfully to explain the anomalous neutron scattering of SnTe and PbTe~\cite{Li2014b}, to study the phase transition of GeTe~\cite{Dangi2021} or the phase diagram of uranium~\cite{Bouchet2017,Ladygin2020} as well as many other phenomena for highly anharmonic systems~\cite{Ding2021,Klarbring2020,Xie2020,He2020}.
Our work provides a theoretical justification for the results of these studies, as well as a generalization, allowing to include quantum effects through the use of PIMD and to include higher order corrections.

\subsection{Markovian approximation and Lorentzian lineshape}

In the limit of small anharmonicity, the spectral lineshapes are often approximated with Lorentzians~\cite{Togo2015,Isaeva2019,Allen2019}.
An interesting conclusion to be taken from the presented theory is that such a solution cannot be an exact representation of the dynamics of the nuclei.
In the Lorentzian approximation, the frequency dependency of the lifetime is neglected and $\Gamma(\omega)$ is approximated with a constant given by $\Gamma_\lambda = \Gamma_\lambda(\Omega_\lambda)$~\cite{Togo2015}.
In the Mori-Zwanzig formalism presented here, such Lorentzian spectra would appear in the context of a Markovian approximation of the memory kernel~\cite{Hijn2010}.
In this case, where it is assumed that $\vec{K}(t)$ decays very quickly compared to $\vec{G}(t)$, the memory kernel can be replaced by a constant friction term $\vec{\Gamma}^{M}(\qvec) = \beta\int_0^\infty dt \vec{K}(\qvec,t)$, so that the Kubo transformed Green's function is written as
\begin{equation}
    \ddot{\vec{G}}(\vec{q}, t) = -\vec{\Omega}^2(\qvec) \vec{G}(\qvec, t) - \vec{\Gamma}^{M}(\qvec) \dot{\vec{G}}(\qvec, t)
\end{equation}
with solutions
\begin{equation}
    \vec{G}(\qvec, t) = \cos \left( \vec{\Omega}(\qvec) t \right) e^{-\vec{\Gamma}^M(\qvec) t} 
\end{equation}
However, it can be observed that this solution is not an even function of time, as requested for the KCF.
For instance, in a Markovian approximation, the property $\frac{d\vec{G}(t)}{dt}\vert_{t=0} = 0$ would not be respected.
This means that a Markovian approximation, and consequently a Lorentzian approximation, will never fully represent the dynamics of the system.

Nevertheless, for midly anharmonic systems, the Markovian limit of the GLE should still be a reliable approximation, and Lorentzian-like lineshapes are frequently observed in spectroscopy measurements.
Consequently, a derivation of the inputs needed for such an approximation is of interest, in order to provide an efficient way to compute the dynamical properties of a system of nuclei from (PI)MD.

\begin{figure*}
    \centering
    \includegraphics[width=\linewidth]{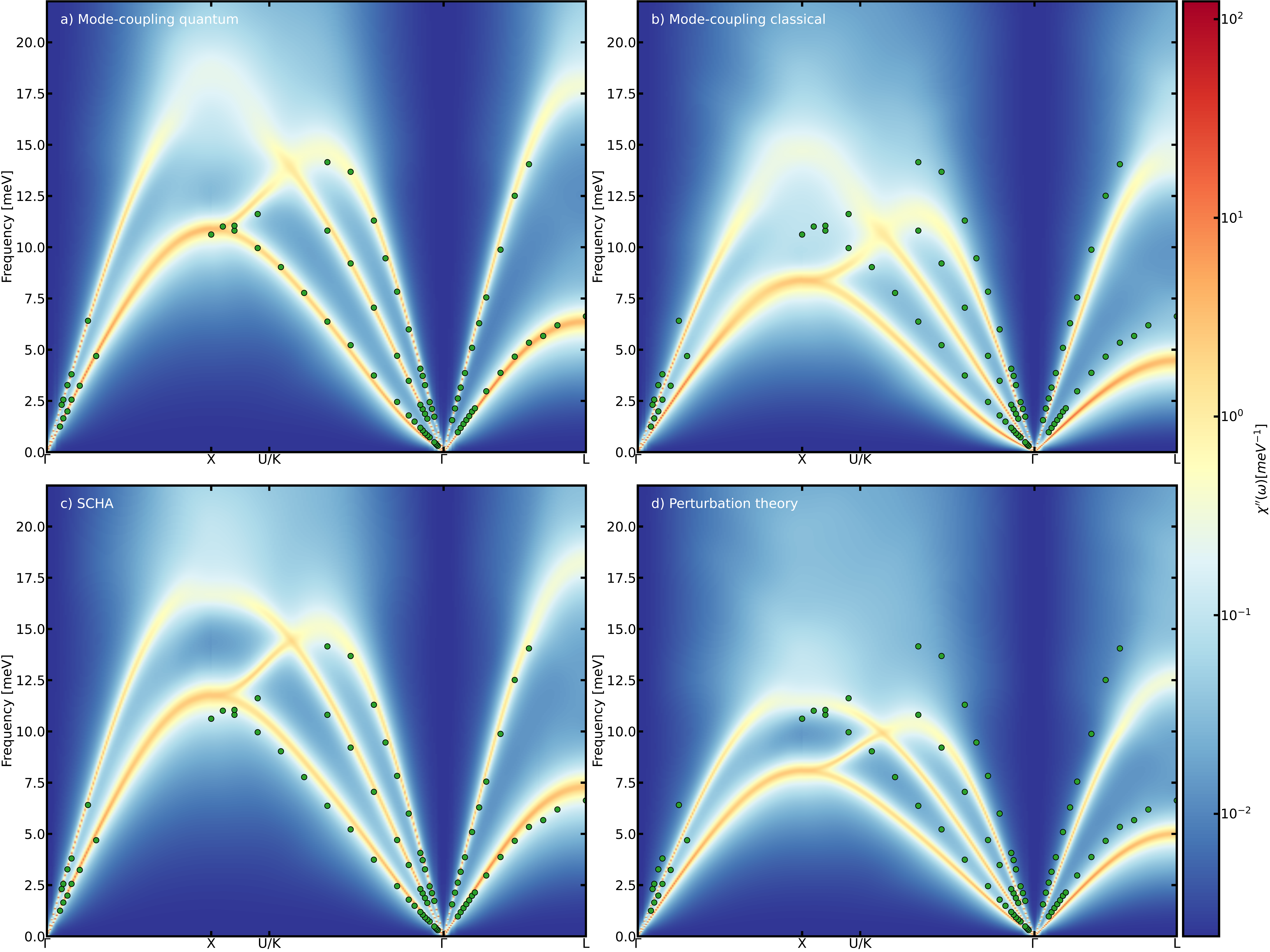}
    \caption{Spectral function $\bigchi''(\omega)$ of fcc \textsuperscript{4}He at $9.03$~cm\textsuperscript{3}/mol and $38$~K computed with the mode-coupling approach in the scattering-like approximation.
    The frequency matrix and second order vertex were computed using PIMD for a), classical MD for b), the SCHA for c) and through perturbation theory for d).
    The green dots are from inelastic neutrons scattering experiment\cite{Eckert1977}.}
    \label{fig:Phonons He fcc}
\end{figure*}

\section{Applications}
\label{sec:Applications}

To demonstrate the accuracy of the theory developed here, we apply it to the fcc phase of \textsuperscript{4}He.
Due to its very small atomic mass, helium is known to show strong nuclear quantum effects (NQE) making it an archetypal example of a quantum crystal~\cite{Cazorla2017}.
Indeed, the zero-point motion in solid helium is large enough to make the system explore the BOS far beyond the range of applicability of the harmonic approximation. This unusual behavior manifests itself in the largest Lindemann ratio (i.e. $\sqrt{\braket{u^2}}/a$ with $\braket{u^2}$ the mean squared atomic displacement and $a$ the lattice parameter) of any material observed at cryogenic temperature~\cite{Cazorla2017}.
The high anharmonicity in helium, which can be found in other rare gas solids, is what started the field of anharmonic lattice dynamics and spawned the development of theories such as the SCHA~\cite{Klein1972,Koehler1966,Werthamer1970,Goldman1968}.

Given these properties, this system represents a perfect case to illustrate that both NQE and strong anharmonicity can be captured with our mode-coupling approach.
To emphasize this point, we performed numerical simulations in a fully quantum setting using PIMD as well as classical MD simulations.
For the sake of comparison, we also performed simulations using both the SCHA and perturbation theory.
To allow for a comparison with inelastic neutron scattering results~\cite{Eckert1977}, all simulations were run at a temperature of $38$~K and with a molar volume of $9.03$ mol\textsuperscript{3}/mol, which correspond to a lattice parameter of $3.915~\mathring{a}$.

To describe the BOS of helium, we use the Aziz pair potential~\cite{Aziz1979}, as it offers an accurate representation of the solid phases of helium at low pressure~\cite{Cazorla2017}.
PIMD and MD simulations were done using the implementation of the Langevin thermostat in LAMMPS~\cite{Thompson2022}, for a duration of $125$~ps with a timestep of $0.5$ fs in both cases and 64 beads for the PIMD.
To compute the frequency matrix and the second order vertex in real space, we used the recursive least-square fit presented in section \ref{sec:Simulations}.
After projecting these tensors on QP following subsection \ref{sec:Quasiparticle space}, we solved the mode-coupling theory equations in the scattering approximation using the greater and lesser Green's function formalism of subsection \ref{sec:quantum Green's function}.
For this step, a $12\times12\times12$ $\qvec$-point grid was sufficient to obtain converged results.
We note that for both PIMD and MD results, the memory kernel is obtained within a quantum formalism, though NQE are included in the frequency matrix and second order vertex only in the PIMD case.
For the SCHA, we used the stochastic TDEP approach presented in subsection \ref{sec:Discussion TDEP} with Bose-Einstein occupations in order to capture NQE.

Before presenting the spectral properties of \textsuperscript{4}He, we first assess the anharmonicity of the system.
To this end, we adopt a recently proposed measure of anharmonicity~\cite{Knoop2020} to our Mori-Zwanzig formalism as
\begin{equation}
    \sigma^\mathrm{A} = \sqrt{\frac{\sum_i\braket{\delta f_i^2}}{\sum_i\braket{f_i^2}}}
\end{equation}
We note, however, that used in this way $\sigma^\mathrm{A}$ is not strictly a measure of anharmonicity, since the frequency matrix already goes beyond the harmonic contribution of the BOS.
It is more accurate to see $\sigma^\mathrm{A}$ as a measure of the dissipative contribution to the dynamics of the system.
For fcc \textsuperscript{4}He, we find $\sigma^\mathrm{A}=0.62$ when computed with PIMD, which would indicate a very dissipative system.
Similarly, using a formulation based on the frequency matrix~\cite{Bottin2020}, we find a Debye temperature of $174$~K, close to the indirect experimental estimation of $154$~K~\cite{Eckert1977}.
At $38$~K, this means that zero-point motion is expected to be important.
Coupled with the high $\sigma^\mathrm{A}$, this confirms that the He system is a stringent test for the ability of the mode-coupling theory to capture both NQE and anharmonicity.

For each of the simulations, the theoretical spectra are presented in Fig.\ref{fig:Phonons He fcc} where they are compared with inelastic neutron scattering results~\cite{Eckert1977}.
Among all the results, PIMD mode coupling presents the best agreement with experiment, with peaks of the theoretical spectral function fully aligning with the experimental phonon frequencies.
On the contrary, the classical MD case severely underestimates phonon frequencies for all reciprocal space directions.
This underestimation is even more important when using (harmonic, 0K based) perturbation theory, showing the need to go to non-perturbative methods for an accurate description of this system.
The spectra computed with the SCHA presents a good agreement with experimental data, even though a slight overestimation is observed compared to the PIMD mode-coupling results, especially for the transverse modes in the $\Gamma-U$ and $\Gamma-L$ directions.

\begin{figure}
    \centering
    \includegraphics[width=\columnwidth]{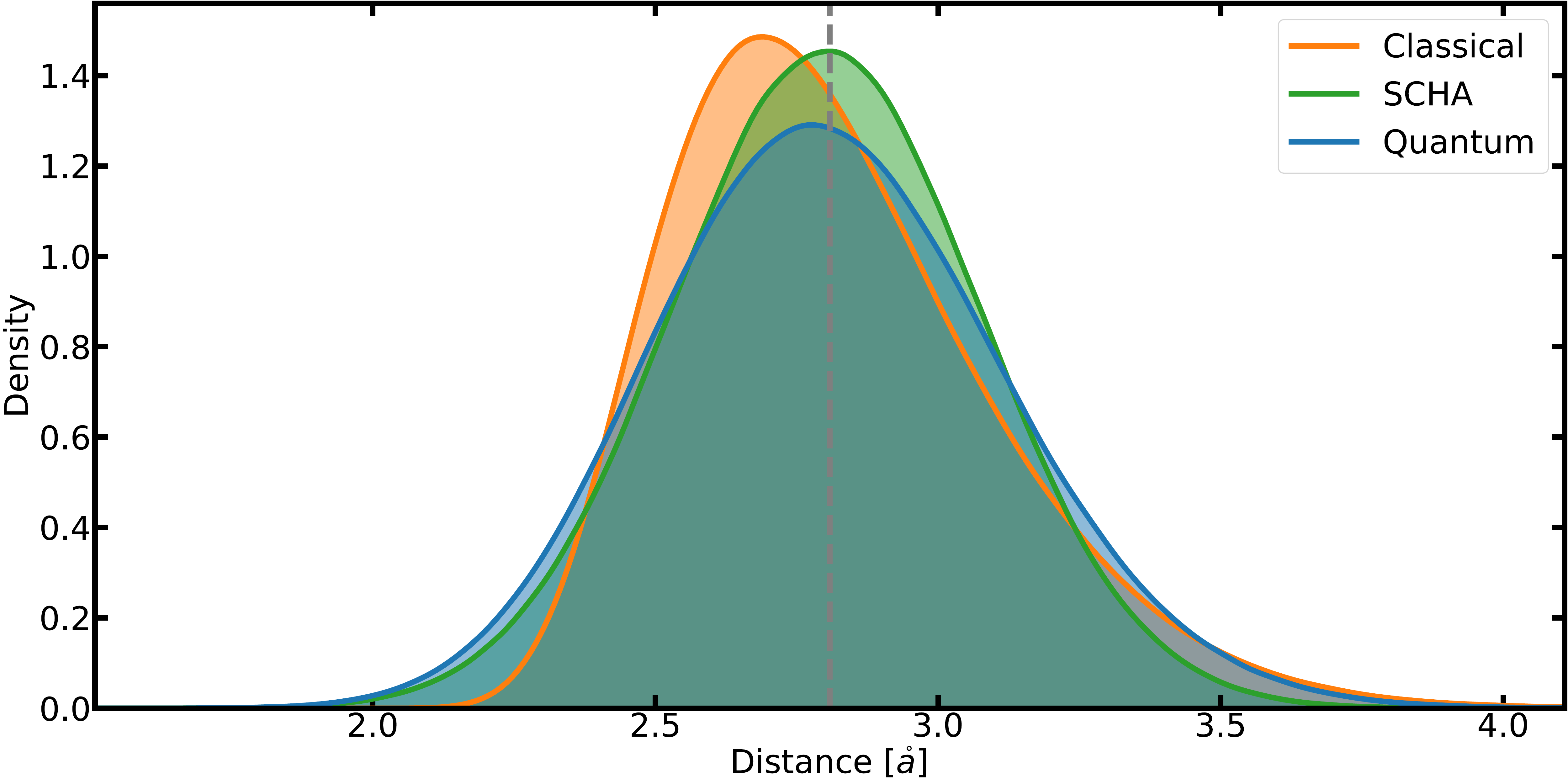}
    \caption{Distribution of first neighbor distances in fcc \textsuperscript{4}He at 38K.
    The vertical grey bar correspond to the equilibrium distance.}
    \label{fig:Neighbor density}
\end{figure}

\begin{figure}[h]
    \centering
    \includegraphics[width=\columnwidth]{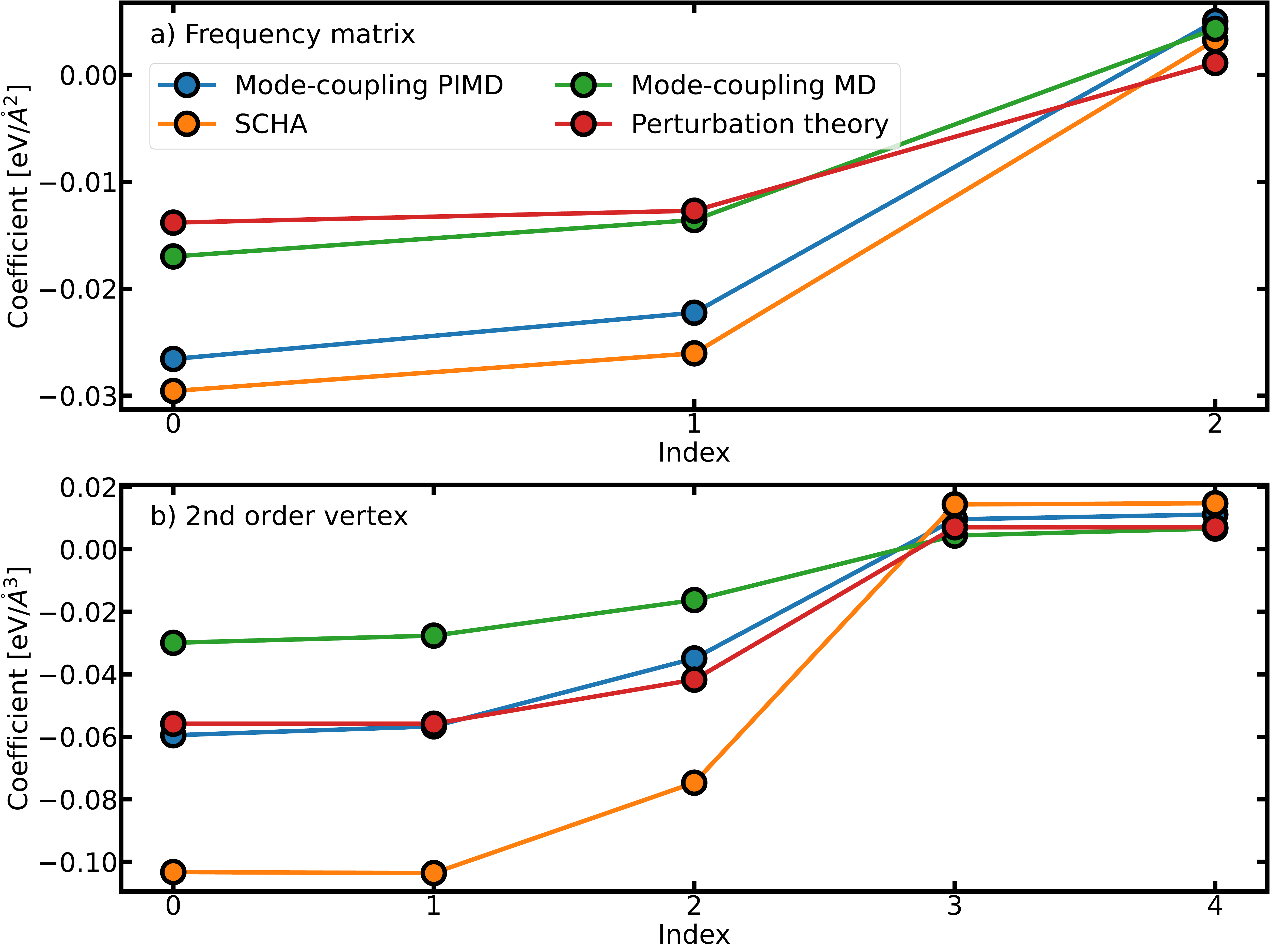}
    \caption{Coefficients entering the a) frequency matrix and b) the 2\textsuperscript{nd} order vertex in the 1\textsuperscript{st} shell of interactions computed with PIMD, MD, SCHA and perturbation theory.}
    \label{fig:Irr. IFC}
\end{figure}

It is interesting to ask why, even with the strong Gaussian approximation, the SCHA still gives an accurate description of the the quasiparticle frequencies, while the classical MD completely disagrees with experimental results.
To answer this question, we plot in Fig. \ref{fig:Neighbor density} the density of first neighbor distances computed with MD, PIMD and the SCHA.
This figure allows us to highlight the effects of NQE on the system.
Indeed, comparing MD and PIMD results, the former presents a high skewness directed to large distances, while the inclusion of zero-point motion in the latter results in a partial re-symmetrization of the density.
Because of this, the tails of the PIMD and SCHA densities are similar, even though their areas differ, indicating a close to symmetric yet non-Gaussian distribution for the displacements in PIMD.
In the end, the differences in exploration of the BOS strongly impact the generalized IFC. This is illustrated in Fig. \ref{fig:Irr. IFC} which compares the symmetry-irreducible coefficients entering the frequency matrix and 2\textsuperscript{nd} order vertex in the first shell of interactions.
If the SCHA and PIMD frequency matrix coefficients are quite close, their magnitudes are almost twice those computed with MD, which explains the softer frequencies observed within the classical mode-coupling theory.
The difference of ensemble sampling used is even more marked for the 2\textsuperscript{nd} order vertex.
Comparing the vertex coefficients for the PIMD and SCHA cases, we observe that the harmonic perturbation theory is coincidentally close to the full PIMD results, while the SCHA has an almost doubled magnitude compared to PIMD. This implies that even if the SCHA has extended validity for effective QP frequencies thanks to NQE, higher order quantities will still be strongly biased by the Gaussian starting point.

From this application, it is clear that the method used to compute the frequency matrix and the interaction vertex significantly impact the resulting generalized susceptibility. 
In future works, it will be interesting to meticulously compare the effects of the sampling strategies in different regimes of anharmonicity.
Moreover, while we limited ourselves to the scattering-like approximation, it remains to test the effects of the self-consistent solution of the mode-coupling equations, especially on systems where the QP picture is broken and the spectral function is strongly non-Lorentzian in shape.

\section{Conclusion}
\label{sec:Conclusion}
In summary, we have developed a non-perturbative theory of lattice dynamics based on the Mori-Zwanzig projection operator formalism.
We derived a Generalized Langevin Equation for the displacement Kubo correlation functions, resulting in an exact description of the dynamics of the system in terms of a conservative part, driven by the frequency matrix $\vec{\Theta}$, and a dissipative part, driven by the memory kernel $\vec{K}(t)$.
From the real space formulation of the dynamics, we were able to project this GLE into a quasiparticle basis, thus giving an equation of motion in reciprocal space.
Using linear response theory, we then derived an exact formulation for the vibrational spectrum, in terms of the spectral function $\vec{\bigchi}''(\omega)$.
In order to approximate the memory kernel, we developed a systematic expansion inspired by mode-coupling theory, in which every new order is constructed to minimize higher-order contributions.
By truncating this expansion to lowest order, we obtain a set of self-consistent equations for the memory kernel.
Within this mode coupling approach, the only inputs needed are the \emph{static} KCFs, and we show how to compute them from path-integral or classical molecular dynamics simulations.

Our Mori-Zwanzig approach to lattice dynamics, being based on the KCF, takes into account the quantum nature of nuclei.
Consequently, this derivation demonstrates how to obtain vibrational properties of quantum systems from PIMD.
Moreover, the classical-like nature of the equations derived allows us to justify the use of classical MD to perform vibrational spectroscopy when quantum effects are not relevant, for instance for systems with heavy nuclei or at high temperature.
We also compare our mode-coupling scheme to more standard perturbation theory, which allows us to derive a scattering-like approximation to the memory kernel.
Our approach gives physical meaning to the temperature-dependent phonons which appear in numerous recent works.
Moreover, we are able to relate the mode-coupling theory to other approaches for anharmonic lattice dynamics.
Notably, we show that in the perturbative limit, the presented theory is equivalent to the usual diagrammatic formulation to lowest order, and that the SCHA can be seen as a Gaussian averaged approximation of it.
We also showed that the TDEP method used in conjunction with (PI)MD is a direct application of the mode-coupling theory, provided that static corrections to the frequencies are \emph{not} applied, as they would double-count interactions.
These comparisons treat all of the main theories for anharmonic lattice dynamics in a unified formalism, and give a better vision of their inter-relationships, strengths and limits.
The Mori-Zwanzig formalism also allows us to show that while the commonly used Lorentzian approximation for lineshapes can be a useful approximation, it can never be an exact description of the lattice dynamics.
Finally, we apply the mode-coupling theory on fcc \textsuperscript{4}He to demonstrate its ability to describe both strong anharmonicity and NQE in quantum crystals, in comparison with other theories of anharmonic lattice dynamics.

The mode-coupling theory of lattice dynamics should apply to a more general class of strongly anharmonic and/or quantum systems, where standard methods are known to fail.
For instance, while we limited ourselves here to crystalline materials, the formalism only requires well defined equilibrium positions so that it would be straightforward to adapt the theory to disordered solids such as alloys or glasses.
This framework can serve as a starting point for new formulations of related properties such as the thermal conductivity~\cite{Simoncelli2019,Isaeva2019,Simoncelli2022,Caldarelli2022,Fiorentino2022}, which will be the focus of future work.

\begin{acknowledgments}
The authors acknowledge the Fonds de la Recherche Scientifique (FRS-FNRS Belgium) and Fonds Wetenschappelijk Onderzoek (FWO Belgium) for EOS project CONNECT (G.A. 40007563), and 
F\'ed\'eration Wallonie Bruxelles and ULiege for funding ARC project DREAMS (G.A. 21/25-11).

Simulation time was awarded by 
the Belgian share of EuroHPC in LUMI hosted by CSC in Finland,
by PRACE on Discoverer at SofiaTech in Bulgaria (optospin project id. 2020225411),
by the CECI (FRS-FNRS Belgium Grant No. 2.5020.11),
and by the Zenobe Tier-1 of the F\'ed\'eration Wallonie-Bruxelles (Walloon Region grant agreement No. 1117545).
\end{acknowledgments}

\bibliography{ModeCoupling}

\providecommand{\noopsort}[1]{}\providecommand{\singleletter}[1]{#1}%
\begin{thebibliography}{97}%
\makeatletter
\providecommand \@ifxundefined [1]{%
 \@ifx{#1\undefined}
}%
\providecommand \@ifnum [1]{%
 \ifnum #1\expandafter \@firstoftwo
 \else \expandafter \@secondoftwo
 \fi
}%
\providecommand \@ifx [1]{%
 \ifx #1\expandafter \@firstoftwo
 \else \expandafter \@secondoftwo
 \fi
}%
\providecommand \natexlab [1]{#1}%
\providecommand \enquote  [1]{``#1''}%
\providecommand \bibnamefont  [1]{#1}%
\providecommand \bibfnamefont [1]{#1}%
\providecommand \citenamefont [1]{#1}%
\providecommand \href@noop [0]{\@secondoftwo}%
\providecommand \href [0]{\begingroup \@sanitize@url \@href}%
\providecommand \@href[1]{\@@startlink{#1}\@@href}%
\providecommand \@@href[1]{\endgroup#1\@@endlink}%
\providecommand \@sanitize@url [0]{\catcode `\\12\catcode `\$12\catcode
  `\&12\catcode `\#12\catcode `\^12\catcode `\_12\catcode `\%12\relax}%
\providecommand \@@startlink[1]{}%
\providecommand \@@endlink[0]{}%
\providecommand \url  [0]{\begingroup\@sanitize@url \@url }%
\providecommand \@url [1]{\endgroup\@href {#1}{\urlprefix }}%
\providecommand \urlprefix  [0]{URL }%
\providecommand \Eprint [0]{\href }%
\providecommand \doibase [0]{https://doi.org/}%
\providecommand \selectlanguage [0]{\@gobble}%
\providecommand \bibinfo  [0]{\@secondoftwo}%
\providecommand \bibfield  [0]{\@secondoftwo}%
\providecommand \translation [1]{[#1]}%
\providecommand \BibitemOpen [0]{}%
\providecommand \bibitemStop [0]{}%
\providecommand \bibitemNoStop [0]{.\EOS\space}%
\providecommand \EOS [0]{\spacefactor3000\relax}%
\providecommand \BibitemShut  [1]{\csname bibitem#1\endcsname}%
\let\auto@bib@innerbib\@empty
\bibitem [{\citenamefont {Born}\ and\ \citenamefont {Huang}(1954)}]{Born1954}%
  \BibitemOpen
  \bibfield  {author} {\bibinfo {author} {\bibfnamefont {M.}~\bibnamefont
  {Born}}\ and\ \bibinfo {author} {\bibfnamefont {K.}~\bibnamefont {Huang}},\
  }\href@noop {} {\emph {\bibinfo {title} {Dynamical theory of crystal
  lattices}}}\ (\bibinfo  {publisher} {Oxford University Press},\ \bibinfo
  {year} {1954})\BibitemShut {NoStop}%
\bibitem [{\citenamefont {Dove}(1993)}]{Dove1993}%
  \BibitemOpen
  \bibfield  {author} {\bibinfo {author} {\bibfnamefont {M.~T.}\ \bibnamefont
  {Dove}},\ }\href {https://doi.org/10.1017/cbo9780511619885} {\emph {\bibinfo
  {title} {Introduction to Lattice Dynamics}}}\ (\bibinfo  {publisher}
  {Cambridge University Press},\ \bibinfo {year} {1993})\BibitemShut {NoStop}%
\bibitem [{\citenamefont {Fultz}(2010)}]{Fultz2010}%
  \BibitemOpen
  \bibfield  {author} {\bibinfo {author} {\bibfnamefont {B.}~\bibnamefont
  {Fultz}},\ }\href {https://doi.org/10.1016/j.pmatsci.2009.05.002} {\bibfield
  {journal} {\bibinfo  {journal} {Prog. in Mater. Sci.}\ }\textbf {\bibinfo
  {volume} {55}},\ \bibinfo {pages} {247} (\bibinfo {year} {2010})}\BibitemShut
  {NoStop}%
\bibitem [{\citenamefont {Klein}\ and\ \citenamefont
  {Horton}(1972)}]{Klein1972}%
  \BibitemOpen
  \bibfield  {author} {\bibinfo {author} {\bibfnamefont {M.~L.}\ \bibnamefont
  {Klein}}\ and\ \bibinfo {author} {\bibfnamefont {G.~K.}\ \bibnamefont
  {Horton}},\ }\href {https://doi.org/10.1007/bf00654839} {\bibfield  {journal}
  {\bibinfo  {journal} {J. Low Temp. Phys.}\ }\textbf {\bibinfo {volume} {9}},\
  \bibinfo {pages} {151} (\bibinfo {year} {1972})}\BibitemShut {NoStop}%
\bibitem [{\citenamefont {Knoop}\ \emph {et~al.}(2020)\citenamefont {Knoop},
  \citenamefont {Purcell}, \citenamefont {Scheffler},\ and\ \citenamefont
  {Carbogno}}]{Knoop2020}%
  \BibitemOpen
  \bibfield  {author} {\bibinfo {author} {\bibfnamefont {F.}~\bibnamefont
  {Knoop}}, \bibinfo {author} {\bibfnamefont {T.~A.~R.}\ \bibnamefont
  {Purcell}}, \bibinfo {author} {\bibfnamefont {M.}~\bibnamefont {Scheffler}},\
  and\ \bibinfo {author} {\bibfnamefont {C.}~\bibnamefont {Carbogno}},\ }\href
  {https://doi.org/10.1103/physrevmaterials.4.083809} {\bibfield  {journal}
  {\bibinfo  {journal} {Phys. Rev. Materials}\ }\textbf {\bibinfo {volume}
  {4}},\ \bibinfo {pages} {083809} (\bibinfo {year} {2020})}\BibitemShut
  {NoStop}%
\bibitem [{\citenamefont {Grimvall}\ \emph {et~al.}(2012)\citenamefont
  {Grimvall}, \citenamefont {Magyari-K\"{o}pe}, \citenamefont
  {Ozoli{\c{n}}{\v{s}}},\ and\ \citenamefont {Persson}}]{Grimvall2012}%
  \BibitemOpen
  \bibfield  {author} {\bibinfo {author} {\bibfnamefont {G.}~\bibnamefont
  {Grimvall}}, \bibinfo {author} {\bibfnamefont {B.}~\bibnamefont
  {Magyari-K\"{o}pe}}, \bibinfo {author} {\bibfnamefont {V.}~\bibnamefont
  {Ozoli{\c{n}}{\v{s}}}},\ and\ \bibinfo {author} {\bibfnamefont {K.~A.}\
  \bibnamefont {Persson}},\ }\href {https://doi.org/10.1103/revmodphys.84.945}
  {\bibfield  {journal} {\bibinfo  {journal} {Rev. Mod. Phys.}\ }\textbf
  {\bibinfo {volume} {84}},\ \bibinfo {pages} {945} (\bibinfo {year}
  {2012})}\BibitemShut {NoStop}%
\bibitem [{\citenamefont {Allen}(2019)}]{Allen2019}%
  \BibitemOpen
  \bibfield  {author} {\bibinfo {author} {\bibfnamefont {P.~B.}\ \bibnamefont
  {Allen}},\ }\href {https://doi.org/10.1142/s0217984920500256} {\bibfield
  {journal} {\bibinfo  {journal} {Modern Phys. Lett. B}\ }\textbf {\bibinfo
  {volume} {34}},\ \bibinfo {pages} {2050025} (\bibinfo {year}
  {2019})}\BibitemShut {NoStop}%
\bibitem [{\citenamefont {Glensk}\ \emph {et~al.}(2015)\citenamefont {Glensk},
  \citenamefont {Grabowski}, \citenamefont {Hickel},\ and\ \citenamefont
  {Neugebauer}}]{Glensk2015}%
  \BibitemOpen
  \bibfield  {author} {\bibinfo {author} {\bibfnamefont {A.}~\bibnamefont
  {Glensk}}, \bibinfo {author} {\bibfnamefont {B.}~\bibnamefont {Grabowski}},
  \bibinfo {author} {\bibfnamefont {T.}~\bibnamefont {Hickel}},\ and\ \bibinfo
  {author} {\bibfnamefont {J.}~\bibnamefont {Neugebauer}},\ }\href
  {https://doi.org/10.1103/PhysRevLett.114.195901} {\bibfield  {journal}
  {\bibinfo  {journal} {Phys. Rev. Lett.}\ }\textbf {\bibinfo {volume} {114}},\
  \bibinfo {pages} {195901} (\bibinfo {year} {2015})}\BibitemShut {NoStop}%
\bibitem [{\citenamefont {Kim}\ \emph {et~al.}(2020)\citenamefont {Kim},
  \citenamefont {Hellman}, \citenamefont {Shulumba}, \citenamefont {Saunders},
  \citenamefont {Lin}, \citenamefont {Smith}, \citenamefont {Herriman},
  \citenamefont {Niedziela}, \citenamefont {Abernathy}, \citenamefont {Li},\
  and\ \citenamefont {Fultz}}]{Kim2020}%
  \BibitemOpen
  \bibfield  {author} {\bibinfo {author} {\bibfnamefont {D.~S.}\ \bibnamefont
  {Kim}}, \bibinfo {author} {\bibfnamefont {O.}~\bibnamefont {Hellman}},
  \bibinfo {author} {\bibfnamefont {N.}~\bibnamefont {Shulumba}}, \bibinfo
  {author} {\bibfnamefont {C.~N.}\ \bibnamefont {Saunders}}, \bibinfo {author}
  {\bibfnamefont {J.~Y.~Y.}\ \bibnamefont {Lin}}, \bibinfo {author}
  {\bibfnamefont {H.~L.}\ \bibnamefont {Smith}}, \bibinfo {author}
  {\bibfnamefont {J.~E.}\ \bibnamefont {Herriman}}, \bibinfo {author}
  {\bibfnamefont {J.~L.}\ \bibnamefont {Niedziela}}, \bibinfo {author}
  {\bibfnamefont {D.~L.}\ \bibnamefont {Abernathy}}, \bibinfo {author}
  {\bibfnamefont {C.~W.}\ \bibnamefont {Li}},\ and\ \bibinfo {author}
  {\bibfnamefont {B.}~\bibnamefont {Fultz}},\ }\href
  {https://doi.org/10.1103/physrevb.102.174311} {\bibfield  {journal} {\bibinfo
   {journal} {Phys. Rev. B}\ }\textbf {\bibinfo {volume} {102}},\ \bibinfo
  {pages} {174311} (\bibinfo {year} {2020})}\BibitemShut {NoStop}%
\bibitem [{\citenamefont {Allen}(2015)}]{Allen2015}%
  \BibitemOpen
  \bibfield  {author} {\bibinfo {author} {\bibfnamefont {P.~B.}\ \bibnamefont
  {Allen}},\ }\href {https://doi.org/10.1103/physrevb.92.064106} {\bibfield
  {journal} {\bibinfo  {journal} {Phys. Rev. B}\ }\textbf {\bibinfo {volume}
  {92}},\ \bibinfo {pages} {064106} (\bibinfo {year} {2015})}\BibitemShut
  {NoStop}%
\bibitem [{\citenamefont {Maradudin}\ and\ \citenamefont
  {Fein}(1962)}]{Maradudin1962}%
  \BibitemOpen
  \bibfield  {author} {\bibinfo {author} {\bibfnamefont {A.~A.}\ \bibnamefont
  {Maradudin}}\ and\ \bibinfo {author} {\bibfnamefont {A.~E.}\ \bibnamefont
  {Fein}},\ }\href {https://doi.org/10.1103/physrev.128.2589} {\bibfield
  {journal} {\bibinfo  {journal} {Phys. Rev.}\ }\textbf {\bibinfo {volume}
  {128}},\ \bibinfo {pages} {2589} (\bibinfo {year} {1962})}\BibitemShut
  {NoStop}%
\bibitem [{\citenamefont {Pathak}(1965)}]{Pathak1965}%
  \BibitemOpen
  \bibfield  {author} {\bibinfo {author} {\bibfnamefont {K.~N.}\ \bibnamefont
  {Pathak}},\ }\href {https://doi.org/10.1103/physrev.139.a1569} {\bibfield
  {journal} {\bibinfo  {journal} {Phys. Rev.}\ }\textbf {\bibinfo {volume}
  {139}},\ \bibinfo {pages} {A1569} (\bibinfo {year} {1965})}\BibitemShut
  {NoStop}%
\bibitem [{\citenamefont {Cowley}(1968)}]{Cowley1968}%
  \BibitemOpen
  \bibfield  {author} {\bibinfo {author} {\bibfnamefont {R.~A.}\ \bibnamefont
  {Cowley}},\ }\href {https://doi.org/10.1088/0034-4885/31/1/303} {\bibfield
  {journal} {\bibinfo  {journal} {Rep. Prog. Phys.}\ }\textbf {\bibinfo
  {volume} {31}},\ \bibinfo {pages} {123} (\bibinfo {year} {1968})}\BibitemShut
  {NoStop}%
\bibitem [{\citenamefont {Broido}\ \emph {et~al.}(2005)\citenamefont {Broido},
  \citenamefont {Ward},\ and\ \citenamefont {Mingo}}]{Broido2005}%
  \BibitemOpen
  \bibfield  {author} {\bibinfo {author} {\bibfnamefont {D.~A.}\ \bibnamefont
  {Broido}}, \bibinfo {author} {\bibfnamefont {A.}~\bibnamefont {Ward}},\ and\
  \bibinfo {author} {\bibfnamefont {N.}~\bibnamefont {Mingo}},\ }\href
  {https://doi.org/10.1103/physrevb.72.014308} {\bibfield  {journal} {\bibinfo
  {journal} {Phys. Rev. B}\ }\textbf {\bibinfo {volume} {72}},\ \bibinfo
  {pages} {014308} (\bibinfo {year} {2005})}\BibitemShut {NoStop}%
\bibitem [{\citenamefont {Togo}\ \emph {et~al.}(2015)\citenamefont {Togo},
  \citenamefont {Chaput},\ and\ \citenamefont {Tanaka}}]{Togo2015}%
  \BibitemOpen
  \bibfield  {author} {\bibinfo {author} {\bibfnamefont {A.}~\bibnamefont
  {Togo}}, \bibinfo {author} {\bibfnamefont {L.}~\bibnamefont {Chaput}},\ and\
  \bibinfo {author} {\bibfnamefont {I.}~\bibnamefont {Tanaka}},\ }\href
  {https://doi.org/10.1103/physrevb.91.094306} {\bibfield  {journal} {\bibinfo
  {journal} {Phys. Rev. B}\ }\textbf {\bibinfo {volume} {91}},\ \bibinfo
  {pages} {094306} (\bibinfo {year} {2015})}\BibitemShut {NoStop}%
\bibitem [{\citenamefont {Li}\ \emph {et~al.}(2014{\natexlab{a}})\citenamefont
  {Li}, \citenamefont {Carrete}, \citenamefont {Katcho},\ and\ \citenamefont
  {Mingo}}]{Li2014}%
  \BibitemOpen
  \bibfield  {author} {\bibinfo {author} {\bibfnamefont {W.}~\bibnamefont
  {Li}}, \bibinfo {author} {\bibfnamefont {J.}~\bibnamefont {Carrete}},
  \bibinfo {author} {\bibfnamefont {N.~A.}\ \bibnamefont {Katcho}},\ and\
  \bibinfo {author} {\bibfnamefont {N.}~\bibnamefont {Mingo}},\ }\href
  {https://doi.org/10.1016/j.cpc.2014.02.015} {\bibfield  {journal} {\bibinfo
  {journal} {Comput. Phys. Commun.}\ }\textbf {\bibinfo {volume} {185}},\
  \bibinfo {pages} {1747} (\bibinfo {year} {2014}{\natexlab{a}})}\BibitemShut
  {NoStop}%
\bibitem [{\citenamefont {Ravichandran}\ and\ \citenamefont
  {Broido}(2020)}]{Ravichandran2020}%
  \BibitemOpen
  \bibfield  {author} {\bibinfo {author} {\bibfnamefont {N.~K.}\ \bibnamefont
  {Ravichandran}}\ and\ \bibinfo {author} {\bibfnamefont {D.}~\bibnamefont
  {Broido}},\ }\href {https://doi.org/10.1103/physrevx.10.021063} {\bibfield
  {journal} {\bibinfo  {journal} {Phys. Rev. X}\ }\textbf {\bibinfo {volume}
  {10}},\ \bibinfo {pages} {021063} (\bibinfo {year} {2020})}\BibitemShut
  {NoStop}%
\bibitem [{\citenamefont {Sun}\ \emph {et~al.}(2010)\citenamefont {Sun},
  \citenamefont {Shen},\ and\ \citenamefont {Allen}}]{Sun2010a}%
  \BibitemOpen
  \bibfield  {author} {\bibinfo {author} {\bibfnamefont {T.}~\bibnamefont
  {Sun}}, \bibinfo {author} {\bibfnamefont {X.}~\bibnamefont {Shen}},\ and\
  \bibinfo {author} {\bibfnamefont {P.~B.}\ \bibnamefont {Allen}},\ }\href
  {https://doi.org/10.1103/physrevb.82.224304} {\bibfield  {journal} {\bibinfo
  {journal} {Phys. Rev. B}\ }\textbf {\bibinfo {volume} {82}},\ \bibinfo
  {pages} {224304} (\bibinfo {year} {2010})}\BibitemShut {NoStop}%
\bibitem [{\citenamefont {Sun}\ and\ \citenamefont {Allen}(2010)}]{Sun2010b}%
  \BibitemOpen
  \bibfield  {author} {\bibinfo {author} {\bibfnamefont {T.}~\bibnamefont
  {Sun}}\ and\ \bibinfo {author} {\bibfnamefont {P.~B.}\ \bibnamefont
  {Allen}},\ }\href {https://doi.org/10.1103/physrevb.82.224305} {\bibfield
  {journal} {\bibinfo  {journal} {Phys. Rev. B}\ }\textbf {\bibinfo {volume}
  {82}},\ \bibinfo {pages} {224305} (\bibinfo {year} {2010})}\BibitemShut
  {NoStop}%
\bibitem [{\citenamefont {Xia}\ \emph {et~al.}(2020{\natexlab{a}})\citenamefont
  {Xia}, \citenamefont {Hegde}, \citenamefont {Pal}, \citenamefont {Hua},
  \citenamefont {Gaines}, \citenamefont {Patel}, \citenamefont {He},
  \citenamefont {Aykol},\ and\ \citenamefont {Wolverton}}]{Xia2020}%
  \BibitemOpen
  \bibfield  {author} {\bibinfo {author} {\bibfnamefont {Y.}~\bibnamefont
  {Xia}}, \bibinfo {author} {\bibfnamefont {V.~I.}\ \bibnamefont {Hegde}},
  \bibinfo {author} {\bibfnamefont {K.}~\bibnamefont {Pal}}, \bibinfo {author}
  {\bibfnamefont {X.}~\bibnamefont {Hua}}, \bibinfo {author} {\bibfnamefont
  {D.}~\bibnamefont {Gaines}}, \bibinfo {author} {\bibfnamefont
  {S.}~\bibnamefont {Patel}}, \bibinfo {author} {\bibfnamefont
  {J.}~\bibnamefont {He}}, \bibinfo {author} {\bibfnamefont {M.}~\bibnamefont
  {Aykol}},\ and\ \bibinfo {author} {\bibfnamefont {C.}~\bibnamefont
  {Wolverton}},\ }\href {https://doi.org/10.1103/physrevx.10.041029} {\bibfield
   {journal} {\bibinfo  {journal} {Phys. Rev. X}\ }\textbf {\bibinfo {volume}
  {10}},\ \bibinfo {pages} {041029} (\bibinfo {year}
  {2020}{\natexlab{a}})}\BibitemShut {NoStop}%
\bibitem [{\citenamefont {Yang}\ \emph
  {et~al.}(2022{\natexlab{a}})\citenamefont {Yang}, \citenamefont {Feng},
  \citenamefont {Li},\ and\ \citenamefont {Ruan}}]{Yang2022b}%
  \BibitemOpen
  \bibfield  {author} {\bibinfo {author} {\bibfnamefont {X.}~\bibnamefont
  {Yang}}, \bibinfo {author} {\bibfnamefont {T.}~\bibnamefont {Feng}}, \bibinfo
  {author} {\bibfnamefont {J.}~\bibnamefont {Li}},\ and\ \bibinfo {author}
  {\bibfnamefont {X.}~\bibnamefont {Ruan}},\ }\href
  {https://doi.org/10.1103/physrevb.105.115205} {\bibfield  {journal} {\bibinfo
   {journal} {Phys. Rev. B}\ }\textbf {\bibinfo {volume} {105}},\ \bibinfo
  {pages} {115205} (\bibinfo {year} {2022}{\natexlab{a}})}\BibitemShut
  {NoStop}%
\bibitem [{\citenamefont {Koehler}(1966)}]{Koehler1966}%
  \BibitemOpen
  \bibfield  {author} {\bibinfo {author} {\bibfnamefont {T.~R.}\ \bibnamefont
  {Koehler}},\ }\href {https://doi.org/10.1103/physrevlett.17.89} {\bibfield
  {journal} {\bibinfo  {journal} {Phys. Rev. Lett.}\ }\textbf {\bibinfo
  {volume} {17}},\ \bibinfo {pages} {89} (\bibinfo {year} {1966})}\BibitemShut
  {NoStop}%
\bibitem [{\citenamefont {Werthamer}(1970)}]{Werthamer1970}%
  \BibitemOpen
  \bibfield  {author} {\bibinfo {author} {\bibfnamefont {N.~R.}\ \bibnamefont
  {Werthamer}},\ }\href {https://doi.org/10.1103/physrevb.1.572} {\bibfield
  {journal} {\bibinfo  {journal} {Phys. Rev. B}\ }\textbf {\bibinfo {volume}
  {1}},\ \bibinfo {pages} {572} (\bibinfo {year} {1970})}\BibitemShut {NoStop}%
\bibitem [{\citenamefont {Ravichandran}\ and\ \citenamefont
  {Broido}(2018)}]{Ravichandran2018}%
  \BibitemOpen
  \bibfield  {author} {\bibinfo {author} {\bibfnamefont {N.~K.}\ \bibnamefont
  {Ravichandran}}\ and\ \bibinfo {author} {\bibfnamefont {D.}~\bibnamefont
  {Broido}},\ }\href {https://doi.org/10.1103/physrevb.98.085205} {\bibfield
  {journal} {\bibinfo  {journal} {Phys. Rev. B}\ }\textbf {\bibinfo {volume}
  {98}},\ \bibinfo {pages} {085205} (\bibinfo {year} {2018})}\BibitemShut
  {NoStop}%
\bibitem [{\citenamefont {Esfarjani}\ and\ \citenamefont
  {Liang}(2020)}]{Esfarjani2020}%
  \BibitemOpen
  \bibfield  {author} {\bibinfo {author} {\bibfnamefont {K.}~\bibnamefont
  {Esfarjani}}\ and\ \bibinfo {author} {\bibfnamefont {Y.}~\bibnamefont
  {Liang}},\ }in\ \href {https://doi.org/10.1088/978-0-7503-1738-2ch7} {\emph
  {\bibinfo {booktitle} {Nanoscale Energy Transport}}}\ (\bibinfo  {publisher}
  {{IOP} Publishing},\ \bibinfo {year} {2020})\ pp.\ \bibinfo {pages}
  {7--1--7--35}\BibitemShut {NoStop}%
\bibitem [{\citenamefont {Souvatzis}\ \emph {et~al.}(2009)\citenamefont
  {Souvatzis}, \citenamefont {Eriksson}, \citenamefont {Katsnelson},\ and\
  \citenamefont {Rudin}}]{Souvatzis2009}%
  \BibitemOpen
  \bibfield  {author} {\bibinfo {author} {\bibfnamefont {P.}~\bibnamefont
  {Souvatzis}}, \bibinfo {author} {\bibfnamefont {O.}~\bibnamefont {Eriksson}},
  \bibinfo {author} {\bibfnamefont {M.}~\bibnamefont {Katsnelson}},\ and\
  \bibinfo {author} {\bibfnamefont {S.}~\bibnamefont {Rudin}},\ }\href
  {https://doi.org/10.1016/j.commatsci.2008.06.016} {\bibfield  {journal}
  {\bibinfo  {journal} {Comput. Mater. Sci.}\ }\textbf {\bibinfo {volume}
  {44}},\ \bibinfo {pages} {888} (\bibinfo {year} {2009})}\BibitemShut
  {NoStop}%
\bibitem [{\citenamefont {Tadano}\ and\ \citenamefont
  {Tsuneyuki}(2015)}]{Tadano2015}%
  \BibitemOpen
  \bibfield  {author} {\bibinfo {author} {\bibfnamefont {T.}~\bibnamefont
  {Tadano}}\ and\ \bibinfo {author} {\bibfnamefont {S.}~\bibnamefont
  {Tsuneyuki}},\ }\href {https://doi.org/10.1103/physrevb.92.054301} {\bibfield
   {journal} {\bibinfo  {journal} {Phys. Rev. B}\ }\textbf {\bibinfo {volume}
  {92}},\ \bibinfo {pages} {054301} (\bibinfo {year} {2015})}\BibitemShut
  {NoStop}%
\bibitem [{\citenamefont {Tadano}\ and\ \citenamefont
  {Tsuneyuki}(2018)}]{Tadano2018}%
  \BibitemOpen
  \bibfield  {author} {\bibinfo {author} {\bibfnamefont {T.}~\bibnamefont
  {Tadano}}\ and\ \bibinfo {author} {\bibfnamefont {S.}~\bibnamefont
  {Tsuneyuki}},\ }\href {https://doi.org/10.7566/jpsj.87.041015} {\bibfield
  {journal} {\bibinfo  {journal} {J. Phys. Soc. Jpn}\ }\textbf {\bibinfo
  {volume} {87}},\ \bibinfo {pages} {041015} (\bibinfo {year}
  {2018})}\BibitemShut {NoStop}%
\bibitem [{\citenamefont {Bianco}\ \emph {et~al.}(2017)\citenamefont {Bianco},
  \citenamefont {Errea}, \citenamefont {Paulatto}, \citenamefont {Calandra},\
  and\ \citenamefont {Mauri}}]{Bianco2017}%
  \BibitemOpen
  \bibfield  {author} {\bibinfo {author} {\bibfnamefont {R.}~\bibnamefont
  {Bianco}}, \bibinfo {author} {\bibfnamefont {I.}~\bibnamefont {Errea}},
  \bibinfo {author} {\bibfnamefont {L.}~\bibnamefont {Paulatto}}, \bibinfo
  {author} {\bibfnamefont {M.}~\bibnamefont {Calandra}},\ and\ \bibinfo
  {author} {\bibfnamefont {F.}~\bibnamefont {Mauri}},\ }\href
  {https://doi.org/10.1103/physrevb.96.014111} {\bibfield  {journal} {\bibinfo
  {journal} {Phys. Rev. B}\ }\textbf {\bibinfo {volume} {96}},\ \bibinfo
  {pages} {014111} (\bibinfo {year} {2017})}\BibitemShut {NoStop}%
\bibitem [{\citenamefont {Monacelli}\ \emph {et~al.}(2021)\citenamefont
  {Monacelli}, \citenamefont {Bianco}, \citenamefont {Cherubini}, \citenamefont
  {Calandra}, \citenamefont {Errea},\ and\ \citenamefont
  {Mauri}}]{Monacelli2021}%
  \BibitemOpen
  \bibfield  {author} {\bibinfo {author} {\bibfnamefont {L.}~\bibnamefont
  {Monacelli}}, \bibinfo {author} {\bibfnamefont {R.}~\bibnamefont {Bianco}},
  \bibinfo {author} {\bibfnamefont {M.}~\bibnamefont {Cherubini}}, \bibinfo
  {author} {\bibfnamefont {M.}~\bibnamefont {Calandra}}, \bibinfo {author}
  {\bibfnamefont {I.}~\bibnamefont {Errea}},\ and\ \bibinfo {author}
  {\bibfnamefont {F.}~\bibnamefont {Mauri}},\ }\href
  {https://doi.org/10.1088/1361-648x/ac066b} {\bibfield  {journal} {\bibinfo
  {journal} {J. Phys.: Condens. Matter}\ }\textbf {\bibinfo {volume} {33}},\
  \bibinfo {pages} {363001} (\bibinfo {year} {2021})}\BibitemShut {NoStop}%
\bibitem [{\citenamefont {van Roekeghem}\ \emph {et~al.}(2021)\citenamefont
  {van Roekeghem}, \citenamefont {Carrete},\ and\ \citenamefont
  {Mingo}}]{van_Roekeghem2021}%
  \BibitemOpen
  \bibfield  {author} {\bibinfo {author} {\bibfnamefont {A.}~\bibnamefont {van
  Roekeghem}}, \bibinfo {author} {\bibfnamefont {J.}~\bibnamefont {Carrete}},\
  and\ \bibinfo {author} {\bibfnamefont {N.}~\bibnamefont {Mingo}},\ }\href
  {https://doi.org/10.1016/j.cpc.2021.107945} {\bibfield  {journal} {\bibinfo
  {journal} {Comput. Phys. Commun.}\ }\textbf {\bibinfo {volume} {263}},\
  \bibinfo {pages} {107945} (\bibinfo {year} {2021})}\BibitemShut {NoStop}%
\bibitem [{\citenamefont {Choquard}(1967)}]{Choquard1967}%
  \BibitemOpen
  \bibfield  {author} {\bibinfo {author} {\bibfnamefont {P.}~\bibnamefont
  {Choquard}},\ }\href@noop {} {\emph {\bibinfo {title} {The anharmonic
  crystal}}}\ (\bibinfo  {publisher} {WA Benjamin},\ \bibinfo {year}
  {1967})\BibitemShut {NoStop}%
\bibitem [{\citenamefont {Samathiyakanit}\ and\ \citenamefont
  {Glyde}(1973)}]{Samathiyakanit1973}%
  \BibitemOpen
  \bibfield  {author} {\bibinfo {author} {\bibfnamefont {V.}~\bibnamefont
  {Samathiyakanit}}\ and\ \bibinfo {author} {\bibfnamefont {H.~R.}\
  \bibnamefont {Glyde}},\ }\href {https://doi.org/10.1088/0022-3719/6/7/009}
  {\bibfield  {journal} {\bibinfo  {journal} {J. Phys. C: Solid State Phys.}\
  }\textbf {\bibinfo {volume} {6}},\ \bibinfo {pages} {1166} (\bibinfo {year}
  {1973})}\BibitemShut {NoStop}%
\bibitem [{\citenamefont {Levy}\ \emph {et~al.}(1984)\citenamefont {Levy},
  \citenamefont {Srinivasan}, \citenamefont {Olson},\ and\ \citenamefont
  {McCammon}}]{Levy1984}%
  \BibitemOpen
  \bibfield  {author} {\bibinfo {author} {\bibfnamefont {R.~M.}\ \bibnamefont
  {Levy}}, \bibinfo {author} {\bibfnamefont {A.~R.}\ \bibnamefont
  {Srinivasan}}, \bibinfo {author} {\bibfnamefont {W.~K.}\ \bibnamefont
  {Olson}},\ and\ \bibinfo {author} {\bibfnamefont {J.~A.}\ \bibnamefont
  {McCammon}},\ }\href {https://doi.org/10.1002/bip.360230610} {\bibfield
  {journal} {\bibinfo  {journal} {Biopolymers}\ }\textbf {\bibinfo {volume}
  {23}},\ \bibinfo {pages} {1099} (\bibinfo {year} {1984})}\BibitemShut
  {NoStop}%
\bibitem [{\citenamefont {Kong}\ \emph {et~al.}(2009)\citenamefont {Kong},
  \citenamefont {Bartels}, \citenamefont {Campa{\~{n}}{\'{a}}}, \citenamefont
  {Denniston},\ and\ \citenamefont {M\"{u}ser}}]{Kong2009}%
  \BibitemOpen
  \bibfield  {author} {\bibinfo {author} {\bibfnamefont {L.~T.}\ \bibnamefont
  {Kong}}, \bibinfo {author} {\bibfnamefont {G.}~\bibnamefont {Bartels}},
  \bibinfo {author} {\bibfnamefont {C.}~\bibnamefont {Campa{\~{n}}{\'{a}}}},
  \bibinfo {author} {\bibfnamefont {C.}~\bibnamefont {Denniston}},\ and\
  \bibinfo {author} {\bibfnamefont {M.~H.}\ \bibnamefont {M\"{u}ser}},\ }\href
  {https://doi.org/10.1016/j.cpc.2008.12.035} {\bibfield  {journal} {\bibinfo
  {journal} {Comput. Phys. Commun.}\ }\textbf {\bibinfo {volume} {180}},\
  \bibinfo {pages} {1004} (\bibinfo {year} {2009})}\BibitemShut {NoStop}%
\bibitem [{\citenamefont {Kong}(2011)}]{Kong2011}%
  \BibitemOpen
  \bibfield  {author} {\bibinfo {author} {\bibfnamefont {L.~T.}\ \bibnamefont
  {Kong}},\ }\href {https://doi.org/10.1016/j.cpc.2011.04.019} {\bibfield
  {journal} {\bibinfo  {journal} {Comput. Phys. Commun.}\ }\textbf {\bibinfo
  {volume} {182}},\ \bibinfo {pages} {2201} (\bibinfo {year}
  {2011})}\BibitemShut {NoStop}%
\bibitem [{\citenamefont {Hellman}\ \emph {et~al.}(2011)\citenamefont
  {Hellman}, \citenamefont {Abrikosov},\ and\ \citenamefont
  {Simak}}]{Hellman2011}%
  \BibitemOpen
  \bibfield  {author} {\bibinfo {author} {\bibfnamefont {O.}~\bibnamefont
  {Hellman}}, \bibinfo {author} {\bibfnamefont {I.~A.}\ \bibnamefont
  {Abrikosov}},\ and\ \bibinfo {author} {\bibfnamefont {S.~I.}\ \bibnamefont
  {Simak}},\ }\href {https://doi.org/10.1103/physrevb.84.180301} {\bibfield
  {journal} {\bibinfo  {journal} {Phys. Rev. B}\ }\textbf {\bibinfo {volume}
  {84}},\ \bibinfo {pages} {180301(R)} (\bibinfo {year} {2011})}\BibitemShut
  {NoStop}%
\bibitem [{\citenamefont {Hellman}\ \emph {et~al.}(2013)\citenamefont
  {Hellman}, \citenamefont {Steneteg}, \citenamefont {Abrikosov},\ and\
  \citenamefont {Simak}}]{Hellman2013a}%
  \BibitemOpen
  \bibfield  {author} {\bibinfo {author} {\bibfnamefont {O.}~\bibnamefont
  {Hellman}}, \bibinfo {author} {\bibfnamefont {P.}~\bibnamefont {Steneteg}},
  \bibinfo {author} {\bibfnamefont {I.~A.}\ \bibnamefont {Abrikosov}},\ and\
  \bibinfo {author} {\bibfnamefont {S.~I.}\ \bibnamefont {Simak}},\ }\href
  {https://doi.org/10.1103/physrevb.87.104111} {\bibfield  {journal} {\bibinfo
  {journal} {Phys. Rev. B}\ }\textbf {\bibinfo {volume} {87}},\ \bibinfo
  {pages} {104111} (\bibinfo {year} {2013})}\BibitemShut {NoStop}%
\bibitem [{\citenamefont {Hellman}\ and\ \citenamefont
  {Abrikosov}(2013)}]{Hellman2013b}%
  \BibitemOpen
  \bibfield  {author} {\bibinfo {author} {\bibfnamefont {O.}~\bibnamefont
  {Hellman}}\ and\ \bibinfo {author} {\bibfnamefont {I.~A.}\ \bibnamefont
  {Abrikosov}},\ }\href {https://doi.org/10.1103/physrevb.88.144301} {\bibfield
   {journal} {\bibinfo  {journal} {Phys. Rev. B}\ }\textbf {\bibinfo {volume}
  {88}},\ \bibinfo {pages} {144301} (\bibinfo {year} {2013})}\BibitemShut
  {NoStop}%
\bibitem [{\citenamefont {Klarbring}\ \emph {et~al.}(2020)\citenamefont
  {Klarbring}, \citenamefont {Hellman}, \citenamefont {Abrikosov},\ and\
  \citenamefont {Simak}}]{Klarbring2020}%
  \BibitemOpen
  \bibfield  {author} {\bibinfo {author} {\bibfnamefont {J.}~\bibnamefont
  {Klarbring}}, \bibinfo {author} {\bibfnamefont {O.}~\bibnamefont {Hellman}},
  \bibinfo {author} {\bibfnamefont {I.~A.}\ \bibnamefont {Abrikosov}},\ and\
  \bibinfo {author} {\bibfnamefont {S.~I.}\ \bibnamefont {Simak}},\ }\href
  {https://doi.org/10.1103/physrevlett.125.045701} {\bibfield  {journal}
  {\bibinfo  {journal} {Phys. Rev. Lett.}\ }\textbf {\bibinfo {volume} {125}},\
  \bibinfo {pages} {045701} (\bibinfo {year} {2020})}\BibitemShut {NoStop}%
\bibitem [{\citenamefont {Mei}\ \emph {et~al.}(2015)\citenamefont {Mei},
  \citenamefont {Hellman}, \citenamefont {Wireklint}, \citenamefont
  {Schlep\"{u}tz}, \citenamefont {Sangiovanni}, \citenamefont {Alling},
  \citenamefont {Rockett}, \citenamefont {Hultman}, \citenamefont {Petrov},\
  and\ \citenamefont {Greene}}]{Mei2015}%
  \BibitemOpen
  \bibfield  {author} {\bibinfo {author} {\bibfnamefont {A.~B.}\ \bibnamefont
  {Mei}}, \bibinfo {author} {\bibfnamefont {O.}~\bibnamefont {Hellman}},
  \bibinfo {author} {\bibfnamefont {N.}~\bibnamefont {Wireklint}}, \bibinfo
  {author} {\bibfnamefont {C.~M.}\ \bibnamefont {Schlep\"{u}tz}}, \bibinfo
  {author} {\bibfnamefont {D.~G.}\ \bibnamefont {Sangiovanni}}, \bibinfo
  {author} {\bibfnamefont {B.}~\bibnamefont {Alling}}, \bibinfo {author}
  {\bibfnamefont {A.}~\bibnamefont {Rockett}}, \bibinfo {author} {\bibfnamefont
  {L.}~\bibnamefont {Hultman}}, \bibinfo {author} {\bibfnamefont
  {I.}~\bibnamefont {Petrov}},\ and\ \bibinfo {author} {\bibfnamefont {J.~E.}\
  \bibnamefont {Greene}},\ }\href {https://doi.org/10.1103/physrevb.91.054101}
  {\bibfield  {journal} {\bibinfo  {journal} {Phys. Rev. B}\ }\textbf {\bibinfo
  {volume} {91}},\ \bibinfo {pages} {054101} (\bibinfo {year}
  {2015})}\BibitemShut {NoStop}%
\bibitem [{\citenamefont {Romero}\ \emph {et~al.}(2015)\citenamefont {Romero},
  \citenamefont {Gross}, \citenamefont {Verstraete},\ and\ \citenamefont
  {Hellman}}]{Romero2015}%
  \BibitemOpen
  \bibfield  {author} {\bibinfo {author} {\bibfnamefont {A.~H.}\ \bibnamefont
  {Romero}}, \bibinfo {author} {\bibfnamefont {E.~K.~U.}\ \bibnamefont
  {Gross}}, \bibinfo {author} {\bibfnamefont {M.~J.}\ \bibnamefont
  {Verstraete}},\ and\ \bibinfo {author} {\bibfnamefont {O.}~\bibnamefont
  {Hellman}},\ }\href {https://doi.org/10.1103/physrevb.91.214310} {\bibfield
  {journal} {\bibinfo  {journal} {Phys. Rev. B}\ }\textbf {\bibinfo {volume}
  {91}},\ \bibinfo {pages} {214310} (\bibinfo {year} {2015})}\BibitemShut
  {NoStop}%
\bibitem [{\citenamefont {Chaney}\ \emph {et~al.}(2021)\citenamefont {Chaney},
  \citenamefont {Castellano}, \citenamefont {Bosak}, \citenamefont {Bouchet},
  \citenamefont {Bottin}, \citenamefont {Dorado}, \citenamefont {Paolasini},
  \citenamefont {Rennie}, \citenamefont {Bell}, \citenamefont {Springell},\
  and\ \citenamefont {Lander}}]{Chaney2021}%
  \BibitemOpen
  \bibfield  {author} {\bibinfo {author} {\bibfnamefont {D.}~\bibnamefont
  {Chaney}}, \bibinfo {author} {\bibfnamefont {A.}~\bibnamefont {Castellano}},
  \bibinfo {author} {\bibfnamefont {A.}~\bibnamefont {Bosak}}, \bibinfo
  {author} {\bibfnamefont {J.}~\bibnamefont {Bouchet}}, \bibinfo {author}
  {\bibfnamefont {F.}~\bibnamefont {Bottin}}, \bibinfo {author} {\bibfnamefont
  {B.}~\bibnamefont {Dorado}}, \bibinfo {author} {\bibfnamefont
  {L.}~\bibnamefont {Paolasini}}, \bibinfo {author} {\bibfnamefont
  {S.}~\bibnamefont {Rennie}}, \bibinfo {author} {\bibfnamefont
  {C.}~\bibnamefont {Bell}}, \bibinfo {author} {\bibfnamefont {R.}~\bibnamefont
  {Springell}},\ and\ \bibinfo {author} {\bibfnamefont {G.~H.}\ \bibnamefont
  {Lander}},\ }\href {https://doi.org/10.1103/physrevmaterials.5.035004}
  {\bibfield  {journal} {\bibinfo  {journal} {Phys. Rev. Materials}\ }\textbf
  {\bibinfo {volume} {5}},\ \bibinfo {pages} {035004} (\bibinfo {year}
  {2021})}\BibitemShut {NoStop}%
\bibitem [{\citenamefont {Hele}(2017)}]{Hele2017}%
  \BibitemOpen
  \bibfield  {author} {\bibinfo {author} {\bibfnamefont {T.~J.~H.}\
  \bibnamefont {Hele}},\ }\href {https://doi.org/10.1080/00268976.2017.1303548}
  {\bibfield  {journal} {\bibinfo  {journal} {Mol. Phys.}\ }\textbf {\bibinfo
  {volume} {115}},\ \bibinfo {pages} {1435} (\bibinfo {year}
  {2017})}\BibitemShut {NoStop}%
\bibitem [{\citenamefont {Schofield}(1960)}]{Schofield1960}%
  \BibitemOpen
  \bibfield  {author} {\bibinfo {author} {\bibfnamefont {P.}~\bibnamefont
  {Schofield}},\ }\href {https://doi.org/10.1103/physrevlett.4.239} {\bibfield
  {journal} {\bibinfo  {journal} {Phys. Rev. Lett.}\ }\textbf {\bibinfo
  {volume} {4}},\ \bibinfo {pages} {239} (\bibinfo {year} {1960})}\BibitemShut
  {NoStop}%
\bibitem [{\citenamefont {Egorov}\ \emph {et~al.}(1999)\citenamefont {Egorov},
  \citenamefont {Everitt},\ and\ \citenamefont {Skinner}}]{Egorov1999}%
  \BibitemOpen
  \bibfield  {author} {\bibinfo {author} {\bibfnamefont {S.~A.}\ \bibnamefont
  {Egorov}}, \bibinfo {author} {\bibfnamefont {K.~F.}\ \bibnamefont
  {Everitt}},\ and\ \bibinfo {author} {\bibfnamefont {J.~L.}\ \bibnamefont
  {Skinner}},\ }\href {https://doi.org/10.1021/jp9919314} {\bibfield  {journal}
  {\bibinfo  {journal} {J. Phys. Chem. A}\ }\textbf {\bibinfo {volume} {103}},\
  \bibinfo {pages} {9494} (\bibinfo {year} {1999})}\BibitemShut {NoStop}%
\bibitem [{\citenamefont {Ram{\'\i}rez}\ \emph {et~al.}(2004)\citenamefont
  {Ram{\'\i}rez}, \citenamefont {L{\'o}pez-Ciudad}, \citenamefont {Kumar~P},\
  and\ \citenamefont {Marx}}]{Ramirez2004}%
  \BibitemOpen
  \bibfield  {author} {\bibinfo {author} {\bibfnamefont {R.}~\bibnamefont
  {Ram{\'\i}rez}}, \bibinfo {author} {\bibfnamefont {T.}~\bibnamefont
  {L{\'o}pez-Ciudad}}, \bibinfo {author} {\bibfnamefont {P.}~\bibnamefont
  {Kumar~P}},\ and\ \bibinfo {author} {\bibfnamefont {D.}~\bibnamefont
  {Marx}},\ }\href {https://doi.org/10.1063/1.1774986} {\bibfield  {journal}
  {\bibinfo  {journal} {J. Chem. Phys.}\ }\textbf {\bibinfo {volume} {121}},\
  \bibinfo {pages} {3973} (\bibinfo {year} {2004})}\BibitemShut {NoStop}%
\bibitem [{\citenamefont {Geng}(2022)}]{Geng2022}%
  \BibitemOpen
  \bibfield  {author} {\bibinfo {author} {\bibfnamefont {H.~Y.}\ \bibnamefont
  {Geng}},\ }\href {https://doi.org/10.1021/acs.jpcc.2c05027} {\bibfield
  {journal} {\bibinfo  {journal} {J. Phys. Chem. C}\ }\textbf {\bibinfo
  {volume} {126}},\ \bibinfo {pages} {19355} (\bibinfo {year}
  {2022})}\BibitemShut {NoStop}%
\bibitem [{\citenamefont {Morresi}\ \emph {et~al.}(2021)\citenamefont
  {Morresi}, \citenamefont {Paulatto}, \citenamefont {Vuilleumier},\ and\
  \citenamefont {Casula}}]{Morresi2021}%
  \BibitemOpen
  \bibfield  {author} {\bibinfo {author} {\bibfnamefont {T.}~\bibnamefont
  {Morresi}}, \bibinfo {author} {\bibfnamefont {L.}~\bibnamefont {Paulatto}},
  \bibinfo {author} {\bibfnamefont {R.}~\bibnamefont {Vuilleumier}},\ and\
  \bibinfo {author} {\bibfnamefont {M.}~\bibnamefont {Casula}},\ }\href
  {https://doi.org/10.1063/5.0050450} {\bibfield  {journal} {\bibinfo
  {journal} {J. Chem. Phys.}\ }\textbf {\bibinfo {volume} {154}},\ \bibinfo
  {pages} {224108} (\bibinfo {year} {2021})}\BibitemShut {NoStop}%
\bibitem [{\citenamefont {Morresi}\ \emph {et~al.}(2022)\citenamefont
  {Morresi}, \citenamefont {Vuilleumier},\ and\ \citenamefont
  {Casula}}]{Morresi2022}%
  \BibitemOpen
  \bibfield  {author} {\bibinfo {author} {\bibfnamefont {T.}~\bibnamefont
  {Morresi}}, \bibinfo {author} {\bibfnamefont {R.}~\bibnamefont
  {Vuilleumier}},\ and\ \bibinfo {author} {\bibfnamefont {M.}~\bibnamefont
  {Casula}},\ }\href {https://doi.org/10.1103/physrevb.106.054109} {\bibfield
  {journal} {\bibinfo  {journal} {Phys. Rev. B}\ }\textbf {\bibinfo {volume}
  {106}},\ \bibinfo {pages} {054109} (\bibinfo {year} {2022})}\BibitemShut
  {NoStop}%
\bibitem [{\citenamefont {Mori}(1965)}]{Mori1965}%
  \BibitemOpen
  \bibfield  {author} {\bibinfo {author} {\bibfnamefont {H.}~\bibnamefont
  {Mori}},\ }\href {https://doi.org/10.1143/ptp.33.423} {\bibfield  {journal}
  {\bibinfo  {journal} {Prog. Theor. Phys.}\ }\textbf {\bibinfo {volume}
  {33}},\ \bibinfo {pages} {423} (\bibinfo {year} {1965})}\BibitemShut
  {NoStop}%
\bibitem [{\citenamefont {Zwanzig}(1961)}]{Zwanzig1961}%
  \BibitemOpen
  \bibfield  {author} {\bibinfo {author} {\bibfnamefont {R.}~\bibnamefont
  {Zwanzig}},\ }\href {https://doi.org/10.1103/physrev.124.983} {\bibfield
  {journal} {\bibinfo  {journal} {Phys. Rev.}\ }\textbf {\bibinfo {volume}
  {124}},\ \bibinfo {pages} {983} (\bibinfo {year} {1961})}\BibitemShut
  {NoStop}%
\bibitem [{\citenamefont {Kubo}(1966)}]{Kubo1966}%
  \BibitemOpen
  \bibfield  {author} {\bibinfo {author} {\bibfnamefont {R.}~\bibnamefont
  {Kubo}},\ }\href {https://doi.org/10.1088/0034-4885/29/1/306} {\bibfield
  {journal} {\bibinfo  {journal} {Rep. Prog. Phys.}\ }\textbf {\bibinfo
  {volume} {29}},\ \bibinfo {pages} {255} (\bibinfo {year} {1966})}\BibitemShut
  {NoStop}%
\bibitem [{\citenamefont {Hij{\'{o}}n}\ \emph {et~al.}(2010)\citenamefont
  {Hij{\'{o}}n}, \citenamefont {Espa{\~{n}}ol}, \citenamefont
  {Vanden-Eijnden},\ and\ \citenamefont {Delgado-Buscalioni}}]{Hijn2010}%
  \BibitemOpen
  \bibfield  {author} {\bibinfo {author} {\bibfnamefont {C.}~\bibnamefont
  {Hij{\'{o}}n}}, \bibinfo {author} {\bibfnamefont {P.}~\bibnamefont
  {Espa{\~{n}}ol}}, \bibinfo {author} {\bibfnamefont {E.}~\bibnamefont
  {Vanden-Eijnden}},\ and\ \bibinfo {author} {\bibfnamefont {R.}~\bibnamefont
  {Delgado-Buscalioni}},\ }\href {https://doi.org/10.1039/b902479b} {\bibfield
  {journal} {\bibinfo  {journal} {Faraday Discuss.}\ }\textbf {\bibinfo
  {volume} {144}},\ \bibinfo {pages} {301} (\bibinfo {year}
  {2010})}\BibitemShut {NoStop}%
\bibitem [{\citenamefont {Carof}\ \emph {et~al.}(2014)\citenamefont {Carof},
  \citenamefont {Vuilleumier},\ and\ \citenamefont {Rotenberg}}]{Carof2014}%
  \BibitemOpen
  \bibfield  {author} {\bibinfo {author} {\bibfnamefont {A.}~\bibnamefont
  {Carof}}, \bibinfo {author} {\bibfnamefont {R.}~\bibnamefont {Vuilleumier}},\
  and\ \bibinfo {author} {\bibfnamefont {B.}~\bibnamefont {Rotenberg}},\ }\href
  {https://doi.org/10.1063/1.4868653} {\bibfield  {journal} {\bibinfo
  {journal} {J. Chem. Phys.}\ }\textbf {\bibinfo {volume} {140}},\ \bibinfo
  {pages} {124103} (\bibinfo {year} {2014})}\BibitemShut {NoStop}%
\bibitem [{\citenamefont {Li}\ \emph {et~al.}(2015)\citenamefont {Li},
  \citenamefont {Bian}, \citenamefont {Li},\ and\ \citenamefont
  {Karniadakis}}]{Li2015}%
  \BibitemOpen
  \bibfield  {author} {\bibinfo {author} {\bibfnamefont {Z.}~\bibnamefont
  {Li}}, \bibinfo {author} {\bibfnamefont {X.}~\bibnamefont {Bian}}, \bibinfo
  {author} {\bibfnamefont {X.}~\bibnamefont {Li}},\ and\ \bibinfo {author}
  {\bibfnamefont {G.~E.}\ \bibnamefont {Karniadakis}},\ }\href
  {https://doi.org/10.1063/1.4935490} {\bibfield  {journal} {\bibinfo
  {journal} {J. Chem. Phys.}\ }\textbf {\bibinfo {volume} {143}},\ \bibinfo
  {pages} {243128} (\bibinfo {year} {2015})}\BibitemShut {NoStop}%
\bibitem [{\citenamefont {Fiorentino}\ and\ \citenamefont
  {Baroni}(2022)}]{Fiorentino2022}%
  \BibitemOpen
  \bibfield  {author} {\bibinfo {author} {\bibfnamefont {A.}~\bibnamefont
  {Fiorentino}}\ and\ \bibinfo {author} {\bibfnamefont {S.}~\bibnamefont
  {Baroni}},\ }\href {https://doi.org/10.48550/ARXIV.2206.01279} {\bibinfo
  {title} {From green-kubo to the full boltzmann kinetic approach to heat
  transport in crystals and glasses}} (\bibinfo {year} {2022})\BibitemShut
  {NoStop}%
\bibitem [{\citenamefont {Bosse}\ \emph {et~al.}(1978)\citenamefont {Bosse},
  \citenamefont {G\"{o}tze},\ and\ \citenamefont {L\"{u}cke}}]{Bosse1978}%
  \BibitemOpen
  \bibfield  {author} {\bibinfo {author} {\bibfnamefont {J.}~\bibnamefont
  {Bosse}}, \bibinfo {author} {\bibfnamefont {W.}~\bibnamefont {G\"{o}tze}},\
  and\ \bibinfo {author} {\bibfnamefont {M.}~\bibnamefont {L\"{u}cke}},\ }\href
  {https://doi.org/10.1103/physreva.17.434} {\bibfield  {journal} {\bibinfo
  {journal} {Phys. Rev. A}\ }\textbf {\bibinfo {volume} {17}},\ \bibinfo
  {pages} {434} (\bibinfo {year} {1978})}\BibitemShut {NoStop}%
\bibitem [{\citenamefont {Reichman}\ and\ \citenamefont
  {Charbonneau}(2005)}]{Reichman2005}%
  \BibitemOpen
  \bibfield  {author} {\bibinfo {author} {\bibfnamefont {D.~R.}\ \bibnamefont
  {Reichman}}\ and\ \bibinfo {author} {\bibfnamefont {P.}~\bibnamefont
  {Charbonneau}},\ }\href {https://doi.org/10.1088/1742-5468/2005/05/p05013}
  {\bibfield  {journal} {\bibinfo  {journal} {J. Stat. Mech.: Theory Exp.}\
  }\textbf {\bibinfo {volume} {2005}}\bibinfo  {number} { (05)},\ \bibinfo
  {pages} {P05013}}\BibitemShut {NoStop}%
\bibitem [{\citenamefont {Markland}\ \emph {et~al.}(2012)\citenamefont
  {Markland}, \citenamefont {Morrone}, \citenamefont {Miyazaki}, \citenamefont
  {Berne}, \citenamefont {Reichman},\ and\ \citenamefont
  {Rabani}}]{Markland2012}%
  \BibitemOpen
\bibfield  {number} {  }\bibfield  {author} {\bibinfo {author} {\bibfnamefont
  {T.~E.}\ \bibnamefont {Markland}}, \bibinfo {author} {\bibfnamefont {J.~A.}\
  \bibnamefont {Morrone}}, \bibinfo {author} {\bibfnamefont {K.}~\bibnamefont
  {Miyazaki}}, \bibinfo {author} {\bibfnamefont {B.~J.}\ \bibnamefont {Berne}},
  \bibinfo {author} {\bibfnamefont {D.~R.}\ \bibnamefont {Reichman}},\ and\
  \bibinfo {author} {\bibfnamefont {E.}~\bibnamefont {Rabani}},\ }\href
  {https://doi.org/10.1063/1.3684881} {\bibfield  {journal} {\bibinfo
  {journal} {J. Chem. Phys.}\ }\textbf {\bibinfo {volume} {136}},\ \bibinfo
  {pages} {074511} (\bibinfo {year} {2012})}\BibitemShut {NoStop}%
\bibitem [{\citenamefont {Janssen}(2018)}]{Janssen2018}%
  \BibitemOpen
  \bibfield  {author} {\bibinfo {author} {\bibfnamefont {L.~M.~C.}\
  \bibnamefont {Janssen}},\ }\href {https://doi.org/10.3389/fphy.2018.00097}
  {\bibfield  {journal} {\bibinfo  {journal} {Front. Phys.}\ }\textbf {\bibinfo
  {volume} {6}},\ \bibinfo {pages} {97} (\bibinfo {year} {2018})}\BibitemShut
  {NoStop}%
\bibitem [{\citenamefont {Kubo}\ \emph {et~al.}(1991)\citenamefont {Kubo},
  \citenamefont {Toda},\ and\ \citenamefont {Hashitsume}}]{Kubo1991}%
  \BibitemOpen
  \bibfield  {author} {\bibinfo {author} {\bibfnamefont {R.}~\bibnamefont
  {Kubo}}, \bibinfo {author} {\bibfnamefont {M.}~\bibnamefont {Toda}},\ and\
  \bibinfo {author} {\bibfnamefont {N.}~\bibnamefont {Hashitsume}},\ }\href
  {https://doi.org/10.1007/978-3-642-58244-8} {\emph {\bibinfo {title}
  {Statistical Physics {II}}}}\ (\bibinfo  {publisher} {Springer Berlin
  Heidelberg},\ \bibinfo {year} {1991})\BibitemShut {NoStop}%
\bibitem [{\citenamefont {Craig}\ and\ \citenamefont
  {Manolopoulos}(2004)}]{Craig2004}%
  \BibitemOpen
  \bibfield  {author} {\bibinfo {author} {\bibfnamefont {I.~R.}\ \bibnamefont
  {Craig}}\ and\ \bibinfo {author} {\bibfnamefont {D.~E.}\ \bibnamefont
  {Manolopoulos}},\ }\href {https://doi.org/10.1063/1.1777575} {\bibfield
  {journal} {\bibinfo  {journal} {J. Chem. Phys.}\ }\textbf {\bibinfo {volume}
  {121}},\ \bibinfo {pages} {3368} (\bibinfo {year} {2004})}\BibitemShut
  {NoStop}%
\bibitem [{\citenamefont {Brown}(1994)}]{Brown1994}%
  \BibitemOpen
  \bibfield  {author} {\bibinfo {author} {\bibfnamefont {E.~B.}\ \bibnamefont
  {Brown}},\ }\href {https://doi.org/10.1103/physrevb.49.4305} {\bibfield
  {journal} {\bibinfo  {journal} {Phys. Rev. B}\ }\textbf {\bibinfo {volume}
  {49}},\ \bibinfo {pages} {4305} (\bibinfo {year} {1994})}\BibitemShut
  {NoStop}%
\bibitem [{Note1()}]{Note1}%
  \BibitemOpen
  \bibinfo {note} {While the choice of prefactor here can seem arbitrary, we
  use it to simplify comparison with other methods of anharmonic lattice
  dynamics}\BibitemShut {NoStop}%
\bibitem [{\citenamefont {Braams}\ and\ \citenamefont
  {Manolopoulos}(2006)}]{Braams2006}%
  \BibitemOpen
  \bibfield  {author} {\bibinfo {author} {\bibfnamefont {B.~J.}\ \bibnamefont
  {Braams}}\ and\ \bibinfo {author} {\bibfnamefont {D.~E.}\ \bibnamefont
  {Manolopoulos}},\ }\href {https://doi.org/10.1063/1.2357599} {\bibfield
  {journal} {\bibinfo  {journal} {J. Chem. Phys.}\ }\textbf {\bibinfo {volume}
  {125}},\ \bibinfo {pages} {124105} (\bibinfo {year} {2006})}\BibitemShut
  {NoStop}%
\bibitem [{\citenamefont {Rossi}\ \emph {et~al.}(2014)\citenamefont {Rossi},
  \citenamefont {Ceriotti},\ and\ \citenamefont {Manolopoulos}}]{Rossi2014}%
  \BibitemOpen
  \bibfield  {author} {\bibinfo {author} {\bibfnamefont {M.}~\bibnamefont
  {Rossi}}, \bibinfo {author} {\bibfnamefont {M.}~\bibnamefont {Ceriotti}},\
  and\ \bibinfo {author} {\bibfnamefont {D.~E.}\ \bibnamefont {Manolopoulos}},\
  }\href {https://doi.org/10.1063/1.4883861} {\bibfield  {journal} {\bibinfo
  {journal} {J. Chem. Phys.}\ }\textbf {\bibinfo {volume} {140}},\ \bibinfo
  {pages} {234116} (\bibinfo {year} {2014})}\BibitemShut {NoStop}%
\bibitem [{\citenamefont {Zwanzig}(2001)}]{Zwanzig2001}%
  \BibitemOpen
  \bibfield  {author} {\bibinfo {author} {\bibfnamefont {R.}~\bibnamefont
  {Zwanzig}},\ }\href@noop {} {\emph {\bibinfo {title} {Nonequilibrium
  statistical mechanics}}}\ (\bibinfo  {publisher} {Oxford university press},\
  \bibinfo {year} {2001})\BibitemShut {NoStop}%
\bibitem [{\citenamefont {Berne}\ and\ \citenamefont {Harp}(1970)}]{Berne1970}%
  \BibitemOpen
  \bibfield  {author} {\bibinfo {author} {\bibfnamefont {B.~J.}\ \bibnamefont
  {Berne}}\ and\ \bibinfo {author} {\bibfnamefont {G.~D.}\ \bibnamefont
  {Harp}},\ }in\ \href {https://doi.org/10.1002/9780470143636.ch3} {\emph
  {\bibinfo {booktitle} {Advances in Chemical Physics}}}\ (\bibinfo
  {publisher} {John Wiley {\&} Sons, Inc.},\ \bibinfo {year} {1970})\ pp.\
  \bibinfo {pages} {63--227}\BibitemShut {NoStop}%
\bibitem [{\citenamefont {Szamel}(2007)}]{Szamel2007}%
  \BibitemOpen
  \bibfield  {author} {\bibinfo {author} {\bibfnamefont {G.}~\bibnamefont
  {Szamel}},\ }\href {https://doi.org/10.1063/1.2759487} {\bibfield  {journal}
  {\bibinfo  {journal} {J. Chem. Phys.}\ }\textbf {\bibinfo {volume} {127}},\
  \bibinfo {pages} {084515} (\bibinfo {year} {2007})}\BibitemShut {NoStop}%
\bibitem [{\citenamefont {Simoncelli}\ \emph {et~al.}(2019)\citenamefont
  {Simoncelli}, \citenamefont {Marzari},\ and\ \citenamefont
  {Mauri}}]{Simoncelli2019}%
  \BibitemOpen
  \bibfield  {author} {\bibinfo {author} {\bibfnamefont {M.}~\bibnamefont
  {Simoncelli}}, \bibinfo {author} {\bibfnamefont {N.}~\bibnamefont
  {Marzari}},\ and\ \bibinfo {author} {\bibfnamefont {F.}~\bibnamefont
  {Mauri}},\ }\href {https://doi.org/10.1038/s41567-019-0520-x} {\bibfield
  {journal} {\bibinfo  {journal} {Nat. Phys.}\ }\textbf {\bibinfo {volume}
  {15}},\ \bibinfo {pages} {809} (\bibinfo {year} {2019})}\BibitemShut
  {NoStop}%
\bibitem [{\citenamefont {Eriksson}\ \emph {et~al.}(2019)\citenamefont
  {Eriksson}, \citenamefont {Fransson},\ and\ \citenamefont
  {Erhart}}]{Eriksson2019}%
  \BibitemOpen
  \bibfield  {author} {\bibinfo {author} {\bibfnamefont {F.}~\bibnamefont
  {Eriksson}}, \bibinfo {author} {\bibfnamefont {E.}~\bibnamefont {Fransson}},\
  and\ \bibinfo {author} {\bibfnamefont {P.}~\bibnamefont {Erhart}},\ }\href
  {https://doi.org/10.1002/adts.201800184} {\bibfield  {journal} {\bibinfo
  {journal} {Adv. Theory Simul.}\ }\textbf {\bibinfo {volume} {2}},\ \bibinfo
  {pages} {1800184} (\bibinfo {year} {2019})}\BibitemShut {NoStop}%
\bibitem [{\citenamefont {Zhou}\ \emph {et~al.}(2019)\citenamefont {Zhou},
  \citenamefont {Nielson}, \citenamefont {Xia},\ and\ \citenamefont
  {Ozoli{\c{n}}{\v{s}}}}]{Zhou2019}%
  \BibitemOpen
  \bibfield  {author} {\bibinfo {author} {\bibfnamefont {F.}~\bibnamefont
  {Zhou}}, \bibinfo {author} {\bibfnamefont {W.}~\bibnamefont {Nielson}},
  \bibinfo {author} {\bibfnamefont {Y.}~\bibnamefont {Xia}},\ and\ \bibinfo
  {author} {\bibfnamefont {V.}~\bibnamefont {Ozoli{\c{n}}{\v{s}}}},\ }\href
  {https://doi.org/10.1103/physrevb.100.184308} {\bibfield  {journal} {\bibinfo
   {journal} {Phys. Rev. B}\ }\textbf {\bibinfo {volume} {100}},\ \bibinfo
  {pages} {184308} (\bibinfo {year} {2019})}\BibitemShut {NoStop}%
\bibitem [{\citenamefont {Bottin}\ \emph {et~al.}(2020)\citenamefont {Bottin},
  \citenamefont {Bieder},\ and\ \citenamefont {Bouchet}}]{Bottin2020}%
  \BibitemOpen
  \bibfield  {author} {\bibinfo {author} {\bibfnamefont {F.}~\bibnamefont
  {Bottin}}, \bibinfo {author} {\bibfnamefont {J.}~\bibnamefont {Bieder}},\
  and\ \bibinfo {author} {\bibfnamefont {J.}~\bibnamefont {Bouchet}},\ }\href
  {https://doi.org/10.1016/j.cpc.2020.107301} {\bibfield  {journal} {\bibinfo
  {journal} {Comput. Phys. Commun.}\ }\textbf {\bibinfo {volume} {254}},\
  \bibinfo {pages} {107301} (\bibinfo {year} {2020})}\BibitemShut {NoStop}%
\bibitem [{\citenamefont {Ram{\'{\i}}rez}\ and\ \citenamefont
  {L{\'{o}}pez-Ciudad}(1999)}]{Ramrez1999}%
  \BibitemOpen
  \bibfield  {author} {\bibinfo {author} {\bibfnamefont {R.}~\bibnamefont
  {Ram{\'{\i}}rez}}\ and\ \bibinfo {author} {\bibfnamefont {T.}~\bibnamefont
  {L{\'{o}}pez-Ciudad}},\ }\href {https://doi.org/10.1063/1.479666} {\bibfield
  {journal} {\bibinfo  {journal} {J. Chem. Phys.}\ }\textbf {\bibinfo {volume}
  {111}},\ \bibinfo {pages} {3339} (\bibinfo {year} {1999})}\BibitemShut
  {NoStop}%
\bibitem [{\citenamefont {Shulumba}\ \emph {et~al.}(2017)\citenamefont
  {Shulumba}, \citenamefont {Hellman},\ and\ \citenamefont
  {Minnich}}]{Shulumba2017}%
  \BibitemOpen
  \bibfield  {author} {\bibinfo {author} {\bibfnamefont {N.}~\bibnamefont
  {Shulumba}}, \bibinfo {author} {\bibfnamefont {O.}~\bibnamefont {Hellman}},\
  and\ \bibinfo {author} {\bibfnamefont {A.~J.}\ \bibnamefont {Minnich}},\
  }\href {https://doi.org/10.1103/physrevb.95.014302} {\bibfield  {journal}
  {\bibinfo  {journal} {Phys. Rev. B}\ }\textbf {\bibinfo {volume} {95}},\
  \bibinfo {pages} {014302} (\bibinfo {year} {2017})}\BibitemShut {NoStop}%
\bibitem [{\citenamefont {Xia}\ \emph {et~al.}(2020{\natexlab{b}})\citenamefont
  {Xia}, \citenamefont {Pal}, \citenamefont {He}, \citenamefont
  {Ozoli{\c{n}}{\v{s}}},\ and\ \citenamefont {Wolverton}}]{Xia2020b}%
  \BibitemOpen
  \bibfield  {author} {\bibinfo {author} {\bibfnamefont {Y.}~\bibnamefont
  {Xia}}, \bibinfo {author} {\bibfnamefont {K.}~\bibnamefont {Pal}}, \bibinfo
  {author} {\bibfnamefont {J.}~\bibnamefont {He}}, \bibinfo {author}
  {\bibfnamefont {V.}~\bibnamefont {Ozoli{\c{n}}{\v{s}}}},\ and\ \bibinfo
  {author} {\bibfnamefont {C.}~\bibnamefont {Wolverton}},\ }\href
  {https://doi.org/10.1103/physrevlett.124.065901} {\bibfield  {journal}
  {\bibinfo  {journal} {Phys. Rev. Lett.}\ }\textbf {\bibinfo {volume} {124}},\
  \bibinfo {pages} {065901} (\bibinfo {year} {2020}{\natexlab{b}})}\BibitemShut
  {NoStop}%
\bibitem [{\citenamefont {Zhu}\ \emph {et~al.}(2020)\citenamefont {Zhu},
  \citenamefont {Xia}, \citenamefont {Wang}, \citenamefont {Sheng},
  \citenamefont {Yang}, \citenamefont {Fu}, \citenamefont {Li}, \citenamefont
  {Zhu}, \citenamefont {Luo}, \citenamefont {Wolverton}, \citenamefont
  {Snyder}, \citenamefont {Liu},\ and\ \citenamefont {Zhang}}]{Zhu2020}%
  \BibitemOpen
  \bibfield  {author} {\bibinfo {author} {\bibfnamefont {Y.}~\bibnamefont
  {Zhu}}, \bibinfo {author} {\bibfnamefont {Y.}~\bibnamefont {Xia}}, \bibinfo
  {author} {\bibfnamefont {Y.}~\bibnamefont {Wang}}, \bibinfo {author}
  {\bibfnamefont {Y.}~\bibnamefont {Sheng}}, \bibinfo {author} {\bibfnamefont
  {J.}~\bibnamefont {Yang}}, \bibinfo {author} {\bibfnamefont {C.}~\bibnamefont
  {Fu}}, \bibinfo {author} {\bibfnamefont {A.}~\bibnamefont {Li}}, \bibinfo
  {author} {\bibfnamefont {T.}~\bibnamefont {Zhu}}, \bibinfo {author}
  {\bibfnamefont {J.}~\bibnamefont {Luo}}, \bibinfo {author} {\bibfnamefont
  {C.}~\bibnamefont {Wolverton}}, \bibinfo {author} {\bibfnamefont {G.~J.}\
  \bibnamefont {Snyder}}, \bibinfo {author} {\bibfnamefont {J.}~\bibnamefont
  {Liu}},\ and\ \bibinfo {author} {\bibfnamefont {W.}~\bibnamefont {Zhang}},\
  }\href {https://doi.org/10.34133/2020/4589786} {\bibfield  {journal}
  {\bibinfo  {journal} {Res.}\ }\textbf {\bibinfo {volume} {2020}},\ \bibinfo
  {pages} {1} (\bibinfo {year} {2020})}\BibitemShut {NoStop}%
\bibitem [{\citenamefont {Zeng}\ \emph {et~al.}(2021)\citenamefont {Zeng},
  \citenamefont {Zhang}, \citenamefont {Xia}, \citenamefont {Fan},
  \citenamefont {Wolverton},\ and\ \citenamefont {Chen}}]{Zeng2021}%
  \BibitemOpen
  \bibfield  {author} {\bibinfo {author} {\bibfnamefont {Z.}~\bibnamefont
  {Zeng}}, \bibinfo {author} {\bibfnamefont {C.}~\bibnamefont {Zhang}},
  \bibinfo {author} {\bibfnamefont {Y.}~\bibnamefont {Xia}}, \bibinfo {author}
  {\bibfnamefont {Z.}~\bibnamefont {Fan}}, \bibinfo {author} {\bibfnamefont
  {C.}~\bibnamefont {Wolverton}},\ and\ \bibinfo {author} {\bibfnamefont
  {Y.}~\bibnamefont {Chen}},\ }\href
  {https://doi.org/10.1103/physrevb.103.224307} {\bibfield  {journal} {\bibinfo
   {journal} {Phys. Rev. B}\ }\textbf {\bibinfo {volume} {103}},\ \bibinfo
  {pages} {224307} (\bibinfo {year} {2021})}\BibitemShut {NoStop}%
\bibitem [{\citenamefont {Yang}\ \emph
  {et~al.}(2022{\natexlab{b}})\citenamefont {Yang}, \citenamefont {Tiwari},\
  and\ \citenamefont {Feng}}]{Yang2022}%
  \BibitemOpen
  \bibfield  {author} {\bibinfo {author} {\bibfnamefont {X.}~\bibnamefont
  {Yang}}, \bibinfo {author} {\bibfnamefont {J.}~\bibnamefont {Tiwari}},\ and\
  \bibinfo {author} {\bibfnamefont {T.}~\bibnamefont {Feng}},\ }\href
  {https://doi.org/10.1016/j.mtphys.2022.100689} {\bibfield  {journal}
  {\bibinfo  {journal} {Mater. Today Phys.}\ }\textbf {\bibinfo {volume}
  {24}},\ \bibinfo {pages} {100689} (\bibinfo {year}
  {2022}{\natexlab{b}})}\BibitemShut {NoStop}%
\bibitem [{\citenamefont {Feng}\ \emph {et~al.}(2017)\citenamefont {Feng},
  \citenamefont {Lindsay},\ and\ \citenamefont {Ruan}}]{Feng2017}%
  \BibitemOpen
  \bibfield  {author} {\bibinfo {author} {\bibfnamefont {T.}~\bibnamefont
  {Feng}}, \bibinfo {author} {\bibfnamefont {L.}~\bibnamefont {Lindsay}},\ and\
  \bibinfo {author} {\bibfnamefont {X.}~\bibnamefont {Ruan}},\ }\href
  {https://doi.org/10.1103/physrevb.96.161201} {\bibfield  {journal} {\bibinfo
  {journal} {Phys. Rev. B}\ }\textbf {\bibinfo {volume} {96}},\ \bibinfo
  {pages} {161201(R)} (\bibinfo {year} {2017})}\BibitemShut {NoStop}%
\bibitem [{\citenamefont {Decoster}(2004)}]{Decoster2004}%
  \BibitemOpen
  \bibfield  {author} {\bibinfo {author} {\bibfnamefont {A.}~\bibnamefont
  {Decoster}},\ }\href {https://doi.org/10.1088/0305-4470/37/39/001} {\bibfield
   {journal} {\bibinfo  {journal} {J Phys. A Math.}\ }\textbf {\bibinfo
  {volume} {37}},\ \bibinfo {pages} {9051} (\bibinfo {year}
  {2004})}\BibitemShut {NoStop}%
\bibitem [{\citenamefont {Li}\ \emph {et~al.}(2014{\natexlab{b}})\citenamefont
  {Li}, \citenamefont {Hellman}, \citenamefont {Ma}, \citenamefont {May},
  \citenamefont {Cao}, \citenamefont {Chen}, \citenamefont {Christianson},
  \citenamefont {Ehlers}, \citenamefont {Singh}, \citenamefont {Sales},\ and\
  \citenamefont {Delaire}}]{Li2014b}%
  \BibitemOpen
  \bibfield  {author} {\bibinfo {author} {\bibfnamefont {C.~W.}\ \bibnamefont
  {Li}}, \bibinfo {author} {\bibfnamefont {O.}~\bibnamefont {Hellman}},
  \bibinfo {author} {\bibfnamefont {J.}~\bibnamefont {Ma}}, \bibinfo {author}
  {\bibfnamefont {A.~F.}\ \bibnamefont {May}}, \bibinfo {author} {\bibfnamefont
  {H.~B.}\ \bibnamefont {Cao}}, \bibinfo {author} {\bibfnamefont
  {X.}~\bibnamefont {Chen}}, \bibinfo {author} {\bibfnamefont {A.~D.}\
  \bibnamefont {Christianson}}, \bibinfo {author} {\bibfnamefont
  {G.}~\bibnamefont {Ehlers}}, \bibinfo {author} {\bibfnamefont {D.~J.}\
  \bibnamefont {Singh}}, \bibinfo {author} {\bibfnamefont {B.~C.}\ \bibnamefont
  {Sales}},\ and\ \bibinfo {author} {\bibfnamefont {O.}~\bibnamefont
  {Delaire}},\ }\href {https://doi.org/10.1103/physrevlett.112.175501}
  {\bibfield  {journal} {\bibinfo  {journal} {Phys. Rev. Lett.}\ }\textbf
  {\bibinfo {volume} {112}},\ \bibinfo {pages} {175501} (\bibinfo {year}
  {2014}{\natexlab{b}})}\BibitemShut {NoStop}%
\bibitem [{\citenamefont {Dangi{\'{c}}}\ \emph {et~al.}(2021)\citenamefont
  {Dangi{\'{c}}}, \citenamefont {Hellman}, \citenamefont {Fahy},\ and\
  \citenamefont {Savi{\'{c}}}}]{Dangi2021}%
  \BibitemOpen
  \bibfield  {author} {\bibinfo {author} {\bibfnamefont {{\DJ}.}~\bibnamefont
  {Dangi{\'{c}}}}, \bibinfo {author} {\bibfnamefont {O.}~\bibnamefont
  {Hellman}}, \bibinfo {author} {\bibfnamefont {S.}~\bibnamefont {Fahy}},\ and\
  \bibinfo {author} {\bibfnamefont {I.}~\bibnamefont {Savi{\'{c}}}},\
  }\bibfield  {journal} {\bibinfo  {journal} {npj Comput. Mater.}\ }\textbf
  {\bibinfo {volume} {7}},\ \href {https://doi.org/10.1038/s41524-021-00523-7}
  {10.1038/s41524-021-00523-7} (\bibinfo {year} {2021})\BibitemShut {NoStop}%
\bibitem [{\citenamefont {Bouchet}\ and\ \citenamefont
  {Bottin}(2017)}]{Bouchet2017}%
  \BibitemOpen
  \bibfield  {author} {\bibinfo {author} {\bibfnamefont {J.}~\bibnamefont
  {Bouchet}}\ and\ \bibinfo {author} {\bibfnamefont {F.}~\bibnamefont
  {Bottin}},\ }\href {https://doi.org/10.1103/physrevb.95.054113} {\bibfield
  {journal} {\bibinfo  {journal} {Phys. Rev. B}\ }\textbf {\bibinfo {volume}
  {95}},\ \bibinfo {pages} {054113} (\bibinfo {year} {2017})}\BibitemShut
  {NoStop}%
\bibitem [{\citenamefont {Ladygin}\ \emph {et~al.}(2020)\citenamefont
  {Ladygin}, \citenamefont {Korotaev}, \citenamefont {Yanilkin},\ and\
  \citenamefont {Shapeev}}]{Ladygin2020}%
  \BibitemOpen
  \bibfield  {author} {\bibinfo {author} {\bibfnamefont {V.}~\bibnamefont
  {Ladygin}}, \bibinfo {author} {\bibfnamefont {P.}~\bibnamefont {Korotaev}},
  \bibinfo {author} {\bibfnamefont {A.}~\bibnamefont {Yanilkin}},\ and\
  \bibinfo {author} {\bibfnamefont {A.}~\bibnamefont {Shapeev}},\ }\href
  {https://doi.org/10.1016/j.commatsci.2019.109333} {\bibfield  {journal}
  {\bibinfo  {journal} {Comput. Mater. Sci.}\ }\textbf {\bibinfo {volume}
  {172}},\ \bibinfo {pages} {109333} (\bibinfo {year} {2020})}\BibitemShut
  {NoStop}%
\bibitem [{\citenamefont {Ding}\ \emph {et~al.}(2021)\citenamefont {Ding},
  \citenamefont {Lanigan-Atkins}, \citenamefont {Calder{\'{o}}n-Cueva},
  \citenamefont {Banerjee}, \citenamefont {Abernathy}, \citenamefont {Said},
  \citenamefont {Zevalkink},\ and\ \citenamefont {Delaire}}]{Ding2021}%
  \BibitemOpen
  \bibfield  {author} {\bibinfo {author} {\bibfnamefont {J.}~\bibnamefont
  {Ding}}, \bibinfo {author} {\bibfnamefont {T.}~\bibnamefont
  {Lanigan-Atkins}}, \bibinfo {author} {\bibfnamefont {M.}~\bibnamefont
  {Calder{\'{o}}n-Cueva}}, \bibinfo {author} {\bibfnamefont {A.}~\bibnamefont
  {Banerjee}}, \bibinfo {author} {\bibfnamefont {D.~L.}\ \bibnamefont
  {Abernathy}}, \bibinfo {author} {\bibfnamefont {A.}~\bibnamefont {Said}},
  \bibinfo {author} {\bibfnamefont {A.}~\bibnamefont {Zevalkink}},\ and\
  \bibinfo {author} {\bibfnamefont {O.}~\bibnamefont {Delaire}},\ }\bibfield
  {journal} {\bibinfo  {journal} {Sci. Adv.}\ }\textbf {\bibinfo {volume}
  {7}},\ \href {https://doi.org/10.1126/sciadv.abg1449}
  {10.1126/sciadv.abg1449} (\bibinfo {year} {2021})\BibitemShut {NoStop}%
\bibitem [{\citenamefont {Xie}\ \emph {et~al.}(2020)\citenamefont {Xie},
  \citenamefont {Feng}, \citenamefont {Li},\ and\ \citenamefont
  {He}}]{Xie2020}%
  \BibitemOpen
  \bibfield  {author} {\bibinfo {author} {\bibfnamefont {L.}~\bibnamefont
  {Xie}}, \bibinfo {author} {\bibfnamefont {J.~H.}\ \bibnamefont {Feng}},
  \bibinfo {author} {\bibfnamefont {R.}~\bibnamefont {Li}},\ and\ \bibinfo
  {author} {\bibfnamefont {J.~Q.}\ \bibnamefont {He}},\ }\href
  {https://doi.org/10.1103/physrevlett.125.245901} {\bibfield  {journal}
  {\bibinfo  {journal} {Phys. Rev. Lett.}\ }\textbf {\bibinfo {volume} {125}},\
  \bibinfo {pages} {245901} (\bibinfo {year} {2020})}\BibitemShut {NoStop}%
\bibitem [{\citenamefont {He}\ \emph {et~al.}(2020)\citenamefont {He},
  \citenamefont {Bansal}, \citenamefont {Winn}, \citenamefont {Chi},
  \citenamefont {Boatner},\ and\ \citenamefont {Delaire}}]{He2020}%
  \BibitemOpen
  \bibfield  {author} {\bibinfo {author} {\bibfnamefont {X.}~\bibnamefont
  {He}}, \bibinfo {author} {\bibfnamefont {D.}~\bibnamefont {Bansal}}, \bibinfo
  {author} {\bibfnamefont {B.}~\bibnamefont {Winn}}, \bibinfo {author}
  {\bibfnamefont {S.}~\bibnamefont {Chi}}, \bibinfo {author} {\bibfnamefont
  {L.}~\bibnamefont {Boatner}},\ and\ \bibinfo {author} {\bibfnamefont
  {O.}~\bibnamefont {Delaire}},\ }\href
  {https://doi.org/10.1103/physrevlett.124.145901} {\bibfield  {journal}
  {\bibinfo  {journal} {Phys. Rev. Lett.}\ }\textbf {\bibinfo {volume} {124}},\
  \bibinfo {pages} {145901} (\bibinfo {year} {2020})}\BibitemShut {NoStop}%
\bibitem [{\citenamefont {Isaeva}\ \emph {et~al.}(2019)\citenamefont {Isaeva},
  \citenamefont {Barbalinardo}, \citenamefont {Donadio},\ and\ \citenamefont
  {Baroni}}]{Isaeva2019}%
  \BibitemOpen
  \bibfield  {author} {\bibinfo {author} {\bibfnamefont {L.}~\bibnamefont
  {Isaeva}}, \bibinfo {author} {\bibfnamefont {G.}~\bibnamefont
  {Barbalinardo}}, \bibinfo {author} {\bibfnamefont {D.}~\bibnamefont
  {Donadio}},\ and\ \bibinfo {author} {\bibfnamefont {S.}~\bibnamefont
  {Baroni}},\ }\href {https://doi.org/10.1038/s41467-019-11572-4} {\bibfield
  {journal} {\bibinfo  {journal} {Nat. Commun.}\ }\textbf {\bibinfo {volume}
  {10}},\ \bibinfo {pages} {3853} (\bibinfo {year} {2019})}\BibitemShut
  {NoStop}%
\bibitem [{\citenamefont {Eckert}\ \emph {et~al.}(1977)\citenamefont {Eckert},
  \citenamefont {Thomlinson},\ and\ \citenamefont {Shirane}}]{Eckert1977}%
  \BibitemOpen
  \bibfield  {author} {\bibinfo {author} {\bibfnamefont {J.}~\bibnamefont
  {Eckert}}, \bibinfo {author} {\bibfnamefont {W.}~\bibnamefont {Thomlinson}},\
  and\ \bibinfo {author} {\bibfnamefont {G.}~\bibnamefont {Shirane}},\ }\href
  {https://doi.org/10.1103/physrevb.16.1057} {\bibfield  {journal} {\bibinfo
  {journal} {Physical Review B}\ }\textbf {\bibinfo {volume} {16}},\ \bibinfo
  {pages} {1057} (\bibinfo {year} {1977})}\BibitemShut {NoStop}%
\bibitem [{\citenamefont {Cazorla}\ and\ \citenamefont
  {Boronat}(2017)}]{Cazorla2017}%
  \BibitemOpen
  \bibfield  {author} {\bibinfo {author} {\bibfnamefont {C.}~\bibnamefont
  {Cazorla}}\ and\ \bibinfo {author} {\bibfnamefont {J.}~\bibnamefont
  {Boronat}},\ }\bibfield  {journal} {\bibinfo  {journal} {Rev. Mod. Phys.}\
  }\textbf {\bibinfo {volume} {89}},\ \href
  {https://doi.org/10.1103/revmodphys.89.035003} {10.1103/revmodphys.89.035003}
  (\bibinfo {year} {2017})\BibitemShut {NoStop}%
\bibitem [{\citenamefont {Goldman}\ \emph {et~al.}(1968)\citenamefont
  {Goldman}, \citenamefont {Horton},\ and\ \citenamefont
  {Klein}}]{Goldman1968}%
  \BibitemOpen
  \bibfield  {author} {\bibinfo {author} {\bibfnamefont {V.}~\bibnamefont
  {Goldman}}, \bibinfo {author} {\bibfnamefont {G.}~\bibnamefont {Horton}},\
  and\ \bibinfo {author} {\bibfnamefont {M.}~\bibnamefont {Klein}},\ }\href
  {https://doi.org/10.1103/physrevlett.21.1527} {\bibfield  {journal} {\bibinfo
   {journal} {Phys. Rev. Lett.}\ }\textbf {\bibinfo {volume} {21}},\ \bibinfo
  {pages} {1527} (\bibinfo {year} {1968})}\BibitemShut {NoStop}%
\bibitem [{\citenamefont {Aziz}\ \emph {et~al.}(1979)\citenamefont {Aziz},
  \citenamefont {Nain}, \citenamefont {Carley}, \citenamefont {Taylor},\ and\
  \citenamefont {McConville}}]{Aziz1979}%
  \BibitemOpen
  \bibfield  {author} {\bibinfo {author} {\bibfnamefont {R.~A.}\ \bibnamefont
  {Aziz}}, \bibinfo {author} {\bibfnamefont {V.~P.~S.}\ \bibnamefont {Nain}},
  \bibinfo {author} {\bibfnamefont {J.~S.}\ \bibnamefont {Carley}}, \bibinfo
  {author} {\bibfnamefont {W.~L.}\ \bibnamefont {Taylor}},\ and\ \bibinfo
  {author} {\bibfnamefont {G.~T.}\ \bibnamefont {McConville}},\ }\href
  {https://doi.org/10.1063/1.438007} {\bibfield  {journal} {\bibinfo  {journal}
  {J. Chem. Phys.}\ }\textbf {\bibinfo {volume} {70}},\ \bibinfo {pages} {4330}
  (\bibinfo {year} {1979})}\BibitemShut {NoStop}%
\bibitem [{\citenamefont {Thompson}\ \emph {et~al.}(2022)\citenamefont
  {Thompson}, \citenamefont {Aktulga}, \citenamefont {Berger}, \citenamefont
  {Bolintineanu}, \citenamefont {Brown}, \citenamefont {Crozier}, \citenamefont
  {in~{\textquotesingle}t~Veld}, \citenamefont {Kohlmeyer}, \citenamefont
  {Moore}, \citenamefont {Nguyen}, \citenamefont {Shan}, \citenamefont
  {Stevens}, \citenamefont {Tranchida}, \citenamefont {Trott},\ and\
  \citenamefont {Plimpton}}]{Thompson2022}%
  \BibitemOpen
  \bibfield  {author} {\bibinfo {author} {\bibfnamefont {A.~P.}\ \bibnamefont
  {Thompson}}, \bibinfo {author} {\bibfnamefont {H.~M.}\ \bibnamefont
  {Aktulga}}, \bibinfo {author} {\bibfnamefont {R.}~\bibnamefont {Berger}},
  \bibinfo {author} {\bibfnamefont {D.~S.}\ \bibnamefont {Bolintineanu}},
  \bibinfo {author} {\bibfnamefont {W.~M.}\ \bibnamefont {Brown}}, \bibinfo
  {author} {\bibfnamefont {P.~S.}\ \bibnamefont {Crozier}}, \bibinfo {author}
  {\bibfnamefont {P.~J.}\ \bibnamefont {in~{\textquotesingle}t~Veld}}, \bibinfo
  {author} {\bibfnamefont {A.}~\bibnamefont {Kohlmeyer}}, \bibinfo {author}
  {\bibfnamefont {S.~G.}\ \bibnamefont {Moore}}, \bibinfo {author}
  {\bibfnamefont {T.~D.}\ \bibnamefont {Nguyen}}, \bibinfo {author}
  {\bibfnamefont {R.}~\bibnamefont {Shan}}, \bibinfo {author} {\bibfnamefont
  {M.~J.}\ \bibnamefont {Stevens}}, \bibinfo {author} {\bibfnamefont
  {J.}~\bibnamefont {Tranchida}}, \bibinfo {author} {\bibfnamefont
  {C.}~\bibnamefont {Trott}},\ and\ \bibinfo {author} {\bibfnamefont {S.~J.}\
  \bibnamefont {Plimpton}},\ }\href {https://doi.org/10.1016/j.cpc.2021.108171}
  {\bibfield  {journal} {\bibinfo  {journal} {Comput. Phys. Commun.}\ }\textbf
  {\bibinfo {volume} {271}},\ \bibinfo {pages} {108171} (\bibinfo {year}
  {2022})}\BibitemShut {NoStop}%
\bibitem [{\citenamefont {Simoncelli}\ \emph {et~al.}(2022)\citenamefont
  {Simoncelli}, \citenamefont {Marzari},\ and\ \citenamefont
  {Mauri}}]{Simoncelli2022}%
  \BibitemOpen
  \bibfield  {author} {\bibinfo {author} {\bibfnamefont {M.}~\bibnamefont
  {Simoncelli}}, \bibinfo {author} {\bibfnamefont {N.}~\bibnamefont
  {Marzari}},\ and\ \bibinfo {author} {\bibfnamefont {F.}~\bibnamefont
  {Mauri}},\ }\href {https://doi.org/10.1103/physrevx.12.041011} {\bibfield
  {journal} {\bibinfo  {journal} {Phys. Rev. X}\ }\textbf {\bibinfo {volume}
  {12}},\ \bibinfo {pages} {041011} (\bibinfo {year} {2022})}\BibitemShut
  {NoStop}%
\bibitem [{\citenamefont {Caldarelli}\ \emph {et~al.}(2022)\citenamefont
  {Caldarelli}, \citenamefont {Simoncelli}, \citenamefont {Marzari},
  \citenamefont {Mauri},\ and\ \citenamefont {Benfatto}}]{Caldarelli2022}%
  \BibitemOpen
  \bibfield  {author} {\bibinfo {author} {\bibfnamefont {G.}~\bibnamefont
  {Caldarelli}}, \bibinfo {author} {\bibfnamefont {M.}~\bibnamefont
  {Simoncelli}}, \bibinfo {author} {\bibfnamefont {N.}~\bibnamefont {Marzari}},
  \bibinfo {author} {\bibfnamefont {F.}~\bibnamefont {Mauri}},\ and\ \bibinfo
  {author} {\bibfnamefont {L.}~\bibnamefont {Benfatto}},\ }\href
  {https://doi.org/10.1103/physrevb.106.024312} {\bibfield  {journal} {\bibinfo
   {journal} {Phys. Rev. B}\ }\textbf {\bibinfo {volume} {106}},\ \bibinfo
  {pages} {024312} (\bibinfo {year} {2022})}\BibitemShut {NoStop}%
\end{thebibliography}%

\end{document}